%% file: main.tex
\documentclass{aa} 

\usepackage{upgreek}
\usepackage{amsmath}
\usepackage{float}
\usepackage{xspace}
\usepackage{pdflscape}
\usepackage{graphicx}
\usepackage{txfonts}
\usepackage[dvipsnames]{xcolor}
\usepackage{rotating}
\usepackage{pdflscape}

\usepackage[normalem]{ulem}

\usepackage{multirow}

\usepackage{hyperref}
\hypersetup{
    colorlinks=true,
    citecolor=blue,
    linkcolor=blue,
    filecolor=blue,
    urlcolor=blue
    }

\usepackage[utf8]{inputenc}
\usepackage[T1]{fontenc}

\DeclareUnicodeCharacter{03BB}{\ensuremath{\lambda}}
\DeclareUnicodeCharacter{03B1}{\ensuremath{\alpha}}
\DeclareUnicodeCharacter{03B2}{\ensuremath{\beta}}
\DeclareUnicodeCharacter{0398}{\ensuremath{\theta}}
\DeclareUnicodeCharacter{03C3}{\ensuremath{\sigma}}
\DeclareUnicodeCharacter{03BC}{\ensuremath{\mu}}
\DeclareUnicodeCharacter{03A3}{\ensuremath{\Sigma}}
\DeclareUnicodeCharacter{2212}{--}

\makeatletter
\renewcommand*\aa@pageof{, page \thepage{} of \pageref*{LastPage}}
\makeatother

\include{macros}

\begin{document}

\authorrunning{D.~Demars et al.}
\titlerunning{ENTROPY -- VLT/UVES Hydrogen line series of Delorme 1 (AB)b}

   \title{ExoplaNeT accRetion mOnitoring sPectroscopic surveY (ENTROPY)}
    \subtitle{II.~Time series of Balmer line profiles of Delorme 1(AB)b}


    \author{
        Dorian~Demars
        \inst{\ref{grenoble}}
        \and
        Micka\"el~Bonnefoy\inst{\ref{grenoble}}
        \and
        Catherine~Dougados\inst{\ref{grenoble}}
        \and
        Gayathri~Viswanath
      \inst{\ref{stockholm}}
        \and
        Simon~C.~Ringqvist
        \inst{\ref{stockholm}}
        \and
        Markus~Janson\inst{\ref{stockholm}}
        \and
        Yuhiko~Aoyama\inst{\ref{tsinghua_adv}, \ref{tsinghua_astro}, \ref{tokyo}}
        \and 
        Thanawuth~Thanathibodee\inst{\ref{chula}, \ref{uboston}}
        \and
        Gabriel-Dominique~Marleau\inst{\ref{bern}, \ref{heidelberg}, \ref{duisburg}}
        \and
        Carlo~F.~Manara \inst{\ref{ESO}}
        \and
        Elisabetta~Rigliaco \inst{\ref{inaf}}
        \and
        Judith~Szul\'agyi\inst{\ref{eth}}
        \and
        Aurora~Sicilia-Aguilar\inst{\ref{dundee}}
        \and
        J\'{e}r\^{o}me~Bouvier\inst{\ref{grenoble}}
        \and
        Evelyne~Alecian\inst{\ref{grenoble}}
        \and
        Simon~Petrus\inst{\ref{santiago1}, \ref{santiago2}}
        \and
        Mathis~Houll\'{e}\inst{\ref{grenoble}}
        }

   \institute{
        Universit\'e Grenoble Alpes, CNRS, IPAG, 38000 Grenoble, France
        \label{grenoble}\\
        \email{dorian.demars@virginia.edu, mickael.bonnefoy@univ-grenoble-alpes.fr}
        \and
        Institutionen f\"{o}r astronomi, Stockholms universitet, AlbaNova universitetscentrum, 106 91, Stockholm, Sweden
        \label{stockholm}
        \and
        Institute for Advanced Study, Tsinghua University, Beijing 100084, PR China 
        \label{tsinghua_adv}
        \and
        Department of Astronomy, Tsinghua University, Beijing 100084, PR China 
        \label{tsinghua_astro}
        \and
        Department of Earth and Planetary Science, The University of Tokyo, 7-3-1 Hongo, Bunkyo-ku, Tokyo 113-0033, Japan 
        \label{tokyo}
        \and
        Department of Physics, Faculty of Science, Chulalongkorn University, 254 Phayathai Rd., Patumwan, Bangkok 10330, Thailand
        \label{chula}
        \and
        Institute for Astrophysical Research and Department of Astronomy, Boston University, 725 Commonwealth Ave., Boston, MA 02215, USA
        \label{uboston}
        \and
        Division of Space Research \&\ Planetary Sciences, Physics Institute, University of Bern, Sidlerstr.~5, 3012 Bern, Switzerland%
        \label{bern}
        \and
        Max-Planck-Institut f\"ur Astronomie, K\"onigstuhl 17, 69117 Heidelberg, Germany
        \label{heidelberg}
        \and
        Fakult\"at für Physik, Universit\"at Duisburg--Essen, Lotharstra\ss{}e 1, 47057 Duisburg, Germany
        \label{duisburg}
        \and
        European Southern Observatory, Karl-Schwarzschild-Stra\ss{}e 2, 85748 Garching bei München, Germany
        \label{ESO}
        \and
        INAF/Osservatorio Astronomico di Padova, Vicolo dell'Osservatorio 5, 35122 Padova, Italy
        \label{inaf}
        \and
        Department of Physics, ETH Z\"urich, Wolfgang-Pauli-Str.~27, CH-8093, Z\"urich, Switzerland
        \label{eth}
        \and
        SUPA, School of Science and Engineering, University of Dundee, Nethergate, DD1 4HN, Dundee, UK
        \label{dundee}
        \and
        Millennium Nucleus on Young Exoplanets and their Moons (YEMS), Santiago, Chile
        \label{santiago1}
        \and
        Instituto de Estudios Astrof\'isicos, Facultad de Ingenier\'ia y Ciencias, Universidad Diego Portales, Av.~Ej\'ercito 441, Santiago, Chile
        \label{santiago2}
         }
        
   \date{Received \dots; accepted \dots}

  \abstract
    {Accretion processes in the planetary-mass regime are still poorly constrained, yet they impact strongly the formation and evolution of planets and the composition of circumplanetary disks (CPDs).} 
   {
   We investigate the resolved Balmer hydrogen emission-line profiles and their variability timescales in the $\sim13$~\Mjup, 30--45~Myr old companion \delorme to derive constraints on the accretion mechanism at play.}
   {With VLT/UVES, we collect 31~new epochs of high-resolution optical (330--680~nm) spectra of the companion at $R=50,000$, probing variability on timescales of hours to years.
   We study the companion's \HI\ emission line shape and flux variability and compare them to two proposed line origins: magnetospheric accretion funnel and localized accretion shock.}
   {We detect 
   \HI\ Balmer lines from \Halpha up to H10 (6564--3799~\AA) as well as the UV continuum excess, signs of ongoing accretion. All lines and the UV excess are variable. The \HI\ lines can be decomposed into two static components that vary only by their flux.
   The broader component in velocity correlates strongly with the UV excess, and its profile is qualitatively reproduced by magnetospheric accretion funnel models but clearly not by shock models. 
   With its strong relative variability, this broad component almost entirely explains the shape variability of the line profiles. The second, narrower, component correlates less with the UV excess and is best reproduced by shock-emission models. Its strong absolute variability makes it responsible for most of the line flux variability. Overall, the lines have low relative flux variability on the hour timescale, but up to $\sim100$~\% on timescales of weeks and beyond, a behavior similar to T Tauri stars.}
   {The properties of the broad component of the \HI lines strongly support magnetospheric accretion. The narrow component could come from an accretion shock but also chromospheric activity. Higher-cadence observations could search for rotational modulations to constrain the object's rotational period and the exact accretion flow geometry.}


   \maketitle
%




\nolinenumbers

\section{Introduction}
\label{sec:introduction}



A few tens of planetary-mass objects (PMOs) and planetary-mass companions (PMCs) have been found to display emission lines at optical and near-infrared wavelengths (e.g., \Halpha\ at 6563~\AA, \PaBeta\ at 1.282~\um; see \citealp{betti_comprehensive_2023} for a recent compilation) indicative of active accretion. These objects are members of star forming regions or nearby-young associations, with masses down to 5~\MJup\ \citep{lodieu_USco_2018, 2023ApJ...949L..36L}, and are found both as free-floating objects \citep[e.g.,][]{muzerolle_measuring_2005, joergens_ots_2013,lodieu_USco_2018}, or as companions to single  stars or binaries of various masses \citep{petrus_new_2020, zhang_13co-rich_2021, chinchilla_2M0249_2021}.
Some are embedded in circumstellar disks \citep[the emblematic and so far unique case of PDS~70~b and c;][]{haffert_PDS70_2019}, while others are found outside these primordial disks \citep[e.g.,][]{seifahrt_near-infrared_2007, santamaria-miranda_accretion_2018,santamaria-miranda_accretion_2019_erratum}.

Recent efforts to derive accretion rates suggest that the \Mdot--M$_*$ correlation derived in the stellar regime flattens out in the planetary-mass range \citep{ayliffe_gas_2009,betti_comprehensive_2023}, therefore suggesting different mechanisms driving accretion in that mass interval. \cite{betti_comprehensive_2023} also found a steepening of the \Mdot--$M$ relation with age in the brown-dwarf mass regime, down to the deuterium-burning limit (approximately 13.6~\Mjup; \citealp{spiegel_2011}), that increases with age above 3~Myr. \cite{almendros-abad_2024_mass_accrate_relationships} provided  similar hints at younger ages ($<3$~Myr).

A few planetary-mass objects in nearby young kinematic associations, have recently been found displaying bright emission lines linked to accretion \citep[2MASS~J02265658-5327032, \delorme, 2MASS~J0249-0557~c;][]{2016ApJ...832...50B,  2020A&A...638L...6E, chinchilla_2M0249_2021}. Their much later estimated ages (20--45~Myr) and the firm detection of disk excess emission indicative of circumplanetary disk (CPD) for one of them (2MASS~J02265658-5327032) suggests that accretion can be sustained longer than on stars in the planetary-mass range.

Emission line properties (flux, line profiles and ratio) can be compared to predictions from nascent accretion models in the planetary mass range. Simulations of planets embedded in circumstellar disks suggest  polar inflow from the circumstellar disk onto the CPD or directly at the planet's poles \citep{2014ApJ...782...65S}, creating line-emitting regions at localized accretion shocks. Shock models have since been developed to predict the flux of the Balmer and Paschen line series \citep{aoyama_theoretical_2018}. Their predicted \Halpha/\Hbeta ratios do not match observational constraints for PDS~70~b \citep{2019ApJ...885L..29A, 2020AJ....159..222H}. \citet{2020AJ....159..222H} proposed extinction along the line of sight caused by  circumstellar or circumplanetary material to match the observations. 
The geometry of the material and gas dynamics around accreting companions is likely to be  complex \citep{krapp_criterion_2024} and may depend on the circumstellar disk properties \citep{lega_simu_2024}. Its effects on the line properties start to be investigated theoretically \citep{2020ApJ...902..126S, marleau_accreting_2022} and has been proposed as a possible explanation of the low detection yield of deep-imaging campaigns seeking for Hydrogen line emissions from protoplanets nested within circumstellar disks \citep{2019A&A...622A.156C, 2020JATIS...6d5004U, 2020A&A...633A.119Z, 2022A&A...668A.138H, 2023AJ....165..225F, 2024AJ....168...70C}.

Instead of PPD-to-planet mass transfer, planets may also accrete directly from their surrounding CPD, either through boundary-layer \citep{szulagyi_accretion_2014,szulagyi_circumplanetary_2016,2016ApJ...819L..14O, 2021ApJ...921...10T}  or magnetospheric accretion.
 Magnetospheric accretion models of T-Tauri stars have very recently been adapted to predict the line emissions of planetary-mass objects \citep{2019ApJ...885...94T}.

Emission-line variability is ubiquitous in the stellar mass range  and can inform on the accretion mechanisms at play \citep[for a review, see][]{2023ASPC..534..355F}. It is presently poorly characterized for planetary-mass companions in spite of its importance for optimizing detection campaigns targeting protoplanet line emissions. \cite{2017AJ....154...26W} found the \PaBeta\ line of DH~Tau~b \citep[a $\sim1$~Myr old, $\sim15$~\MJup\ companion;][]{2005ApJ...620..984I} to vanish within five weeks, although the observations were based on only four epochs. \cite{2021AJ....161..244Z} set an upper limit of 30\% for the variability of the \Halpha\ line of PDS~70~b over timescales of days to months,  based on 8 epochs. While the line of PDS~70~c was not detected in the 2020 data, it was detected later in 2024, reaching approximately twice the 2020 upper limit \citep{zhou_evidence_2025}.
\cite{2023AJ....166..143W} conducted a six-night monitoring campaign of FU~Tau~B (19~\MJup) to search for rotational modulations of \Halpha as suggested by magnetospheric accretion models, but their conclusions were limited by the sensitivity of the data.
Additionally, \cite{demars_emission_2023} collected J-band spectra (1.1-1.35$\mu$m) of the 14~\Mjup\ and 10--30~\MJup\ companions \gsc\ and \gqlupb\ probing the variability of the \PaBeta\ line over timescales ranging from minutes to a decade.
They find variability on all these timescales but a higher amplitude of variability on timescales greater than typical rotation periods (a few hours to several days on companions), a behavior commonly observed in T Tauri stars \cite{costigan_temperaments_2014,venuti_multicolor_2021,zsidi_accretion_2022}.

Line profiles can also inform on the gas kinematics, properties of the accreting material near the shock front, line-of-sight extinction, and on the accretion mechanisms \citep{marleau_accreting_2022}. The few single-epoch spectra at high spectral resolution ($R>20,000$) gathered on bright 1--10~Myr and 10--20~\Mjup\ free-floating objects reveal significant diversity in line shapes \citep{muzerolle_measuring_2005, 2005ApJ...626..498M}, suggesting a complex picture. 

The characterization of emission line profiles on planetary-mass companions is just beginning. \cite{santamaria-miranda_accretion_2018,santamaria-miranda_accretion_2019_erratum} obtained medium-resolution ($R_{\lambda}$ from 3890 to 5400) spectra of the line emission produced by the planetary-mass companion SR~12~c.
The \Halpha~line profile showed a possibly asymmetrical peak structure that they attributed to magnetospheric accretion.  
\cite{demars_emission_2023} modeled \PaBeta\ emission line profiles obtained at $R_{\lambda}=1800-2360$ of \gsc\ and \gqlupb\ with both 1D shock and magnetospheric accretion models and found that the blue-shifted profiles of \gqlupb\ could only be reproduced by magnetospheric accretion models while those of \gsc\ can be reproduced by both sets of models.
Finally, \cite{ringqvist_resolved_2023} presented a near-UV spectrum at $R\sim50,000$ of the companion \delorme\ showing complex profiles that can be modeled using multiple Gaussian components. Comparison to shock models implies a small line-emitting area of $\sim 1\%$ relative to the planetary surface that is consistent with magnetospheric accretion. This motivates a more in-depth characterization of these objects to confirm whether this mechanism is truly operating and universal in the planetary-mass regime.

We present VLT/UVES (Very Large Telescope / Ultraviolet and Visual Echelle Spectrograph) observations obtained as part of the ExoplaNeT accRetion mOnitoring sPectroscopic surveY (ENTROPY) survey which aims to characterize accretion processes of planetary mass objects through the monitoring of their emission lines at high spectral resolution. The first study presented two epochs of observations of the 7--21~\Mjup free-floating object 2MASS J11151597+1937266, showing \Halpha line profiles compatible with magnetospheric accretion and variability between the two epochs of observations \citep{viswanath2024_entropy_I}.
 We focus in this study on follow-up observations of \delorme\ using an identical technical setup as in \cite{ringqvist_resolved_2023} to investigate variability in line profiles, fluxes, and ratios in an attempt to constrain the accretion mechanism operating in this object. 

This paper is organized as follows. We describe our target of interest in Sect. \ref{sec:td}. We detail in Sect.~\ref{sec:observations} the observations and data reduction procedures. We present in Sect.~\ref{sec:results} the main spectral features of \delorme, the line variability and our decomposition. We model the lines and their sub-components with  accretion models in Sect.~\ref{sec:modeling}. We discuss the results in Sect.~\ref{sec:discussion}, and summarize our findings in Sect.~\ref{sec:conclusion}.

\section{Target description}
\label{sec:td}
\delorme\ (also known as 2MASS J01033563-5515561 (AB)b) is a wide-separation $\sim13$~\Mjup\ companion \citep[$\sim1.77''$, i.e., 84~au;][]{delorme_direct-imaging_2013}, orbiting a tight pair ($\sim0.25''$, i.e., 12~au) of M5--M6 \citep{2014AJ....147..146K, eriksson_strong_2020} stars (0.19~\Msun\ and 0.17~\Msun) displaying emission lines as well. \delorme\ has a reported $\sim\textrm{L0}$ spectral type \citep{eriksson_strong_2020}. It is the nearest of all presently known accreting PMCs, located at $47.2\pm3.1$~pc \citep{riedel_solar_2014}. The binary stars are proposed members of the Tucana-Horologium or Carina associations \citep{riedel_solar_2014}, with estimated ages of $\sim30$--45~Myr old \citep{gagne_banyan_2018, 2023MNRAS.520.6245G}. While relatively young, \delorme is much older than commonly studied young accreting companions.  
  
\cite{eriksson_strong_2020} first reported strong \Halpha, \Hbeta, helium and Ca\,\textsc{ii} emission lines in the medium-resolution ($R\approx1740$--3750) VLT/MUSE optical (480--930~nm) spectrum of \delorme. 
Using stellar relations \citep{alencar_accretion_2012,alcala_x-shooter_2017}, they derived accretion rates of $\log(\Mdot/\MdotUJ) \sim -9.4$ to $-9.8$.  
\PaBeta, \PaGamma\ and \BrGamma\ lines were then identified on that target \citep{betti_near-infrared_2022,betti_erratum_2022}, yielding accretion rates of $\log{\Mdot} \sim -8.2$ to $-8.8$~\MdotUJ using the same scalings, and $\log{\Mdot} \sim -7.30 \pm 0.30$~\MdotUJ using \Mdot--L$_{\textrm{line}}$ relationships from \cite{aoyama_theoretical_2018}. The higher end of the Balmer series (\Hgamma, \Hdelta, \Hepsilon, ...) were then detected in \cite{ringqvist_resolved_2023} from which they derived an accretion rate of $\log{\Mdot} \sim -7.70$~\MdotUJ using planetary scaling, compatible with the value derived by \cite{betti_erratum_2022} from near-infrared (NIR) lines. The line profiles constrain the dynamical mass of the companion to $13\pm5\Mjup$. \cite{betti_erratum_2022} suggests the presence
of a long-lived ‘Peter Pan’ disc, which may explain why the planet is still actively accreting. Both dust and gas signatures have now been detected around the companion by JWST (priv. comm.) and confirm the existence of a mass reservoir. 

Smoothed-particle hydrodynamic (SPH) simulations of the gas in the system were conducted to understand the formation mechanism of \delorme\ \citep{2024MNRAS.533.2294T}, but they could not account for both the accretion rate onto the companion and its location in the system.
\delorme\ is a unique testbed for understanding the formation mechanisms and accretion in the planetary-mass regime. With its bright lines, and a reasonable angular separation, \delorme\ allows for collecting spectra with seeing-limited, slit-based spectrographs, such as UVES. 

\begin{figure*}[ht]
    \centering
    \begin{minipage}[c]{0.5\textwidth}
        \centering
        \includegraphics[width=\linewidth]{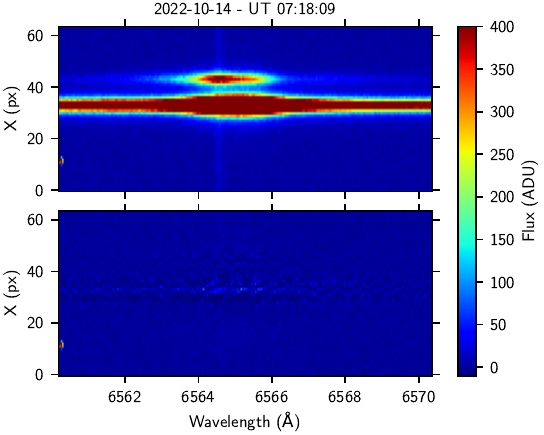}
    \end{minipage}%
    \hfill
    \begin{minipage}[c]{0.5\textwidth}
        \centering
        \includegraphics[width=\linewidth]{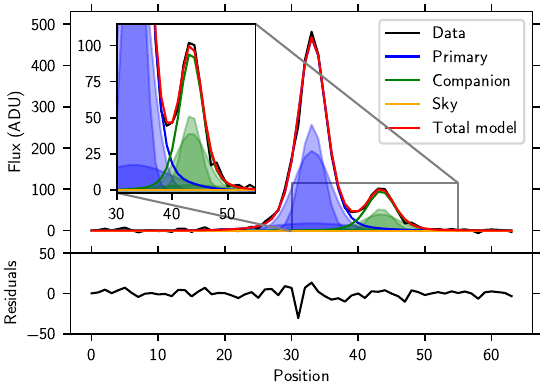}
    \end{minipage}
    \caption{Illustration of the extraction process for one of the exposures, in the RED arm. Left: 2D reconstructed slit, before (top) and after (bottom) subtraction of the PSF models for both the primary and companion. Right: illustration of the extraction process at a single wavelength bin within \Halpha. The blue and green shaded regions are the various Gaussian components used in the PSF fitting process for the primary and companion respectively (3 Gaussian components here for the RED arm). In both cases, the lower panel shows the modeling residuals.}
    \label{fig:extraction}
\end{figure*}

\section{Observations and data reduction}
\label{sec:observations}
\delorme\ was observed from October 13, 2022 to January 2, 2023 -- approximately one year after the observations presented in \cite{ringqvist_resolved_2023} -- using the UVES instrument \citep{2000SPIE.4008..534D} at VLT/UT2.
The observations probe timescales of $\sim$20 minutes (time between successive exposures), days, months and a year.
The light was split with a dichroic (DIC\#1, 390+580~nm) and sent into two arms (RED and BLUE) containing 0.8$''$ wide slits and echelle spectrographs which yielded spectra at resolving powers $R_{\lambda}\approx 50 000$. The RED arm has a mosaic of two chips with a physical gap of 96mm.  The final spectra span 3300--4520~\AA\ in the BLUE arm and 4780--5750~\AA\ and 5830--6800~\AA\ for the two detectors of the RED arm. The slits were aligned along the position angle of the companion to place both the unresolved pair of stars and \delorme\ in it. The setup was identical to the one used by \cite{ringqvist_resolved_2023}. Table~\ref{tab:obs} gives the observation log.

The data were processed with the release 5.10.13 of the ESO/UVES data handling pipeline \citep{2000Msngr.101...31B} through the ESO automatic reduction workflow with EsoReflex \citep{esoreflex} in default settings. The pipeline uses day-time calibration exposures to correct the science exposures from the bias and dark current levels,  pixel-to-pixel sensitivity variations, and the blaze function. Orders are located and a wavelength calibration is performed for each one using arc lamp exposures. The telescope, instrument, and detector efficiencies are estimated per wavelength using the observations of a standard star taken the closest in time to our observations. The pipeline produces 2D images -- hereafter ``trace'' -- of the dispersed slit allowing for the spatial extraction of the companion signal at each wavelength. We present in Section~\ref{sec:obs:extraction_process} the extraction process, while Section~\ref{sec:flux_calibration} details how the spectra were flux-calibrated.

\begin{figure*}[t]
    \centering
    \includegraphics[width=\textwidth]{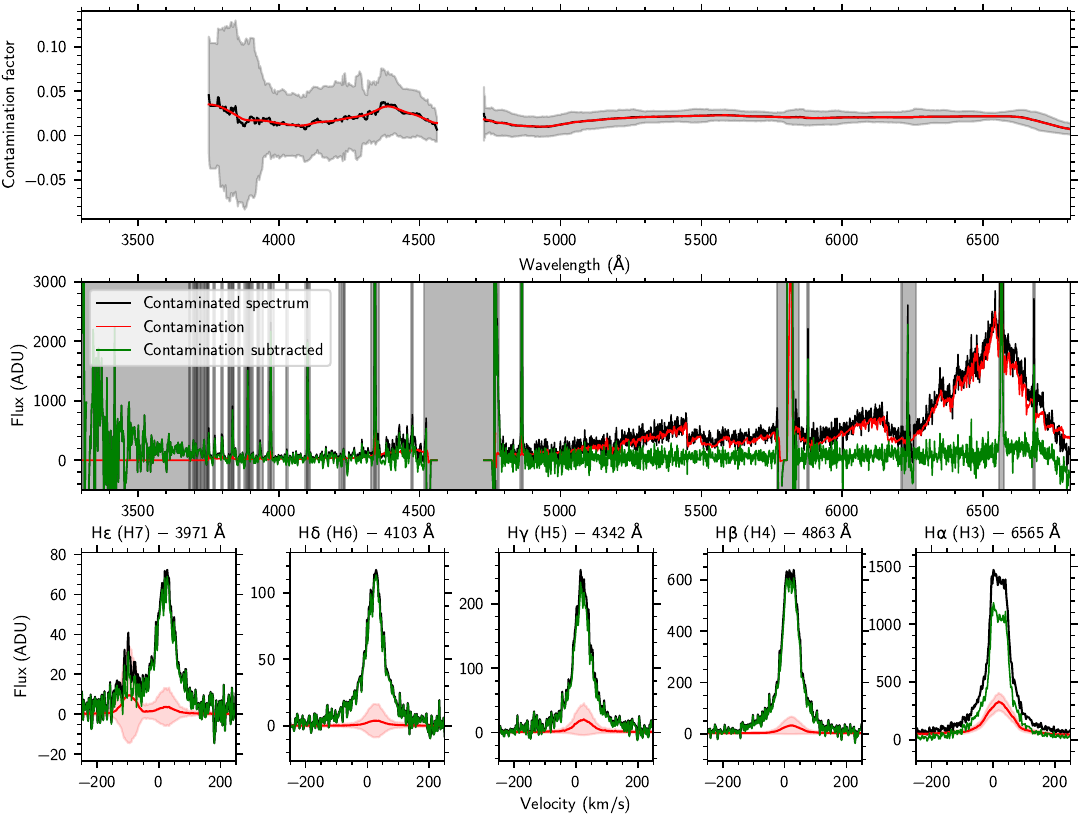}
    \caption{Illustration of the residual contamination correction, for the exposure 2022-11-05 UT 01:22:25. This corresponds to the case of highest contamination.
    Top panel: residual contamination factor (black) smoothed with a 10-px wide moving Gaussian box (red) and its associated error bars (black shaded region).
    Middle panel: companion spectrum as the output of the extraction process (black) with the contamination contribution (red) and contamination-corrected companion spectrum (green). All black-shaded regions were excluded from the fit, they correspond to emission lines and detector edges.
    Third row: contamination contribution within emission lines (same colors).
    The contamination removal process is able to remove most of the residual primary flux in the companion spectrum, which accounts for up $\sim10$~\% of the lines flux depending on the exposure.}
    \label{fig:contamination}
\end{figure*}

\begin{figure*}[t]
    \centering
    \includegraphics[width=\linewidth]{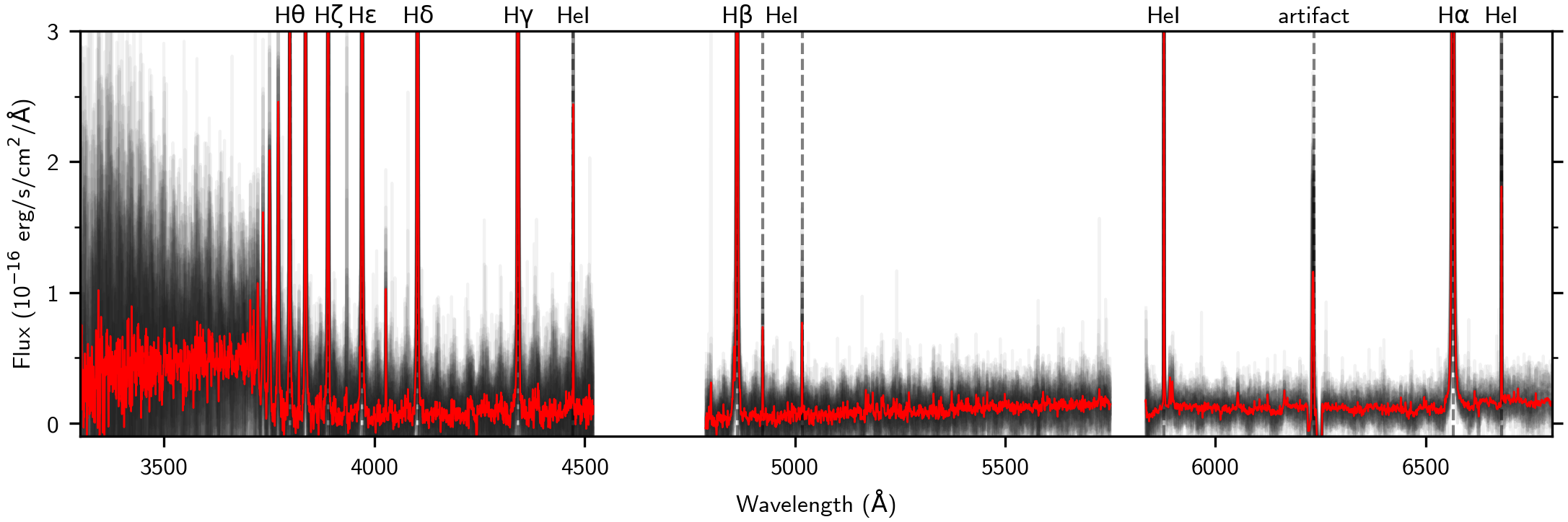}
    \caption{Final spectra over the full spectral range. All \delorme\ spectra, overlaid in black, with the mean spectrum in red. All spectra were smoothed to $R=5000$ for clarity. The UV excess of \delorme\ is clearly visible below $\sim$3700~\AA, as well as the generally flat continuum shape.
    The line marked as an artifact is due to a bad pixels row of the detector.}
    \label{fig:full_spectrum}
\end{figure*}

\begin{figure*}[t]
    \centering
    \includegraphics[width=\linewidth]{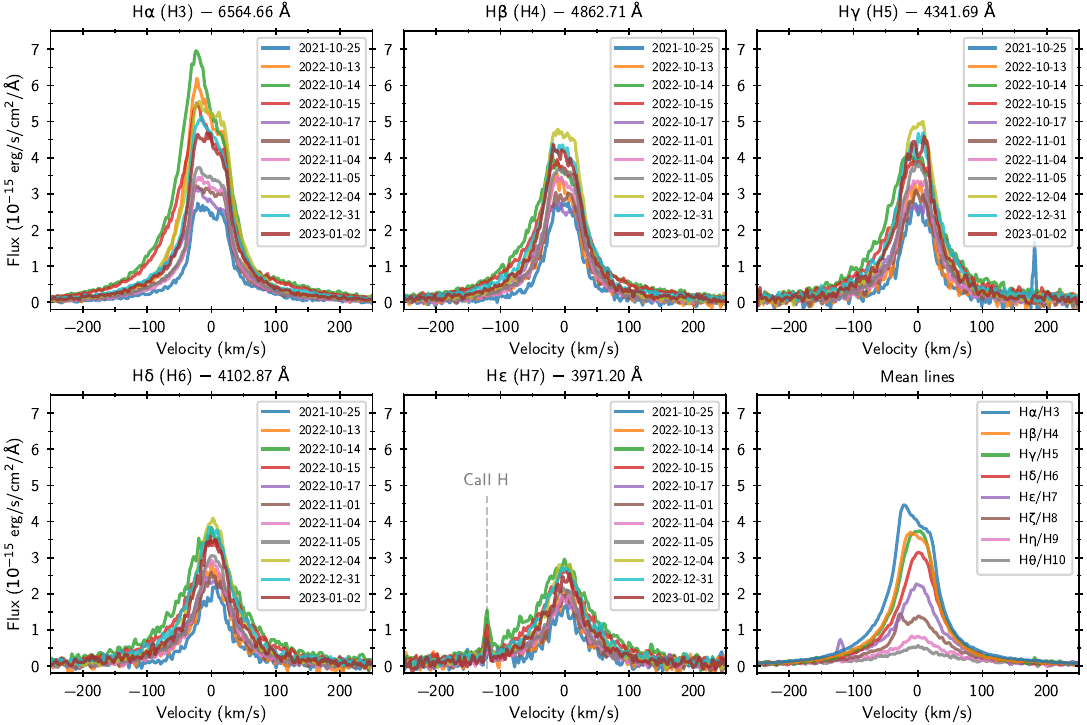}
    \caption{Balmer lines of \dlrb. Each panel shows the mean line of each observing epoch (11 epochs), from \Halpha\ to \Hepsilon. The last panel shows the mean line (over all epochs) for each transition order, all the way up to H$_\uptheta$/H10.
    All lines are plotted after subtracting the local continuum baseline, averaged between $(-750;-400)$ and $(+400;+750)$~\kms. The lines show great profile and amplitude variability across epochs.}
    \label{fig:all_lines}
\end{figure*}

\subsection{Spectra extraction}
\label{sec:obs:extraction_process}

The companion is visible at the location of emission lines through the 2D traces at all epochs, but is affected by contamination from the halo of the host binary.
This is shown in Fig.~\ref{fig:extraction}: the left panel shows the signal around \Halpha, while the right panel shows the signal in a single wavelength-bin centered on \Halpha.
Such contamination is common in direct imaging (especially in seeing-limited observations) and must be accounted for to prevent blending of the companion and primary signals.

We developed a two-stage process to subtract the halo of the host binary:
\begin{itemize}
    \item[--] Multi-Gaussian fitting and subtraction of the primary signal at each wavelength within the trace (one-dimention flux profile) (Sect.~\ref{sec:obs:psf_fitting}),
    \item[--] Estimation and removal of any residual stellar contamination in the extracted companion spectrum that the first step may have missed (Sect.~\ref{sec:obs:contamination}).
\end{itemize}

\subsubsection{PSF fitting}
\label{sec:obs:psf_fitting}

The subtraction of the binary signal was performed separately on the BLUE and RED arms.
The data was binned by a factor of 41 in wavelength to increase the S/N in each bin, yielding an effective spectral resolution of $\sim1200$. These binned spectra are only used for the modeling of the PSF. All analysis is performed on the full resolution spectra (R$\sim$50\,000).

The extraction follows an iterative process in a CLEAN-like approach \citep{1974A&AS...15..417H}: a monochromatic 1D PSF model is fitted iteratively on the primary and companion. The companion+primary model is subtracted to the data and we repeat the PSF construction until the residuals are minimized (least-square). We reached convergence in $\sim 4$ iterations.

The initial PSF model is obtained following a method adapted from \cite{hynes_optimal_2002}. They proposed two astronomical signals observed with a wide-slit instrument could be disentangled by fitting a PSF model at all wavelength bins, and regularizing its parameters over the spectral dimension.
We used a PSF model made of one Gaussian for the BLUE arm, and three Gaussians for the RED arm, with identical centroid but varying contrasts and widths (see Fig.~\ref{fig:extraction}, right panel). The difference is due to the lesser S/N in the BLUE arm, which did not require a more complex model.

A regularization of the PSF model is introduced along the wavelength axis, fitting the wavelength-dependence of the gaussian parameters with a polynomial of degree 1 for the BLUE arm and 5 for the RED arm.
The difference in degree is due to both the lesser S/N in the BLUE arm, and the broader wavelength coverage of the RED arm which required a higher order polynomial to properly capture the chromatic dependence of the parameters. 

The contribution of the sky emission and residual background from the detector were accounted for in the fit at each wavelength  as a constant baseline. Much like the normalization factor of the primary and companion, they were never regularized because they directly represent their respective spectrum. The separation between the primary and the companion was set as a detector-dependent constant, obtained from their respective measured position within the bright emission lines.

Finally, we went back to the un-binned traces and used the determined wavelength-dependent PSF model to fit both the primary and companion simultaneously in each wavelength bin  and retrieve their individual spectra. The final error bars on the extracted spectra are computed from the root-mean-square of the residuals within the PSF full-width-at-half-maximum at the position of each object.

\begin{figure*}[t]
    \includegraphics[width=\linewidth]{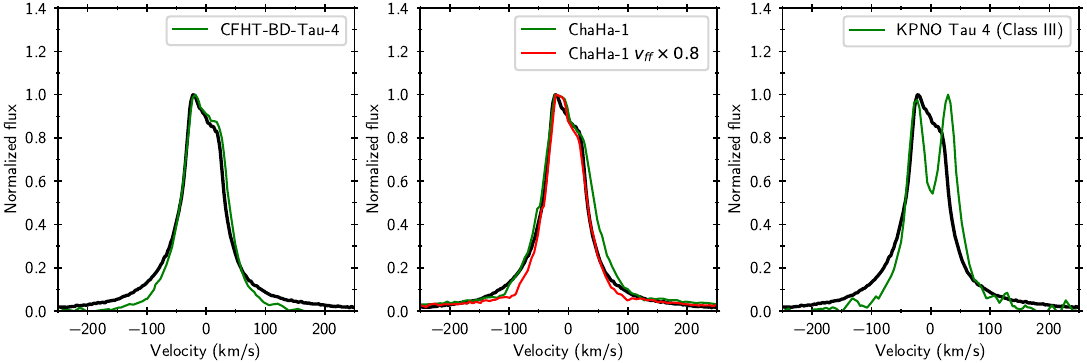}
    \caption{Comparison of the mean \Halpha line profile of \delorme\ to that of two young sub-stellar objects with disks (left: CFHT~BD~Tau~4, center: ChaHa-1) and of a Class III 15~\Mjup brown dwarf (right: KPNO Tau 4). We scaled (in red) the profile of Ca Ha 1 to the free-fall velocities expected for \delorme. The profiles are similar to that of CFHS~BD~Tau~4 and ChaHa-1, but not to that of KPNO~Tau~4.}
    \label{fig:empirical}
\end{figure*}

\subsubsection{Residual contamination}
\label{sec:obs:contamination}

Despite the apparent quality of the extraction in most epochs, we found the PSF to be slightly asymmetric, a feature reported in UVES's manual. In fact, we found this asymmetry to vary between epochs. This leads to a residual contamination of the primary in the companion spectrum, particularly during epochs with pronounced asymmetry. We note that we did not identify any correlation between seeing and the PSF asymmetry.

Figure~\ref{fig:contamination} shows, in the middle panel, one epoch where the companion spectrum is strongly contaminated by the star. This is apparent from the spectral features that strongly follow that of the primary (see Fig.~\ref{fig:appendix:prim_overlap} for the full primary spectrum). The residual contamination was subtracted as a post-processing step, by fitting the companion's local continuum with the primary spectrum, and an offset that accounts for both the contrast and baseline continuum shape.

The contamination in each exposure is modeled as:
\begin{equation}
    F(\lambda) = k \cdot M(\lambda) + \alpha(\lambda) \cdot P(\lambda),
\end{equation}
where
\begin{itemize}
    \item[--] $F(\lambda)$ is the extracted spectrum of the companion for that exposure,
    \item[--] $P(\lambda)$ is the spectrum of the primary for that exposure,
    \item[--] $M(\lambda)$ is a master spectrum of the companion, computed as the average of the epochs which are least affected by contamination,
    \item[--] $k$ is a general scaling factor to account for that exposure's calibration,
    \item[--] $\alpha(\lambda)$ is the contamination factor from the primary to the companion.
\end{itemize}
The algorithm solves for $k$ and $\alpha(\lambda)$. Because it involves both a general scaling factor ($k$) and a wavelength-dependent parameter ($\alpha(\lambda)$), the fit is performed simultaneously on all wavelength bins. Each $\alpha(\lambda)$ is computed by solving the equation within a $\pm 5000$~\kms\ window centered on $\lambda$.

The fit was performed using \texttt{scipy}'s implementation of the limited-memory Broyden--Fletcher--Goldfarb--Shanno (L-BFGS-B) algorithm \citep{lbfgsb1995,lbfgsb1997,2020SciPy-NMeth}.
To avoid over-fitting noise features, the final $\alpha(\lambda)$ parameter is smoothed with a Gaussian kernel on a 10-px window (see Fig.~\ref{fig:contamination}, first panel, red line).
The 1$\sigma$ error bars were computed by setting $k$ to the best fit value, and computing the $\chi^2_r$ map for all $\alpha(\lambda)$ values. Then, the $\chi^2_r +1$ contours serve to compute the location of 1$\sigma$ error values on the $\alpha(\lambda)$ parameters. This is performed independently at each wavelength. The error-bars are shown as the shaded region in Fig.~\ref{fig:contamination}, first panel.

The resulting contamination-free spectrum is shown in Fig.~\ref{fig:contamination}, second panel, in green, following subtraction of the contamination component (red) from the extracted spectrum (black). The contribution of the contamination to the emission lines is shown in the last row of Fig.~\ref{fig:contamination}.

Contamination correction was not applied below $\sim3700$~\AA, due to the UV excess of the companion which introduces significant continuum variability that the master spectrum $M(\lambda)$ cannot reproduce. This does not have a significant impact on the results as the contrast between the primary and the companion is much better at these wavelengths, which translates into a minor contamination relative to the companion intrinsic flux. It should also be noted that while the contamination factor error-bar increases at lower wavelengths, this is only because the flux of the binary decreases: the "true" error on the contamination stays mostly constant, but the error-bar on the contamination factor increases to make up for the decrease of binary flux.

\subsection{Absolute flux and spectral calibration}\label{sec:flux_calibration}


All spectra were corrected for barycentric velocity (BERV) at the time of observations, and radial velocity of the Delorme~1 system. We used the radial velocity shift of $7.3 \pm 2.6$~km/s \citep{2014AJ....147..146K}. The orbital velocity of \delorme is unknown, so we did not correct for it. However, at its projected separation (88~AU), and given the mass of the central binary, even if the planet was seen along its orbital plane, the keplerian velocity for circular orbit would be at most $\sim2$~\kms.

The slit width ($0.8''$) of UVES can lead to slit losses depending on the seeing. However, the binary is known to display very little flux variability, both in the NIR \citep[$\sim$3\%;][their Fig.~20]{bowler_rotation_2023} and in the optical \citep[GAIA DR3, $\sim$1\%;][]{gaia_collaboration_gaia_2018,gaia_collaboration_gaia_2023} and therefore provides a reliable baseline for flux calibration of the companion spectra. We therefore normalized all spectra of the binary to the same flux level around \Halpha\ to construct a mean spectrum of the binary, which was then scaled to the GAIA DR3 \GBP magnitude ($\GBP=15.690\pm0.005$~mag) to provide a spectro-photometric calibration of each individual epoch. The \GBP filter is only slightly broader (by $\sim$50~\AA) in the near-UV part but benefits from multiple epochs of observations obtained at a high accuracy level, making it particularly suitable for absolute flux calibration.

There are no apparent discrepancies between the spectral slopes of the spectra of the primary, as shown in Fig.~\ref{fig:appendix:prim_overlap} from one epoch to the other. There are no correlations between the spectral slope and seeing, airmass, or other observational parameters (note that the airmass correction is automatically performed by the pipeline), which implies no strong systematics from epoch to epoch. We also find a good match of the binary mean spectrum with the spectra of young PMS stars from \cite{manara_x-shooter_2013} (Appendix~\ref{sec:fluxcal}) supporting the lack of significant bias in our data.

The spectra of Delorme 1 AB are globally consistent with flux densities corresponding to published APASS $B$, SkyMapper $v$, SDSS $g$, APASS $V$, and SDSS $r$ magnitudes of Delorme 1 AB \citep{2016yCat.2336....0H, onken_skymapperDR4_2024, 2022ApJS..259...35A} converted to flux densities using the VOSA tool \citep{2008A&A...492..277B} and SVO Filter Profile Service \citep{2012ivoa.rept.1015R, 2020sea..confE.182R}. We find a slight shift of the APASS $B$ and SkyMapper $v$ photometry with respect to the flux level of the BLUE arm spectra although we refrained from using that photometry for additional corrections because the star is either accreting or active and that part of the photometry might be variable as well (especially given that the GAIA \GBP filter spans a broader wavelength range and is more stable across epochs). 

The mean MUSE spectrum of Delorme 1 AB published in \cite{eriksson_strong_2020} is significantly brighter than the whole set of reference photometry and our spectra. More details on the comparison can be found in Appendix \ref{sec:fluxcal}.

\section{Results}
\label{sec:results}


\subsection{The spectrum of \delorme}
The derived spectrum of \delorme\ is shown in Fig.~\ref{fig:full_spectrum}, with all epochs overlaid in black, and the mean smoothed spectrum ($R=5000$) in red. The mean spectrum of \delorme\ resembles that of SR~12~c, a young ($\sim$2~Myr) companion with a similar quoted spectral type \citep[XSHOOTER;][]{santamaria-miranda_accretion_2018,santamaria-miranda_accretion_2019_erratum}. We find that \delorme\ has a GAIA \GBP magnitude of $\GBP =20.6 \pm 0.2$. It should be noted that the apparent spread of the black shaded region only represents the S/N of individual spectra. The noise is much higher in the BLUE arm (lower wavelengths) than it is in the RED arm (higher wavelengths). Continuum variability is observed (see Sect.~\ref{sec:continuum_variability}), but is not obvious at R = 5\,000 in Fig.~\ref{fig:full_spectrum}.

A UV excess is evident at wavelengths below $\sim$3600~\AA\ (Fig.~\ref{fig:full_spectrum}). In T Tauri stars, UV excesses are believed to originate from the accretion shock at the object surface. Such a feature is commonly interpreted as a signpost of ongoing accretion \citep{valenti_t_1993,gullbring_disk_1998,herczeg_uv_2008,ingleby_accretion_2013,hartmann_accretion_2016}. Only a handful of PMCs have shown detectable UV excesses to date \citep{zhou_accretion_2014,zhou_hubble_2021}.

\subsection{H {\sc i} Balmer emission lines}

All Balmer lines of \dlrb\ can be found in Fig.~\ref{fig:all_lines} and lines of the host binary are shown in Fig.~\ref{fig:appendix:primary_lines}.
The \Halpha, \Hbeta, \Hgamma, and \Hdelta\ line profiles for each individual exposure (companion only) can be found in Fig.~\ref{fig:appendix:all_halpha_lines}, \ref{fig:appendix:all_hbeta_lines}, \ref{fig:appendix:all_hgamma_lines}, \ref{fig:appendix:all_hdelta_lines}, along with the standard deviation of the lines around the mean line. The integrated line fluxes are given in Table.~\ref{tab:full_lines_fluxes}.

While higher-order lines are quite symmetrical, the \Halpha\ and \Hbeta\ lines have asymmetrical profiles with extended wings (up to $\pm500$~\kms) and a core with a tilted plateau profile.
Although this could reflect partial saturation in the line emission, we instead explain it through a 2-components decomposition (see Sect.~\ref{sec:decomposition}).
This plateau is most tilted on epochs 2022 October 13, 14 and 15 and coincides with an extended blue wing seen on October 14 and 15. This epoch coincides with an outburst of accretion, as seen from the increased UV excess flux at the same dates (see Sect.~\ref{sec:decomposition:results_2}).
The first epoch of our series (2021-10-25) corresponds to the one already published by \cite{ringqvist_resolved_2023}. They had only presented the BLUE arm, which covers \Hgamma and higher order lines. Here, we introduce \Halpha\ and \Hbeta\ for the first time.


\subsection{Comparison to free-floating objects}
\label{sec:freefloating}

\begin{figure*}[t]
    \centering
    \includegraphics[width=\linewidth]{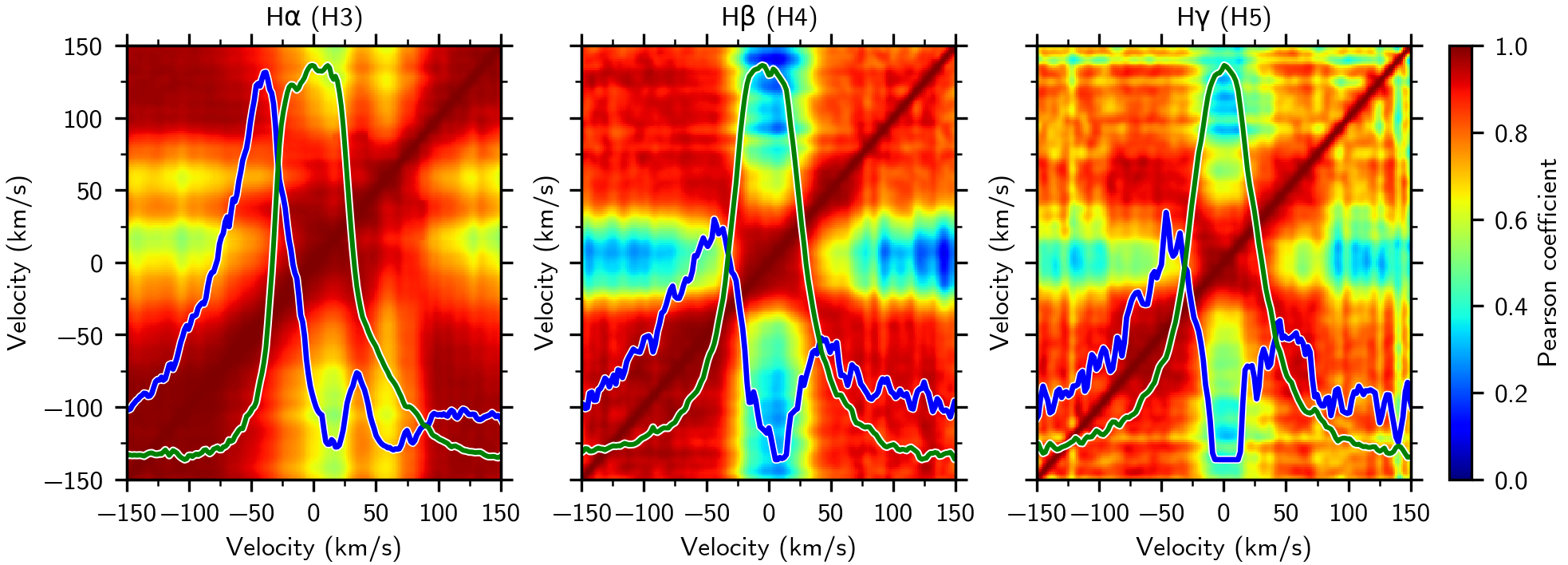}
    \caption{Auto-correlations of the lines \Halpha, \Hbeta, and \Hgamma. The color indicates the value of the Pearson correlation coefficient. The lines were smoothed to $R\sim40,000$ to remove some of the noisy features. The lack of correlation between the wings and the core suggests the lines may be made of two sub-components. Over-plotted in blue and green are the two sub-components decomposition obtained in Section~\ref{sec:decomposition}, which are able to explain the features of the auto-correlation. The correlation diagrams for higher order lines have the same overall shape as \Hbeta\ and \Hgamma.}
    \label{fig:auto_correlations}
\end{figure*}

In Figure \ref{fig:empirical}, we compare the \Ha  line profiles of substellar accretors from \cite{mohanty_accretion_2005} and \cite{muzerolle_measuring_2005} to the mean \Ha profile of \delorme. The plateau of the \Ha profile of \delorme\ closely resembles those of CFHT-BD-Tau 4 and Cha~Ha~1, two young (1--3~Myr) brown dwarfs with estimated masses of 60~\Mjup and 35~\Mjup, respectively (Fig.~\ref{fig:empirical}). The core of CFHT-BD-Tau 4's line is also comparable to that of \delorme\ while Cha~Ha~1's is noticeably broader. We rescaled the profiles of Cha~Ha~1 to match the expected free-fall velocities of \delorme\ assuming a similar inclination of the two objects.
A rescaling by a factor 0.8 provides the best fit (corresponding to a mass of 20~\Mjup). However, the wings of \delorme\ are still more extended, in particular on the blue side of the line. Both objects have clear infrared excess, as identified by the VOSA SED analyser \citep{2008A&A...492..277B}.

We also show in Fig \ref{fig:empirical} the profile of the 15~\MJup\ brown dwarf from Taurus KPNO Tau 4. The object is known to be devoid of infrared excess \citep{2014A&A...562A.127B}. As such, it is the only class III object in the mass range of \delorme\ with profiles available at a resolution close to UVES. It displays a double-lobe profile also retrieved in active M dwarfs \citep[e.g.,][]{1998A&A...336..613S, mohanty_accretion_2005}.



\begin{figure*}
    \includegraphics[width=\linewidth]{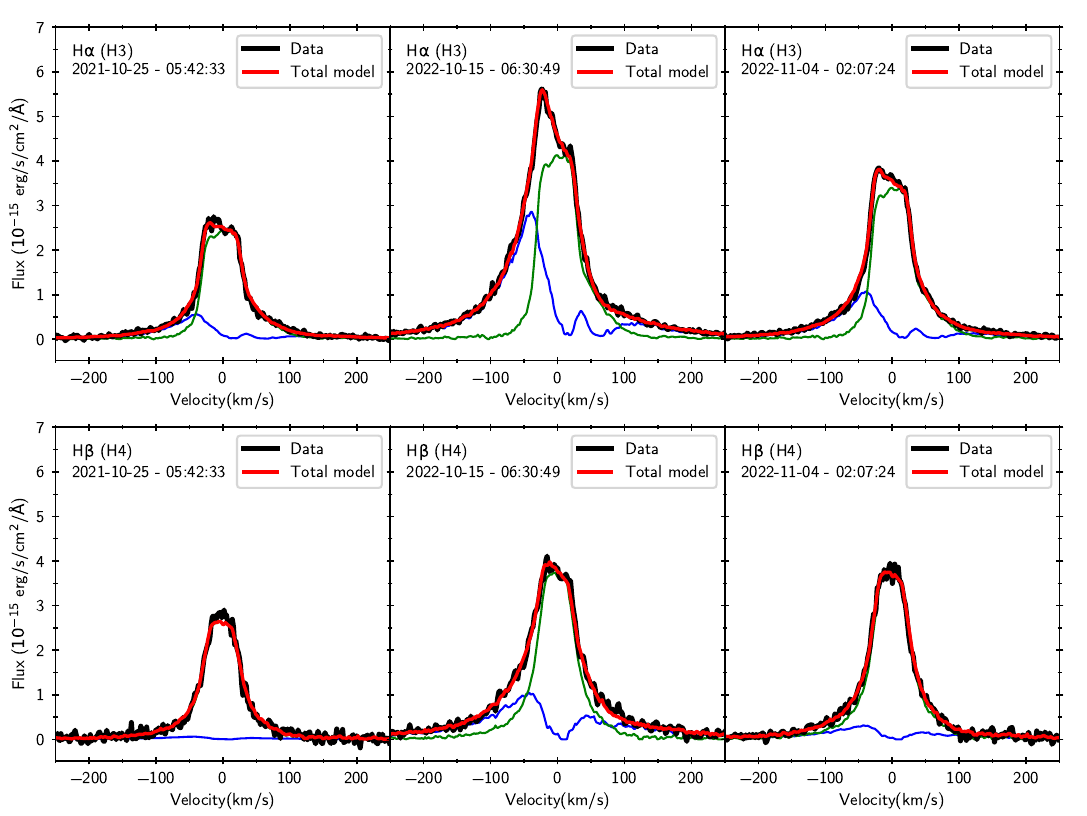}
    \caption{Illustration of the decomposition for three different exposures, for \Halpha\ (first row) and \Hbeta\ (second row). The two components are shown in blue and green, the total model in red, and the corresponding data in black. The decompositions for all exposures are shown in Figs.~\ref{fig:appendix:all_halpha_lines}, \ref{fig:appendix:all_hbeta_lines}, \ref{fig:appendix:all_hgamma_lines} and \ref{fig:appendix:all_hdelta_lines} for \Halpha, \Hbeta, \Hgamma\ and \Hdelta\ respectively. This decomposition provide a great fit to all exposures despite the strong constraint of a constant profile.}
    \label{fig:results:decomposition}
    \label{fig:two_components_plot}
\end{figure*}

\subsection{H {\sc i} emission line auto-correlation diagrams}
\label{subsec:autocor}

Figure~\ref{fig:auto_correlations} shows the auto-correlation (Pearson coefficient) diagrams for
(\Halpha, \Hbeta\ and \Hgamma).
Over-plotted in blue and green are the respective two-components decomposition for the line (see next Section~\ref{sec:decomposition}).
While auto-correlation diagrams are commonly used in studying time-series of accretion lines in T-Tauri stars \citep[e.g.,][]{alencar_accretion_2012}, we extend their usage to the planetary-mass regime.
Auto-correlation diagrams represent the Pearson correlation coefficient between different velocity channels of the lines. They help show how different parts of the lines evolve with respect to each other. They have been used, for example, as a mean to hint at the magnetic field topology, by comparing observed auto-correlation matrices to that predicted from MHD simulations \citep{alencar_accretion_2012}.

In all three lines, the core and wings of the profiles appear un-correlated over a central region $\sim50~$\kms\ wide.
\Halpha\ shows an additional uncorrelated region around $\sim60$~\kms. This may be interpreted in that the \Halpha\ line is likely composed of up to three independent components (wings, core and $\sim60$~\kms\ component), while \Hbeta\ and \Hgamma\ are likely formed of two components. It remains unclear whether the $\sim60$~\kms\ component belongs to the core or wing structure, or constitutes a distinct third component. Alternatively, correlated velocity channels may originate from different physical regions but still have common evolution.

In the case of \Hbeta\ and \Hgamma, the correlation coefficient between the core and the wing reaches $\sim$0, which means both components are completely independent. However, for \Halpha, the correlation coefficient between the core and the wings only reaches a minimum of 0.5.
This implies that if \Halpha's core and wings components were to be truly independent (as seen in \Hbeta\ and \Hgamma), the contrast between the components should be lower than for \Hbeta's or \Hgamma's.



\begin{figure*}[t]
    \includegraphics[width=0.5\linewidth]{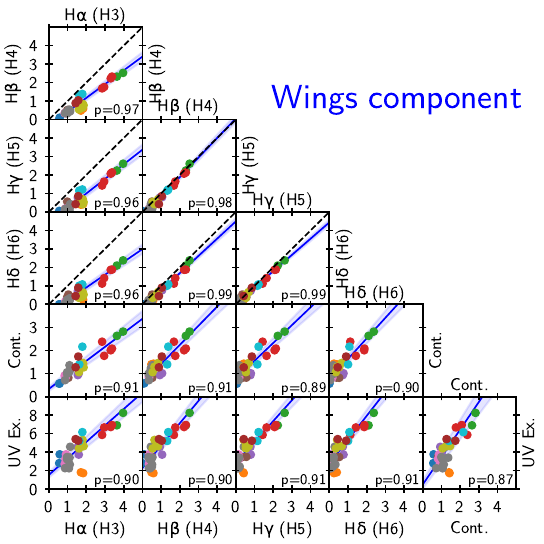}%
    \includegraphics[width=0.5\linewidth]{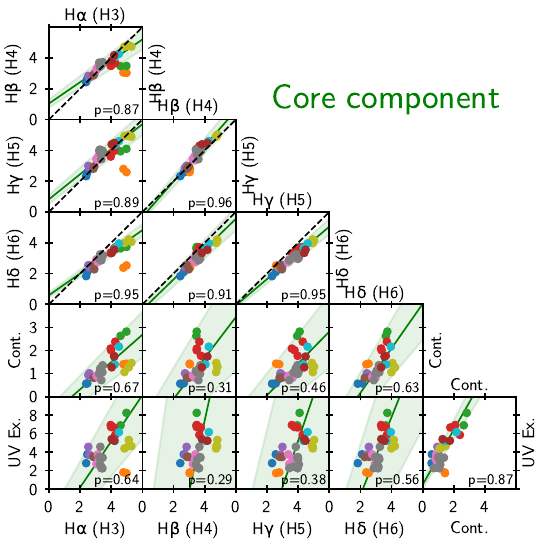}%
    \caption{Correlations between the wings/core components over multiple lines (\Halpha\ to \Hdelta), as well as the mean continuum level (cont) around \Halpha\ (in $10^{-17}$~\ergsang), and the mean level of the UV excess (in $10^{-17}$~\ergsang). The Pearson correlation coefficient is given as text in each panel. The black dashed line is the y=x line. The blue/green lines and shaded regions are the best affine fit. The color coding of the data points are the same as in Fig.~\ref{fig:all_lines}. The x/y units are normalized scaling factors of each component. The epoch on 2022 October 13 (orange) was excluded from the analysis as it is so far off the correlation in the core \Halpha\ component. The wings and core components are strongly correlated with themselves. However, only the wings components has a strong correlation with the UV excess, which suggests an accretion origin. The wings component and the UV excess are both strongly correlated with the continuum level around \Halpha, which suggests the apparent continuum is veiled by the extension of the apparent continuum.}
    \label{fig:shapes_correlations}
\end{figure*}

\subsection{Decomposition of the H {\sc i} line profiles}
\label{sec:decomposition}

Emission line profiles are usually decomposed as multiple Gaussians \citep[e.g.,][]{ringqvist_resolved_2023,viswanath2024_entropy_I} that can trace different emitting areas or mechanisms.
However, most recent accretion models indicate that the line profiles should deviate significantly from Gaussians shapes when complex emitting regions are involved, or co-exist. While shock-emitted lines have a more symmetrical profile \citep{aoyama_theoretical_2018} that Gaussians may be able to reproduce, lines emitted in a magnetospheric accretion funnel tend to be affected by complex absorption features \citep{thanathibodee_magnetospheric_2019}.
Alternatively, the lines may also originate from a mixture of accretion and chromospheric activity \citep[e.g., ][]{2020ApJ...892...81T}.

Due to the complex shape of the auto-correlation diagrams (Fig.~\ref{fig:auto_correlations}), we chose to retrieve an empirical decomposition of the lines that does not assume Gaussian components.
We developed a custom method based on the L-BFGS-B algorithm to disentangle multiple components within the lines.
This approach is made possible by the large volume of time-series data available. The method is described in Sect.~\ref{sec:decomposition:method}, and the results in Sect.~\ref{sec:decomposition:results_1}~\&~\ref{sec:decomposition:results_2}.


\subsubsection{Method}
\label{sec:decomposition:method}

We developed a novel method to decompose the lines into multiple components. This method makes no assumption as to what shape these components should have, and effectively infers the components that best reproduce the observed profiles across all epochs and exposures.

Using the fact that multiple components of a line may be related to different emitting mechanisms, we make the following assumptions: (i) each line is the sum of multiple emission components of arbitrary shape, (ii) the shape of each component does not vary between exposures, only their relative amplitude does.
This assumption corresponds physically to distinct emitting regions that would have negligible variability of their respective line profile.
In the following, we set the number of components to 2, as more components would have the algorithm jump between multiple local minima. We also tried adding an absorption component (that would correspond to an absorption region on top of an emitting region), but this did not improve convergence.

The model treats the observed flux F(v,t) as a linear combination of velocity-dependent components scaled by exposure-dependent factors, $(C_i(v), \, k_i(t))$, such that the total line model is:
\begin{equation}
    F(v,t) = k_1(t) C_1(v) + k_2(t) C_2(v) 
\end{equation}
where $C_i(v)$ is the value of the component $i$ at the velocity $v$, and $k_i(t)$ is the scaling parameter for the component $i$ at exposure $t$.

The fit is performed iteratively, by solving for the values of $C(v)$ for one of the components, and $k(t)$ for both components. The $C_i(v)$ are alternatively solved for each component along the iterations.
At each iteration, fit was performed using \texttt{Scipy}'s implementation of the limited-memory Broyden--Fletcher--Goldfarb--Shanno (L-BFGS-B)  algorithm \citep{lbfgsb1995,lbfgsb1997,2020SciPy-NMeth}.

The process does involve many free parameters, but this is still far below the number of data points that are actually involved in the fit.
The total number of parameters is given by $2(2N_t + N_v)$, with $N_t$ the number of exposures (33), and $N_v$ the number of velocity bins (322 over $(-250,\, +250)$~\kms). This yields a ratio of data-to-parameters equal to 13.7.


While the decomposition was performed using a flat prior for the shape of the components, we did try different initial guesses for the shapes and we find the results to be insensitive to the choice of initial parameters. This confirms that the local minimas in the parameter space (if they exist) are all located within proximity of each other, with only slight deviations to the profile of each component.

We also investigated an alternative decomposition based on the Non-Negative Matrix Factorization (NMF), used by the direct imaging community to build PSF models. We retrieve decompositions of the \Halpha profiles close to the ones achieved with our method and the results confirm the L-BFGS-B two-component decomposition, although with some slight differences (see Fig.~\ref{fig:appendix:compare_nmf_to_mine} for a comparative discussion of the two methods). However, the NMF requires all data to be positive, which is not verified at low SNR and requires excluding the faint wings of the lines, in particular in higher order lines (e.g., \Hgamma\ and higher). Therefore,  we adopted in the following the decomposition gathered from our custom method.

\subsubsection{The shape of the two sub-components}
\label{sec:decomposition:results_1}

The final decomposition for \Halpha\ and \Hbeta\ is shown in Fig.~\ref{fig:results:decomposition} (blue and green lines) for 3 different exposures.
The shape of both components at the end of each iteration is shown in Fig.~\ref{fig:shapes_evolution}, in transparent lines (blue and green). Each panel corresponds to the results of the decomposition of \Halpha, \Hbeta, and \Hgamma\ for the October 15, 2022  epoch at 06:30:49~UT. The results for all epochs are presented in Figs.~\ref{fig:two_comp_Halpha_allepochs}, \ref{fig:two_comp_Hbeta_allepochs}, \ref{fig:two_comp_Hgamma_allepochs} and \ref{fig:two_comp_Hdelta_allepochs}.
Due to the respective contribution of the two components within the wings and the core of the line, we hereafter call them \textit{wings} (blue) and \textit{core} (green) components.

The wings component rapidly converges within a few tens of iterations to a stabilized shape with deep absorption features, broadly centered at $\sim15$~\kms\ for \Halpha, and $\sim3$~\kms\ for \Hbeta, \Hgamma\ and \Hdelta. The second (green) component converges to a narrow asymmetric profile centered on zero velocity.

The two components are shown overlaid on the auto-correlation diagrams in Fig.~\ref{fig:auto_correlations}.
For \Halpha, the core component correlates almost perfectly with the core component from the auto-correlation, as well as the reduced correlation in the 50-100~\kms range. The correlated wings are attributed to the wings component. The reduced correlation between $-50$ and 0~\kms\ corresponds to the blending of the wings component peak with the green wing. The double-peaked feature explains the correlation peak at $\sim40$~\kms\ in the auto-correlation diagrams. This wings profile is explained by magnetospheric accretion components (see Sect.~\ref{sec:modeling}).

Both \Hbeta\ and \Hgamma\ auto-correlation shapes are perfectly explained by their two-components decomposition as well. Similarly to \Halpha, the uncorrelated core and wings are explained by the two independent components.
Their blue wings have a stronger auto-correlation than their red wings, which is explained by the shape of the wings component: its blue wing is brighter than its red wing. This asymmetry enhances contrast between the wings and core components, increasing the correlation strength in the blue wing.

\begin{figure*}[t]
    \centering
    \includegraphics[width=0.5\linewidth]{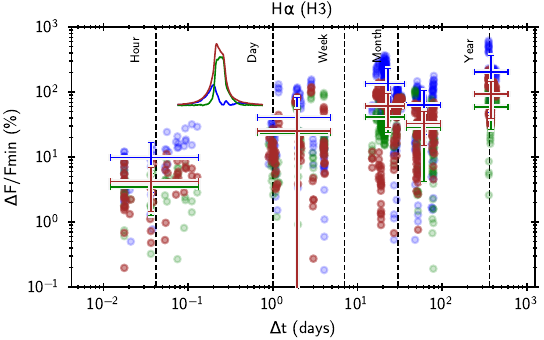}%
    \includegraphics[width=0.5\linewidth]{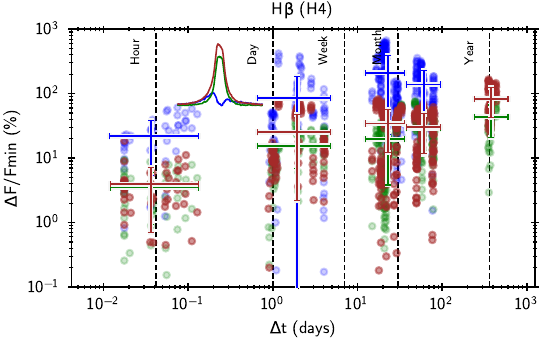}
    \caption{Variability diagrams of the \Halpha\ (left) and \Hbeta\ (right): the relative flux variability for all pairs, as a function of the timescale, following the same method as \cite{demars_emission_2023}.
    We show the variability amplitude of the wing (blue) and core (green) components and of the total profile (dark red). The error bars represent the spread of the variability within a given timescale bin, weighted by the significance of the measurement: the $\upsigma$-distance between the two lines corresponding to the measurement.
    Higher order lines have variability diagrams similar to that of \Hbeta, with slightly increased variability amplitude at $\sim$hour timescales (see Fig.~\ref{fig:hour_timescale_variability}).
    The full-line variability amplitude is driven by that of the core component, because it contains most of the line flux, whereas the wings component is quite faint in comparison.}
    \label{fig:variability_timescales}
\end{figure*}

\subsubsection{Correlations between sub-components}
\label{sec:decomposition:results_2}

The insets in Fig.~\ref{fig:shapes_evolution} show the correlations between the scaling factors of the wings and core components. Observations from October 14, 15, and 17 of 2022 correspond to brighter emission lines possibly tracing an outburst, and are plotted as crosses. In \Halpha, the wings and core component amplitudes are strongly correlated, except within this ``outburst'' epoch, which corresponds to a strong wings component. Still, the wings and core component remain correlated during the outburst. However, \Hbeta, \Hgamma\ and \Hdelta\ exhibit weaker correlation between their wings and core components, but the outburst epoch consistently corresponds to a brighter wings component.

Figure~\ref{fig:shapes_correlations} shows the correlation of each component (wings or core) across different lines, with Pearson correlations labeled in each panel. Both components are well correlated from one line to another. The mean flux level around \Halpha (500-5000 \kms on each side of the line), and within the UV excess (mean flux level in the 3350-3600~\AA range) are also shown.

The continuum level around \Halpha\ (Fig.~\ref{fig:shapes_correlations}, last row, last column) is strongly correlated with the UV excess. This hints that the apparent continuum may instead represent a blend of atmospheric emission and veiling extending at these wavelengths.
The wings components are strongly correlated with both the continuum level (around \Halpha) and the UV excess level ($p \sim 0.90$ in all panels). The core components, however, are only slightly correlated with the continuum and UV excess ($p\sim0.4$), except for the \Halpha\ line ($p\sim0.65$).

Given the similarity of the shapes of the wings and core components from one line to another, the correlations are not necessarily surprising. However, since the \Halpha\ line profile is distinct from that of other lines, it was not so evident that its sub-components (both wings and core) would be so strongly correlated from one line to another. Still, this hints at a common mechanism for the wings (blue) components of all lines, and core (green) components of all lines.

\subsubsection{Limits to the decomposition}
Although the method described here provides a decent fit to the line profiles, residuals are clearly observed in Fig.~\ref{fig:two_comp_Halpha_allepochs} to \ref{fig:two_comp_Hgamma_allepochs}. The residuals have either a single or double-peak feature. These residuals might be the sign of variable-shape components, missed by our method.

It should be noted that both the highest and smallest residuals surprisingly appear within and around the outburst epoch.
The strongest residuals are observed on epoch 2022-10-13, that precedes the outburst epoch, while he smallest residuals appear within the last two exposures of epoch 2022-10-15. These two exposures are taken an hour after the initial two, which shows that variability in the line profile likely occurs on timescales of $<1$h. In fact, the wings component flux on that epoch are comparable to that of non-outburst epochs, while also appearing as an outlier in the correlation plots in Fig.~\ref{fig:shapes_correlations}. We find in Sect.~\ref{sec:modeling} that the wings component is very likely due to magnetospheric accretion, a scenario in which line profiles are intrinsically variable. Therefore, the residuals are likely due to the intrinsic profile variability of the wings component, which is missed by our static components approach.

\subsection{Line variability}
\begin{figure}
    \includegraphics[width=1.0\columnwidth]{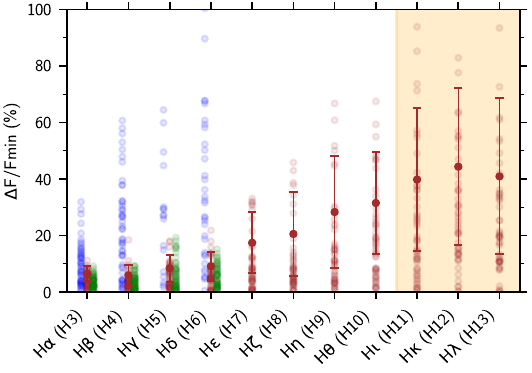}
    \caption{Variability of the Balmer lines at the hour timescale.The color of the points represents the component (blue/green for wings/core) or the full line (brown). Error bars represent the spread of the variability (weighted by significance), i.e., the brown crosses in Fig.~\ref{fig:variability_timescales}. The shaded region corresponds to lines where the detection is compatible with contamination from the host binary. The hour timescale variability seems to be increasing with the order of the transition. The wings component has a much higher variability amplitude than the green component.}
    \label{fig:hour_timescale_variability}
\end{figure}

All lines present both profile and amplitude variability across epochs with minimal variability within individual nights.
Figure~\ref{fig:variability_timescales} shows the variability of the line flux of both \Halpha\ and \Hbeta\ as a function of timescale: all possible line-pairs were associated with their relative flux ratio (normalized to the fainter of the pair) and the time-difference between the two observations.
This follows the same methodology as \cite{demars_emission_2023}, and is similar to that of \cite{costigan_temperaments_2014} and \cite{zsidi_accretion_2022}.
The blue/green points correspond to the sub-components evidenced in the previous Section, while the brown points correspond to the full line.
Note that the $\sim$year timescale is driven by the 2021 October 25 epoch presented in \cite{ringqvist_resolved_2023}. Figure~\ref{fig:hour_timescale_variability} shows the variability amplitude at $\sim$hour timescale, i.e., the lower timescales in Fig.~\ref{fig:variability_timescales} and with the same color-scheme. 
Notably, the wings component has a greater relative variability than the core one.


Both \Halpha\ and \Hbeta\ have low variability amplitudes at $\sim$hour timescales ($\sim$8~\%), and reach a near-plateau from days to months timescales.
Higher order lines have variability diagrams similar to that of \Hbeta\ (not shown here) at $>$days timescales. In contrast, \Halpha\ shows several peaks in variability amplitude, in particular slightly below the month timescale, likely driven by the blue-shifted ``outburst'' on 2022 October 14, 15, and 17 (see Fig.~\ref{fig:all_lines}). This is linked to the increased amplitude of the wings (blue) component.

The core component has a variability amplitude of a few tens of \%, and a shape that is mostly in line with the overall profile, except for the month timescale where its amplitude is lesser. It makes sense given that this timescale is driven by the burst epoch, reflecting the increased importance of the wings component.
However, the wings component has a variability amplitude that is greater (a few hundred percent) than both the core component and the overall profile, at all timescales, and so is the case for all lines. This is mostly due to the almost complete disappearance of the wings component in some epochs, consequently amplifying the spread of variability relative across epochs.

The variability amplitude at $\sim$hour timescales seems to be increasing with the line transition order (see Fig.~\ref{fig:hour_timescale_variability}). This trend is absent at longer timescales. This is broadly consistent with the much higher variability amplitudes seen at the hour timescale in \PaBeta\ for \gqlupb\ and \gsc\ \citep[$\sim$50~\%,][]{demars_emission_2023}.



\subsection{Line ratios}

\begin{figure}
    \centering
    \includegraphics[width=1.0\columnwidth]{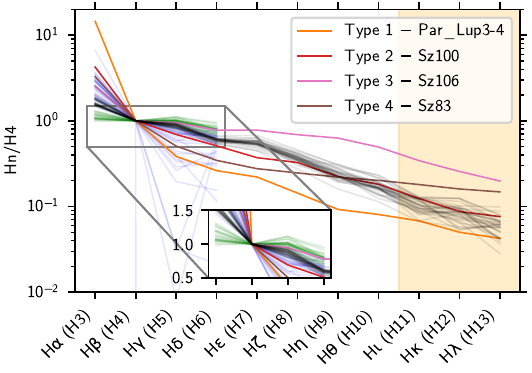}
    \caption{Line ratios decrements. For each line, the line ratio is computed with respect to \Hbeta\ (H4). In black, the line ratios of the full-lines in this study. Colored lines correspond to the four decrement types identified in \cite{antoniucci_x-shooter_2017} (see their Fig.~3). Blue/green lines represent the sub-components (wings/core) obtained in Sect.~\ref{sec:decomposition}. The shaded region corresponds to lines which could be due to contamination from the host binary (as in Fig.~\ref{fig:hour_timescale_variability}).
    The full-line decrement (black) is broadly compatible with Type 2 case (optically thin emission and low accretion rate). The wings component is not clearly compatible with any decrement type. However, the core component is compatible with the Type 3 sources, which happen not to be easily reproduced by emission line models \citep{antoniucci_x-shooter_2017}.}
    \label{fig:lines_decrements}
\end{figure}

Figure~\ref{fig:lines_decrements} shows the \HI\ line ratios (all lines normalized to \Hbeta) of the full profiles (in black), as well as for the wings and core sub-components. 
The typical line ratios found in T~Tauri stars \citep{antoniucci_x-shooter_2017} are also shown. Their study focuses on line decrements in the Lupus region ($\sim 3$--4~Myr), for a sample of 36 stars with masses in the range 30~\Mjup ~--~1.5~\Msun, with only 4 objects below 100~\Mjup. They identify four classes of decrements. The lowest accretion sources in the sample show mostly Types 1 \& 2 decrements.

The overall line ratios of \delorme\ (black lines) are consistent with the general trend of \cite{antoniucci_x-shooter_2017} decrements but they are not a perfect match to any of the four classes.  
They appear most compatible with the Type 2 case, which corresponds to optically thin emission and low accretion rate sources in \cite{antoniucci_x-shooter_2017}.
Surprisingly, the \Halpha/\Hbeta\ ratio, with values ranging from 1 to 3, is consistently below all of the T Tauri classes.
We also note that our \Halpha/\Hbeta\ ratios are inconsistent with the one obtained from MUSE/VLT on October 18, 2018 \citep[\Halpha/\Hbeta$=9.8$;][]{eriksson_strong_2020}. However, the MUSE data are also incompatible with the published photometry of the primary, suggesting that they may suffer from residual bias.

The wings component's line decrements display a more dispersed behavior across epochs, likely due to the faint emission level of that component in \Hgamma\ and \Hdelta. However, the core component shows lower dispersion and resembles the decrement profile of Type 3 sources from \cite{antoniucci_x-shooter_2017}, except for its \Halpha/\Hbeta, which remains significantly lower than all classes.
\cite{antoniucci_x-shooter_2017} mentions that Class 3 decrements cannot be well reproduced by the line emission processes (their Case B: optically thin emission and the local excitation model developed by \cite{kwan_origins_2011} for radial flows around young stars) and remain the most difficult to interpret. The observed differences could reflect the complexity of the line-emitting region.


\begin{table*}[ht]
    \caption{Parameters range for the magnetospheric accretion models \citep{thanathibodee_magnetospheric_2019}, and best fit results.}
    \label{tab:magacc_parameters}      
    \centering
        \begin{tabular}{l c c c c c c c c c c}
        \hline
        & \Rin & Width & \Mdot & \Tcol & $i$ & \Mp & \Rp & \Lacc \\
        & (\Rp) & (\Rp) & (\Msun/yr) & (K) & (\deg) & (\Mjup )& (\Rjup) & (L$_\odot$)\\
        \hline
        Range & 1.5--6.0 & 0.5--2.5& $9.5\times10^{-(13-10)}$ & 6\,000--10\,000 & 15--75 & 11--15 & 1.4--1.6 & \\
        \hline
        \Halpha\ Wings $\pm50$~\kms & 2.0 & 2.0 & $6.02 \times 10^{-12}$ & 9\,250 & 60 & 15 & 1.4   & $1.88 \times 10^{-5}$ \\
        \Halpha\ Wings $\pm250$~\kms & 1.5 & 1.0 & $1.51 \times 10^{-12}$ & 10\,000 & 75 & 15 & 1.4 & $4.72 \times 10^{-6}$ \\
        \Halpha\ Core $\pm50$~\kms & 3.0 & 0.5 & $1.51 \times 10^{-12}$ & 8\,750 & 30 & 15 & 1.6  & $4.13 \times 10^{-6}$ \\
        \Halpha\ Core $\pm250$~\kms & 1.5 & 0.5 & $3.80 \times 10^{-10}$ & 6\,000 & 75 & 11 & 1.6 & $7.62 \times 10^{-4}$ \\
        
        \hline
        
        \Hbeta\ Wings $\pm50$~\kms & 1.5 & 2.5 & $1.51 \times 10^{-10}$ & 8\,000 & 75 & 13 & 1.5   & $3.82 \times 10^{-4}$ \\
        \Hbeta\ Wings $\pm250$~\kms & 2.0 & 2.5 & $9.55 \times 10^{-10}$ & 10\,000 & 75 & 15 & 1.6 & $2.61 \times 10^{-3}$ \\
        \Hbeta\ Core $\pm50$~\kms & 2.0 & 0.5 & $9.55 \times 10^{-13}$ & 10\,000 & 30 & 13 & 1.4 & $2.59 \times 10^{-3}$ \\
        \Hbeta\ Core $\pm250$~\kms & 2.0 & 0.5 & $9.55 \times 10^{-10}$ & 6\,250 & 75 & 11 & 1.5 & $2.04 \times 10^{-3}$ \\
        \hline
        \end{tabular}
        \tablefoot{Results are shown for the wings and core components, in the $(-50,\, +50)$~\kms and $(-250,\, +250)$~\kms cases, for both \Halpha\ and \Hbeta. The accretion luminosity \Lacc\ was computed as $\Lacc = G \Mdot \Mp / \Rp$. For the full $\chi^2_r$ space, see Fig.~\ref{fig:modeling_Halpha_corner_plots} and \ref{fig:modeling_Hbeta_corner_plots}.}
\end{table*}

\begin{figure*}
    \includegraphics[width=\linewidth]{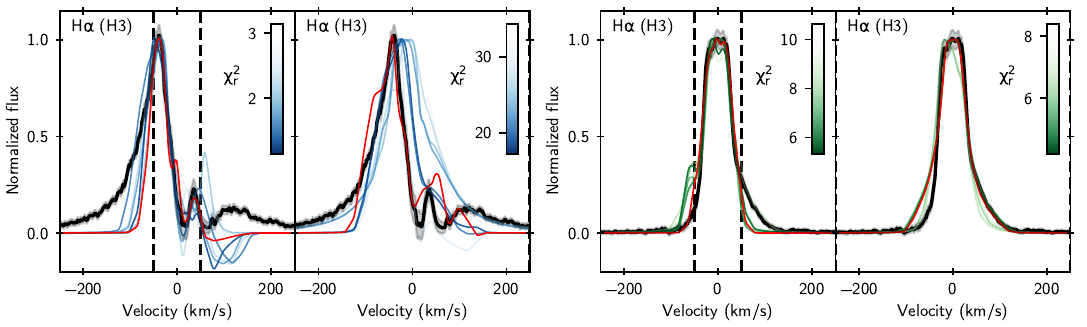}\\
    \includegraphics[width=\linewidth]{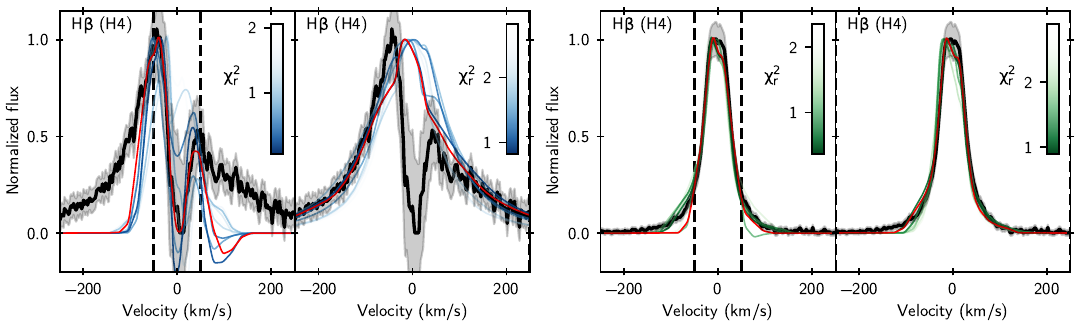}
    \caption{Modeling results for the magnetospheric accretion models \citep{thanathibodee_magnetospheric_2019}. Left: results for the wings sub-component, right: results for the core sub-component. First row: \Halpha, second row: \Hbeta. Each left sub-panel correspond to the fit performed on the [-50;+50]~\kms range, while right sub-panels correspond to the fit on the full the line width ([-250;+250]~\kms). The red line is the best model, while colored lines represent a sample of the lowest $\chi^2_r$ models. The models are able to reproduce the deep absorption features of the wings component, and reproduce the general shape of the core component.}
    \label{fig:modeling_thanathibodee_best_fits}
\end{figure*}



\section{Comparison to models}
\label{sec:modeling}

We investigate two different \HI\ line formation scenarios: 1) formation in magnetospheric accretion columns, 2) formation in a shock front. These two mechanisms are not mutually exclusive and may contribute simultaneously to the observed emission.
We use the magnetospheric accretion models from \cite{thanathibodee_magnetospheric_2019}, and the shock models from \cite{aoyama_theoretical_2018}. Magnetospheric accretion models are available for both \Halpha\ and \Hbeta, while shock models are available up to \Hdelta.
 
The analysis is similar to that of \cite{demars_emission_2023} for the \PaBeta\ line of \gqlupb\ and \gsc.
The line profiles are fitted with the two models for both \Halpha\ and \Hbeta, while \Hgamma\ and \Hdelta\ are only fitted with shock models.
For both lines, the wings component has deep absorption features reminiscent of that from infalling material (e.g., an accretion column).
The core component, however, has no obvious feature that could help discriminate between the two models.

\subsection{Magnetospheric accretion models}
\subsubsection{Model description}
The model assumes that the planet has  an aligned dipolar magnetic field strong enough to truncate the disk. The \HI\ lines form in the axisymmetrical funnel flow following the closed dipolar magnetic field lines connecting the disk to the central object.

The parameters of the model are given in Table~\ref{tab:magacc_parameters}. Respectively, they correspond to the inner truncation radius (\Rin), width of the column (width), accretion rate (\Mdot), column temperature (\Tcol), inclination of the system ($i$), planet mass (\Mp), radius (\Rp) and effective temperature (\Teff). Only a single value of $\Teff=1800$~K is considered as it mostly affects the continuum level within the synthetic lines, which is subtracted before fitting.

It should be noted that geometrical parameters are given in terms of planetary radius, and the planet radius mostly serves as a scaling parameter for the truncation radius and column width. They should therefore not be interpreted as a robust determination of the planet mass or radius. These may also serve as scaling parameters for the location and intensity of various absorption features. This is the case in particular for parameters with low sampling rates, for example the inclination, which is only sampled in 15\deg\ increments.


\subsubsection{Modeling method}

The synthetic lines sometimes have absorptions that extend below the continuum, which makes sense in the context of CTTS. However, the models cannot reproduce the line flux of our observations, where the continuum is barely visible. We therefore subtract the photospheric contribution from the models using a linear fit on either side of the line.
This approach introduces negative values where modeled absorptions exceed the photospheric continuumm. We therefore exclude negative values from the fit to mitigate this issue. We also normalize the lines to 1 around $\sim0$~\kms and focuses solely on fitting the shape of the line components.

As done by \cite{demars_emission_2023}, the fits are performed by computing the $\chi^2_r$ over the whole grid.
The modeling is carried out over both the full line width and the core region, as the key features (wings and central absorption) are differently reproduced depending which velocity range considered.
The wings and core components are modeled independently. The best fit models, for both \Halpha\ and \Hbeta, are shown in Fig.~\ref{fig:modeling_thanathibodee_best_fits}. The $\chi^2_r$ parameter space is shown in Appendix~\ref{sec:appendix:modeling_results}, along with a sketch of the best fit models. Table~\ref{tab:magacc_parameters} shows the parameter values of the best fits.


\subsubsection{\texorpdfstring{\Halpha}{Halpha} modeling}

As shown in Fig.~\ref{fig:modeling_thanathibodee_best_fits}, the models are able to reproduce the absorption features in the wings component (blue), both when fitting the core and the full line.
However, the best-fit model appears to yield a much narrower magnetosphere in the full-range fitting than in the core-fitting. This is explained by the degeneracy between the accretion rate and the column width seen in the full-range fit (Fig.~\ref{fig:modeling_Halpha_corner_plots}): a higher accretion rate with a wider column broadly yields a similar column density, with similarly shaped absorption features.
Both the core and full-line cases broadly yield two minima: a low accretion rate with high column temperature, and high accretion rate with low column temperature (Figs.~\ref{fig:modeling_Halpha_corner_plots}). Both the core and full-line cases profile types yield high inclinations, in the $60$--$75$\deg\ range.

When fitting the core ($\pm50$~\kms) of \Halpha's core component, the model reproduces the blue wing, but not the red wing tail. However, when the fit is extended to the full range of velocities ($\pm250$~\kms), the red wing is reproduced, but the main peak of the line is slightly narrower than the data. In this case, the blue wing is not reproduced by the model.
In both cases, the modeling of \Halpha's core component yield magnetospheres that are relatively thin. However, the core-fit yields a large truncation radius, while the full-line fit yields a low truncation radius. This difference might also be explained by degeneracies between multiple minima at truncation radii of 1.5~\Rjup\ and 3.0~\Rjup.
The core and full-range modeling yield different inclinations: $\sim30$\deg\ for the core-fitting, and $\sim75$° for the full-range fitting.

\subsubsection{\texorpdfstring{\Hbeta}{Hbeta} modeling}

\Hbeta's wings component has a very different behavior. Models can reproduce the dip at the core of the lines when the fit is performed from $\pm50$~\kms, but fail to capture the wings simultaneously. Conversely,  the fit on the larger velocity interval ($\pm250$~\kms) reproduces the wings, but not the central absorption.
In fact, some of \Halpha's best models also have a similar behavior, as shown by some of the blue-shaded profiles.
Both \Hbeta\ fits yield magnetospheres that have similar extents and mainly differ in terms of accretion rates and column temperature. However, the $\chi^2_r$ maps show that when the line core alone is fitted ($\pm50$~\kms), the magnetosphere geometry remains poorly constrained. The accretion rate and column temperatures are also highly degenerate.
When the full profile is fitted, the geometry is best reproduced with a small launching radius and large column width, but no constraints on the inclination are derived.

\begin{figure*}
    \includegraphics[width = \linewidth]{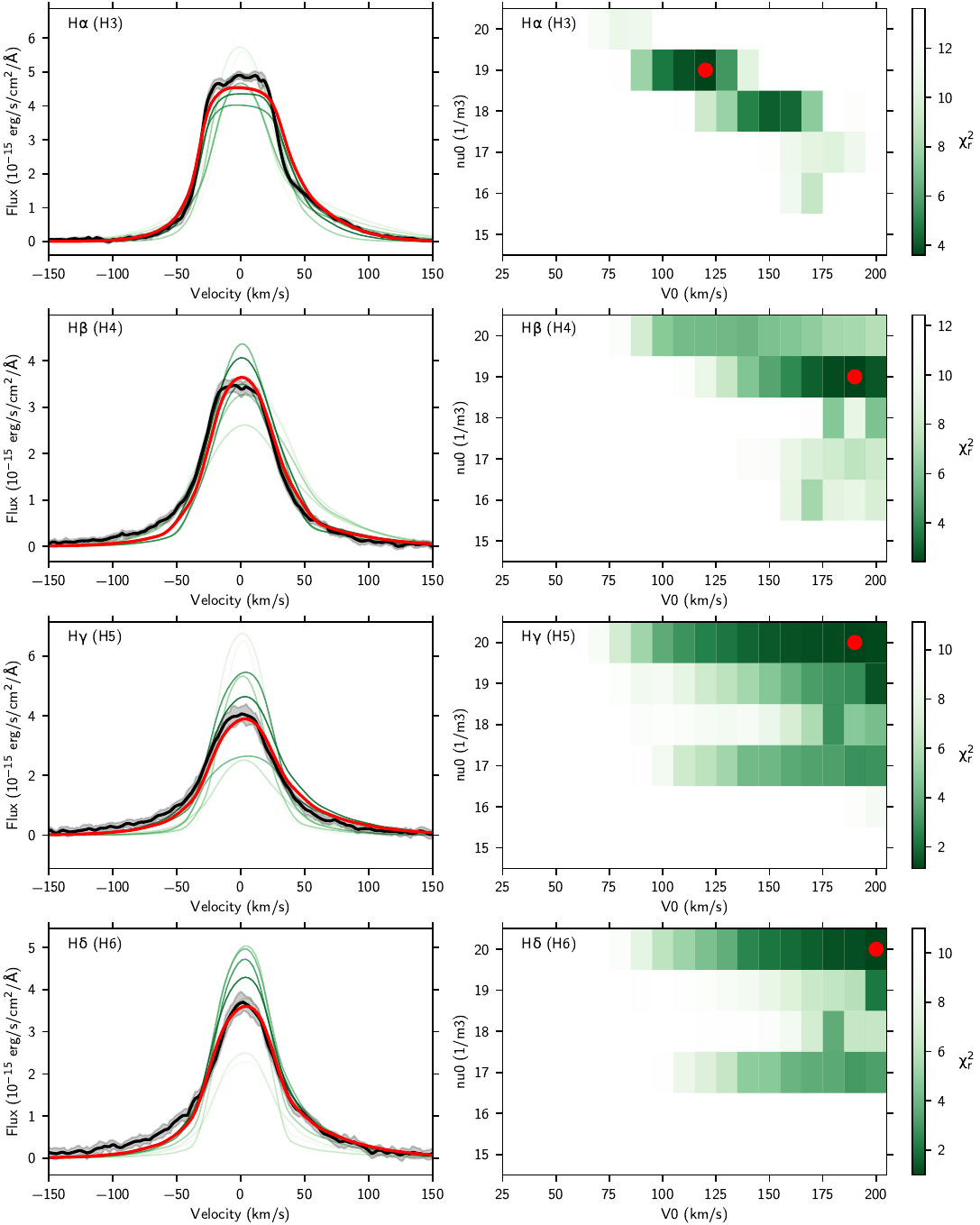}
    \caption{Modeling results for the shock models \citep{aoyama_theoretical_2018}, with the flux scaling corresponding to epoch 2022-10-14 -- 06:52:50). On each row, the left panel shows the core component for that line (black line), a sample of the best fits (green lines) and the best fit (red line). The right panel represents the corresponding $\chi^2_r$ map. The models provide an increasingly good fit as the order of the line-transition increases. The shift between data and models for \Hbeta and \Hgamma are likely due to the lower accuracy on the reference wavelength for these two lines in NIST.}
    \label{fig:modeling_shock}
\end{figure*}

\begin{figure*}
\centering
    \includegraphics[width = \linewidth]{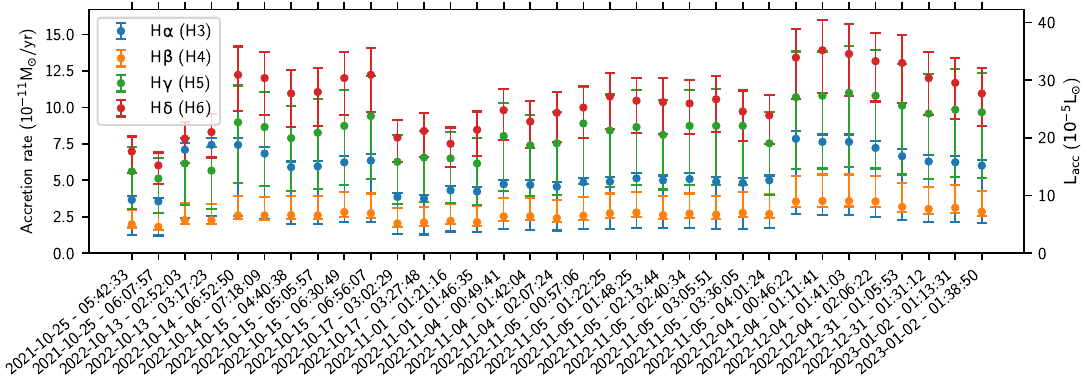}
    \caption{Evolution of the accretion rate inferred from shock modeling of the core component, depending on the line. The accretion rate was computed as $\Mdot = S n_0 v_0 \mu$. The right axis shows the corresponding $\Lacc = G \Mdot \Mp / Rp$, with $\Mp = 13$~\Mjup and $\Rp = 1.5$~\Rjup. The accretion rates inferred from different lines are systematically ordered, which may point to a bias from deriving accretion rates from a single line.}
    \label{fig:modeling_shock_accretion_rate}
\end{figure*}

\Hbeta's core component has a similar behavior to that of \Halpha, both when considering the fit over the wide and narrow velocity intervals. When modeling the line core, the width of the line is well reproduced, as well as its red wing, but not its blue wing. The slope at the top of line, however, does not match that of the data.
In the full-line modeling case, both wings are fairly well reproduced, along with the width of the line.
The magnetosphere extent is similar in both cases, and in agreement with that of \Halpha's: a narrow and compact funnel.
It is, however, slightly degenerate with a wider and extended magnetosphere.
The inclination differs slightly between the core and full-line modeling cases.
The core-fittig yields two inclination minima, one at low inclination ($<30$\deg) and one at high inclination (75\deg). This degeneracy is lifted by the full-range modeling, which favors a highly inclined system.
The column temperature and accretion rate are in both cases degenerate between two regimes: low-\Mdot/high-\Tcol or high-\Mdot/low-\Tcol, the same trend that is observed for \Halpha's core component.




\begin{table}
    \caption{Core-component shock-modeling results\citep{aoyama_theoretical_2018}.}
    \label{tab:shock_best_fit_params}
    \centering
        \begin{tabular}{c c c c c}
        \hline
        Parameter & \Halpha & \Hbeta & \Hgamma & \Hdelta \\
        \hline
        $v_0$ (\kms) & 120 & 190 & 190 & 200 \\
        $\log(n_0)$ (1/m$^3$) & 19 & 19 & 20 & 20 \\
        \hline
        Date & \multicolumn{4}{c}{$\ffill \times1000$} \\
        \hline
        2021-10-25 - 05:42:33 & 5.77 & 1.98 & 0.56 & 0.66 \\
        2021-10-25 - 06:07:57 & 5.61 & 1.80 & 0.51 & 0.57 \\
        2022-10-13 - 02:52:03 & 11.20 & 2.27 & 0.61 & 0.74 \\
        2022-10-13 - 03:17:23 & 11.73 & 2.25 & 0.56 & 0.79 \\
        2022-10-14 - 06:52:50 & 11.73 & 2.59 & 0.90 & 1.16 \\
        2022-10-14 - 07:18:09 & 10.79 & 2.59 & 0.86 & 1.14 \\
        2022-10-15 - 04:40:38 & 9.31 & 2.61 & 0.79 & 1.04 \\
        2022-10-15 - 05:05:57 & 9.40 & 2.59 & 0.82 & 1.05 \\
        2022-10-15 - 06:30:49 & 9.84 & 2.81 & 0.87 & 1.14 \\
        2022-10-15 - 06:56:07 & 10.03 & 2.73 & 0.94 & 1.16 \\
        2022-10-17 - 03:02:29 & 6.09 & 2.03 & 0.63 & 0.75 \\
        2022-10-17 - 03:27:48 & 5.93 & 2.11 & 0.65 & 0.79 \\
        2022-11-01 - 01:21:16 & 6.81 & 2.21 & 0.65 & 0.71 \\
        2022-11-01 - 01:46:35 & 6.68 & 2.13 & 0.61 & 0.80 \\
        2022-11-04 - 00:49:41 & 7.47 & 2.52 & 0.80 & 0.93 \\
        2022-11-04 - 01:42:04 & 7.40 & 2.54 & 0.74 & 0.86 \\
        2022-11-04 - 02:07:24 & 7.19 & 2.40 & 0.75 & 0.91 \\
        2022-11-05 - 00:57:06 & 7.67 & 2.56 & 0.89 & 0.95 \\
        2022-11-05 - 01:22:25 & 7.75 & 2.73 & 0.84 & 1.02 \\
        2022-11-05 - 01:48:25 & 8.11 & 2.78 & 0.86 & 0.99 \\
        2022-11-05 - 02:13:44 & 7.89 & 2.61 & 0.81 & 0.98 \\
        2022-11-05 - 02:40:34 & 8.04 & 2.71 & 0.87 & 0.97 \\
        2022-11-05 - 03:05:51 & 7.75 & 2.63 & 0.87 & 1.00 \\
        2022-11-05 - 03:36:05 & 7.60 & 2.76 & 0.87 & 0.92 \\
        2022-11-05 - 04:01:24 & 7.89 & 2.68 & 0.75 & 0.90 \\
        2022-12-04 - 00:46:22 & 12.40 & 3.54 & 1.07 & 1.27 \\
        2022-12-04 - 01:11:41 & 12.06 & 3.57 & 1.08 & 1.32 \\
        2022-12-04 - 01:41:03 & 12.06 & 3.57 & 1.10 & 1.30 \\
        2022-12-04 - 02:06:22 & 11.41 & 3.54 & 1.08 & 1.25 \\
        2022-12-31 - 01:05:53 & 10.50 & 3.17 & 1.01 & 1.24 \\
        2022-12-31 - 01:31:12 & 9.93 & 3.02 & 0.96 & 1.14 \\
        2023-01-02 - 01:13:31 & 9.84 & 3.11 & 0.98 & 1.11 \\
        2023-01-02 - 01:38:50 & 9.49 & 2.84 & 0.96 & 1.04 \\
        \hline
        \end{tabular}
        \tablefoot{The best-fit filling factors are given in units of $10^{-3}$, assuming a planetary radius $\Rp = 1.5$~\Rjup.}
\end{table}

\subsection{Shock models}
Despite the small correlation factor of the core components to the UV excess, their shape resembles that of the prediction from \cite{aoyama_theoretical_2018} accretion shock models. However, the models do not predict lines similar to the wings component. Therefore, we only consider the core component for the fit.

\subsubsection{Model description}

The shock model includes two free parameters: the infall velocity $v_0$ and the post-shock density $n_0$.
An additional parameter, the shock surface, is introduced as a scaling factor for the line intensity.
Since the core component varies only in intensity and not in shape, the shock surface is the sole epoch-dependent parameter.
As with the magnetospheric accretion models, we compute the $\chi^2_r$ value across the entire parameter grid.

\begin{figure*}
    \centering
    \includegraphics[width=\linewidth]{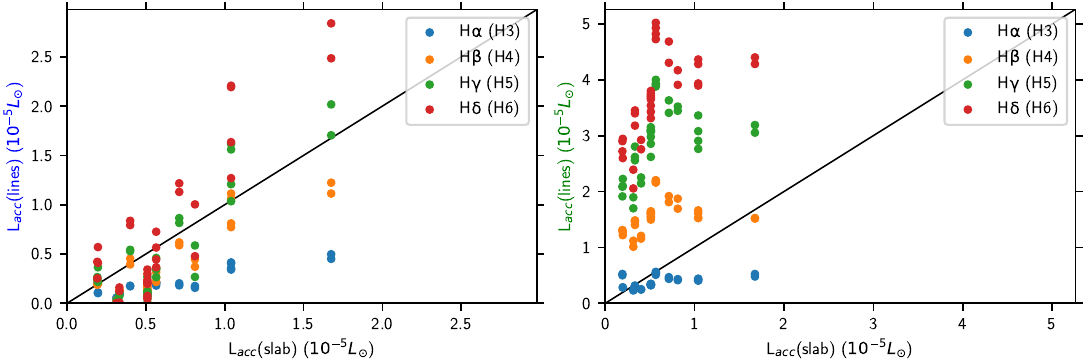}
    \caption{Correlations between the accretion luminosity (\Lacc) obtained from the UV excess fitting (slab-modeling) and the wings (left) and core (right) components fluxes \citep{alcala_x-shooter_2017}. The black line shows the $Y=X$ curve. The average accretion luminosity derived from the wings component of various lines is correlated with the accretion luminosity derived from that of slab-modeling (Sect.~\ref{sec:UV_modeling}), while the accretion luminosity derived from the core component is well above it.}
    \label{fig:UV_exess_Lacc_comparison}
\end{figure*}

\subsubsection{Core component modeling}

Figure~\ref{fig:modeling_shock} shows the best fit models to the core components of \Halpha\ to \Hdelta, along with the $\chi^2_r$ space. Figure~\ref{fig:modeling_shock_accretion_rate} shows the accretion rate computed from the models, depending on the epoch.
The accretion rate is computed using the relation $\Mdot = S \mu' n_0 v_0$, where $\mu'=2.3 \times 10^{-24}$~g is the mean molecular weight per hydrogen nucleus and $S=4\pi \Rp^2 \ffill$ is the emitting area given the filling factor \ffill \citep{aoyama_constraining_2019,aoyama_spectral_2020}.
We assume a planetary radius $\Rp=1.5$~\Rjup. The best-fit parameters are given in Table~\ref{tab:shock_best_fit_params}.

For \Halpha, the models broadly reproduce the width of the line and the red-wing asymmetry.
The central feature is too broad to fully match the line peak.
However, the model does reproduce the flat-topped line observed in the data. The best models are those for $v_0\sim120$~\kms\ and $n_0\sim10^{19}$~m$^{-3}$, which are slightly degenerate with $v_0\sim150$~\kms\ and 
$n_0\sim10^{18}$~m$^{-3}$.

For \Hbeta, \Hgamma\ and \Hdelta, the core components are much more symmetrical, which are well matched by the shock models, as shown by the corresponding $\chi^2_r$ values. For all three lines, however, the blue side of the line is consistently underestimated, while the red-side is well fitted. All three lines have a preference for high infall velocities $v_0>180$~\kms. \Hbeta\ fits at $n_0=10^{19}$~m$^{-3}$, similar to \Halpha, while \Hgamma\ and \Hdelta\ fit at $n_0=10^{20}$~m$^{-3}$. Both \Hgamma\ and \Hdelta\ converge near the grid limits, a trend also noted for the \PaBeta\ line of \gsc\ \citep{demars_emission_2023}.

As shown in Fig.~\ref{fig:modeling_shock_accretion_rate}, the accretion rate inferred from the core-component modeling varies between lines. They follow a consistent ordering: \Hdelta$>$\Hgamma$>$\Halpha$>$\Hbeta. Accretion rates derived from \Halpha\ and \Hbeta\ are compatible with each other within 1$\upsigma$, and so are \Hgamma\ and \Hdelta.
The \Mdot\ obtained from \Halpha\ modeling is systematically higher than that of \Hbeta, despite its lower $n_0$ ($10^{19}$ vs $10^{20}$ m$^{-3}$). This is due to the larger effective shock-surface, as highlighted by the values of $f_{\textrm{fill}}$ shown in Table~\ref{tab:shock_best_fit_params}. $f_{\textrm{fill}}(\Halpha)$ is on the order of $\sim1$\%, whereas $f_{\textrm{fill}}(\Hbeta)$ is on the order of $\sim0.2$\%. \Hgamma\ and \Hdelta\ are approximately half that of \Hbeta: $f_{\textrm{fill}}(\Hgamma) \sim f_{\textrm{fill}}(\Hdelta) \sim 0.1$\%.
Although all values are compatible within 3$\upsigma$, the systematic ordering across epochs hints at a persistent bias. It may be due to the model, or to the flux calibration, which remains challenging to achieve with UVES. However, \Hgamma\ and \Hdelta\ are located at nearby wavelengths (4341~\AA\ and 4103~\AA) and therefore within a single detector and arm of UVES, which makes relative calibration errors unlikely.

\begin{table}
    \caption{UV excess best-fit results for the slab models.}             
    \label{tab:accslab}      
    \centering
        \begin{tabular}{c c c}
        \hline
        Date & $\log(\Lacc/\Lsun)$ & \Mdot  \\
          & (dex) & (\MdotUS)   \\
        \hline           
        2021-10-25 & $-5.504$ & $1.24 \times 10^{-12}$ \\
        2022-10-13 & $-5.718$ & $7.57 \times 10^{-13}$ \\
        2022-10-14 & $-4.776$ & $6.63 \times 10^{-12}$ \\
        2022-10-15 & $-4.983$ & $4.11 \times 10^{-12}$ \\
        2022-10-17 & $-5.398$ & $1.58 \times 10^{-12}$ \\
        2022-11-01 & $-5.709$ & $7.73 \times 10^{-13}$ \\
        2022-11-04 & $-5.479$ & $1.31 \times 10^{-12}$ \\
        2022-11-05 & $-5.293$ & $2.01 \times 10^{-12}$ \\
        2022-12-04 & $-5.248$ & $2.23 \times 10^{-12}$ \\
        2022-12-31 & $-5.148$ & $2.81 \times 10^{-12}$ \\
        2023-01-02 & $-5.091$ & $3.21 \times 10^{-12}$ \\
        \hline
        \end{tabular}
\end{table}

\subsection{Analysis of the UV excess with slab models}
\label{sec:UV_modeling}

We modeled the UV excess as the emission of a slab of hydrogen following \cite{manara2013_accurate_accretion} under the assumption of local thermal equilibrium (LTE). To do so, we considered the X-Shooter spectrum of the class III young M9.5 dwarf TWA29 \citep{manara_x-shooter_2013} as a template of photospheric emission for \delorme\ and fitted simultaneously for visual extinction. 
To ensure a sufficient S/N, the modeling was performed on the mean spectrum of each night, at low spectral resolution ($R\sim1000$, convolved with a Gaussian kernel).
The accretion luminosity ($\mathrm{log(L_{acc})}$) and accretion rate (\Mdot) derived from the models are reported in Table~\ref{tab:accslab}. All fits converged to $A_V=0$.

\subsubsection[Comparison of Lacc and Mdot]{Comparison of \Lacc and \Mdot}

Figure~\ref{fig:UV_exess_Lacc_comparison} shows a comparison between accretion luminosities derived from slab-modeling ($\Lacc(\textrm{slab})$) and lines fluxes ($\Lacc(\textrm{line})$), obtained with the relationships from \cite{alcala_x-shooter_2017}.
The derived accretion luminosities from the slab model are broadly in line with that obtained from the wings component of all lines, and not consistent with that from the core component.
This further highlights a strong connection between the wings component and the UV excess, whereas the core component is not correlated to the accretion shock, as was shown from correlations in Fig.~\ref{fig:shapes_correlations}.
However, accretion luminosity derived from the wings component of different lines have different behaviors.
$\Lacc(\Halpha)$ and $\Lacc(\Hbeta)$ are consistently below  $\Lacc(\textrm{slab})$, whereas $\Lacc(\Hgamma)$ and $\Lacc(\Hdelta)$ are consistently above $\Lacc(\textrm{slab})$.
This is in line with the findings of \cite{alcala_x-shooter_2017} for stellar and sub-stellar objects, where the average \Lacc from all lines is in very good agreement with that obtained from the UV excess modeling, suggesting that the same phenomena may be at play.

The \Lacc and \Mdot computed here are compatible with that obtained from modeling the wings components with magnetospheric accretion models. The slab modeling yield results of a few $10^{-12}$~\Msun/yr, which is consistent with wings-component modeling for \Halpha\ as well. For \Hbeta\ however, the best fit results are obtained with higher accretion rate ($\sim 10^{-(10-11)}$~\Msun/yr). Still, the core-fitting of \Hbeta's wings component is highly degenerated with a much lower accretion rates, of a few $10^{-12}$~\Msun/yr, in line with the slab modeling. Yet the wings are only reproduced under higher accretion rate conditions. As for the core component, it consistently requires a high accretion rate, and is therefore not compatible with what we find from the UV excess fitting. We find the same results with the computed \Mdot\ and \Lacc\ from the shock-modeling (Fig.~\ref{fig:modeling_shock_accretion_rate}), which yield accretion rates an order of magnitude higher than those from slab modeling.

In summary, excess parameters derived from slab modeling align with that obtained from the wings components, both from empirical behaviors and magnetospheric accretion modeling. Instead, the core component's corresponding \Mdot\ and \Lacc\ are much higher than that from slab-modeling.

\subsubsection{UV and continuum variability}
\label{sec:continuum_variability}

\begin{figure*}
    \centering
    \includegraphics[width=\linewidth]{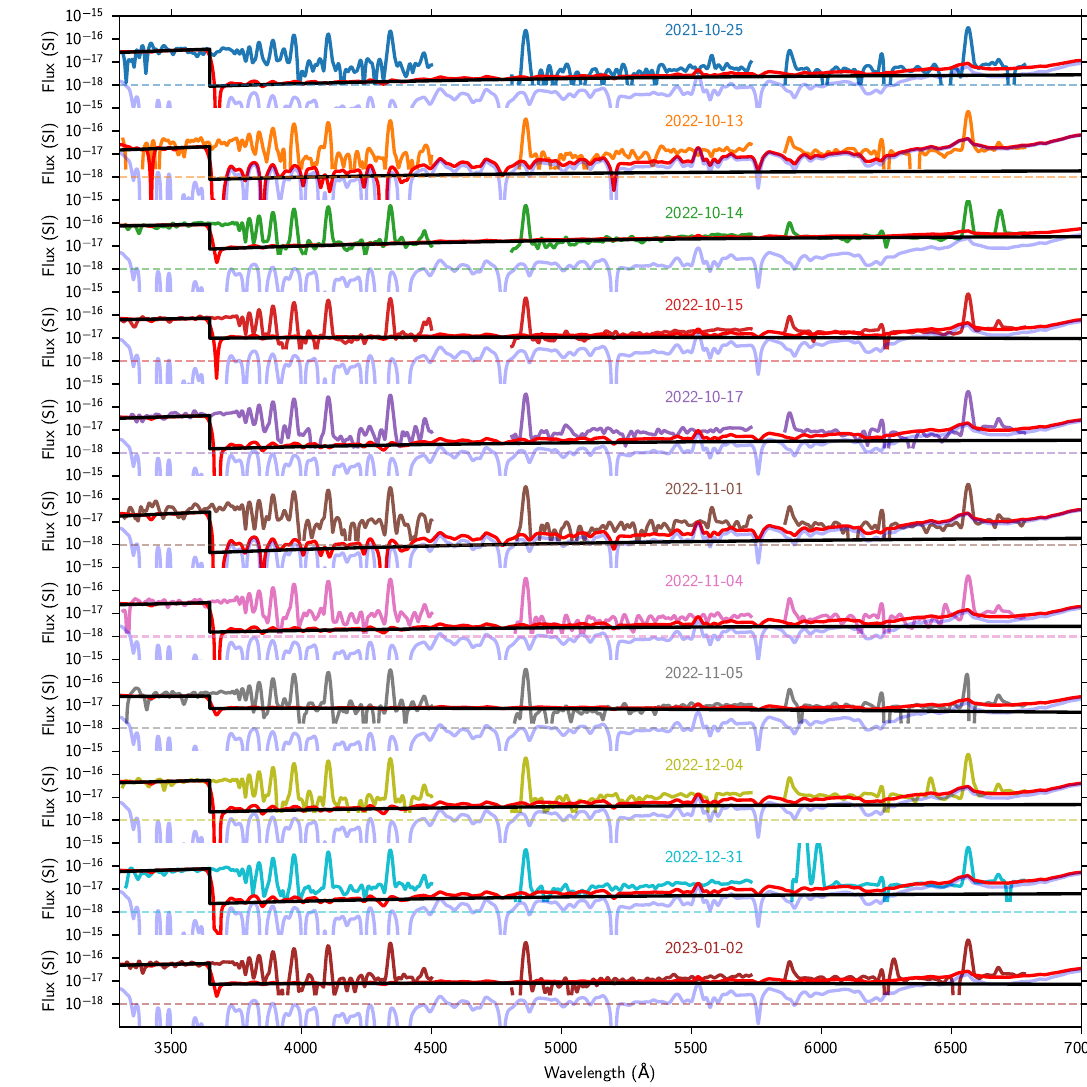}
    \caption{Best-fit results to the UV excess and general continuum level, with slab models. The colored lines represent the individual mean epochs. The best fit components are shown for the UV excess (black, slab), photospheric template (light blue, class III) and total model (red). The data were smoothed to $R\sim250$, points below 20\% the median level are masked, and a dashed line at $10^{-18}$~\ergsang\ is plotted at each epoch. All spectra are given in \ergsang\ units. The modeling provide a good fit to the UV excess, and show that almost all epochs have their continuum dominated by the shock-component extending to higher wavelengths. This is particularly evident in outburst epochs (2022-10-14/15).}
    \label{fig:slab_modeling}
\end{figure*}

Figure~\ref{fig:slab_modeling} shows the best-fit slab-models for all epochs.
While the UV excess is well reproduced, the model does not provide a good fit to the shape of the continuum. However, it's general flux level is 

The slab models are able to reproduce the height of UV excess ($<3700$~\AA), but not the location of the jump itself. This is a well known behavior as high-order Balmer lines blend near the true location of the jump, shifting its apparent location, while the slab model only accounts for continuum emission.
The general shape of the continuum (4000-7000~\AA) is not systematically reproduced by the models.
The modeled UV excess appears to extend all the way up to \Halpha, and dominates over the photospheric contribution in most epochs.
The best fits are for the outburst epochs (2022-10-14/15), where the residuals are only up to 25\% for certain wavelengths, and average to 0 overall. In other epochs, the residuals go up to 50\% in the RED arm, while the BLUE flux is not reproduced.
Overall, even with the residuals, the results are consistent with the correlation observed between the UV excess level, and the continuum level around \Halpha\ (Fig.~\ref{fig:shapes_correlations}).
This suggests that a fraction (if not most in some epochs) of the optical continuum on \delorme\ is actually due to the extension of the UV excess, which is strongly linked to the accretion rate.



The difference between our modeling and the data may have multiple origins. \cite{espaillat_measuring_2021} had used different slab components to account for the density structure of the accretion shock on GM Aurigae. They found best fits are achieved with 3 shock components. It is possible that our assumption of a single shock component may prove to be too simplistic, possibly explaining at least the lack of flux in the models in the BLUE arm. Alternatively, our spectra may have residual spectral shape distortions, but the stability of the calibration to the host binary (see Fig.~\ref{fig:appendix:primary_lines}) tend to suggest that the calibration should be fairly stable. It is hard to conclude whether the modeling or data should be improved. Still, the general conclusion remains that the UV excess extends to longer wavelengths, at least up to \Halpha (6564~\AA) where we observe the strongest continuum variability.


\section{Discussion}
\label{sec:discussion}

\subsection{Line variability origin and impact}

\cite{eriksson_strong_2020} had already given a glimpse of the hour-scale line variability of \delorme. Here, we explore hour to year timescales for the first time on that object.
The variability diagrams (Fig.~\ref{fig:variability_timescales}) show that, similarly to \gqlupb\ and \gsc, the variability at hour timescale is significantly lower than at longer timescales.
However, while \gqlupb\ and \gsc\ had non-negligible variability at the $\sim$hour timescale, \delorme\ lines only vary by a few \% (Fig.~\ref{fig:hour_timescale_variability}).
Higher order lines do reach 10--40\% variability, similar to that of the \PaBeta\ line of \gqlupb\ and \gsc. However, the line-flux variability is mostly driven by its core component, whereas the wings component does reach several $\sim10$~\% variability at \Halpha\ and higher transition orders.
The wings component variability is responsible for the changing shape of the line profile, in particular with the excess seen in the outburst epochs of October 13-17, 2022.
The general shape of the variability amplitude diagram is similar to that of \gsc, which may hint at a common emitting mechanism.

Previous studies have proposed that line variability could explain the low detection yield of \Halpha\ imaging campaigns \citep[e.g.,][]{follette_giant_2023}. However, if \delorme\ properties are typical of most PMCs, its moderate variability range at the hour ($<\sim10$~\%) to month ($<\sim50$~\%) timescales seems insufficient to account for the low detection rates.
The variability at the year timescale does reach factors of 2--3, but it is only driven by the 2021 epoch presented in \cite{ringqvist_resolved_2023}, and might not represent the full variability at this timescale.
\cite{close_three_years_2025} and \cite{zhou_evidence_2025} have studied the \Halpha\ line variability of PDS~70. They both found similar results, where the line varies by a factor of a few over years timescales, but never reaches a full order of magnitude.

Given the large sample of lines and epochs, we searched for periodicity in the data but found no statistically significant signal. In fact, any rotationally-modulated effects that may be present at days timescales are lost to aliases inherent to the random time-sampling of the data (obtained in service mode without pre-defined time constraints).
Future observations, if looking for rotational modulation of the lines, should focus on time-series of 4-5 hours over 3-6 days, as they typically cover the rotation periods of these objects, and the likely funnels that may periodically obscure the line of sight.

\subsection{Origin of the H {\sc i} Balmer lines in \delorme}

\subsubsection{Accretion origin: wings component}

Magnetospheric accretion emission models are able to reproduce the main features of both \Halpha\ components. In higher-order lines, the models provide a great fit to the central absorption of the wings components, and are also able to reproduce its wings, although not simultaneously.  
Such features are strongly reminiscent of magnetospheric accretion profiles predicted when funnel flows intersect the line of sight to hot spots on the stellar surface.

The autocorrelation maps of \Hbeta and \Hgamma in Figure \ref{fig:auto_correlations} are also qualitatively similar to the predicted maps from \cite{alencar_accretion_2012} based on the MHD simulations of magnetospheric accretion funnels launched from an inclined dipole from \cite{2011MNRAS.411..915R}.

Still, the ability of fixed-profile components to adequately fit all epochs suggests either a well-aligned magnetic dipole, persistent visibility of at least one accretion column, or a pole-on viewing angle of a more complex magnetospheric geometry.
In the latter scenario, the whole observed variability would be caused by changes in the accretion rate and/or column temperature.
However, observations of magnetospheres in T Tauri stars show that the dipole is usually moderately inclined with respect to the stellar rotation axis \citep{mcginnis_magnetic_2020}, and the modeling shows that both the wings and core components are best reproduced by magnetospheric accretion models at intermediate to high inclination.
The magnetosphere may still be axisymmetrical, with models converging to high-inclination to simulate a non axisymmetrical column seen at an angle.
This may lead to substantial rotational modulation of the line profiles, if the magnetosphere were not axisymmetrical. The identification of the rotation period of Delorme 1 (AB)b coupled to a determination of the v.sin(i) will be important to better constrain the models. In addition, a dense spectroscopic monitoring on timescales of the rotation period will be critical to further test the magnetospheric accretion scenario.

Overall, it seems likely that the wings component is linked to ongoing accretion, both from the model-fitting and the strong correlation with the UV excess, the derived \Mdot\ and \Lacc.
Although the wings component has wings extending to velocities reminiscent of winds, no evident wind or jet tracers (e.g., OI lines) are detected, suggesting they are either absent or below the detection threshold.
With broad values of accretion rate of $10^{-12}$ to $10^{-11}$ ~\Msun/yr, the planet would only accrete about 0.03--0.3~Mjup over its 35~Myr of age, suggesting that it is in the late phases of accretion, or that accretion may occur in bursts rather than a homogeneous process.

\subsubsection{Accretion origin: core component}

The magnetospheric accretion models provide a reasonable fit to the core component, although there are no definitive absorption features as there are for the wings component. If the core component is in fact also due to magnetospheric accretion, it could not originate from the same emitting region. \cite{thanathibodee_complex_2019} proposed an onion-like funnel structure for the low-accretor CVSO~1335, where two funnels are launched from two different truncation radii.
This scenario may explain the goodness of the magnetospheric accretion fits, as well as the relative shape of the two components.
\cite{thanathibodee_complex_2019} also suggest that a companion in the accretion circum-stellar disk may open a gap, causing two distinct accretion flows.
Having a more complex magnetospheric accretion geometry would certainly better reproduce both the central absorption and wing structures of the wings component of \delorme. 
The core component is seen in emission broadly where the wings component has its main absorption feature, which could be a sign of the relative projection to the observer of each emitting region. However, the physical parameters of the fits do not yield a core-component funnel that is strictly encapsulating the wings component funnel. This might be due to the simplicity of assuming an axisymmetrical flow.
Further modeling, for example with inclined dipole models, is required.

Alternatively, if the magnetospheric accretion scenario holds, at least for the wings component, the ability of magnetospheric accretion models are able to separately model the wings and the core of the line (Fig.~\ref{fig:modeling_thanathibodee_best_fits}) may also be explained by this onion-like structure. In this scenario, a line decomposition with three components might have unveiled the structure of the accretion columns. However, this was not be done because a three-component decomposition would be too unstable (in an algorithmic, numerical sense) given the method described in Appendix.~\ref{sec:appendix:nmf}.

\subsubsection{The shock scenario}
The correlations between wings components and UV excess (Fig.~\ref{fig:shapes_correlations}) suggest that the wings component emission emerges only past a certain threshold level of the UV excess emission. Although it is possible that the correlation may not extend linearly to a zero origin, there are still data-points very close to zero, for instance the wings component of all lines on October 25, 2021.
As for the core component, the correlation could similarly be interpreted as the UV excess requiring a certain threshold of core component emission for UV excess to appear. However, the poor correlation likely means that if the core component and UV excess emission are related, this is only to a moderate degree.
Still, the width of the core component fits surprisingly well with shock models \citep{aoyama_theoretical_2018}, and the retrieved parameters have an infall velocity ($>170$~\kms) compatible with that one might expect from magnetospheric accretion: shock at planetary surface ($R\sim1.5$~\Rjup) coming from a truncation radius of $\sim10$~\Rjup.
This would suggest that the core component is in fact emitted at the shock, but then one might expect a strong correlation with the UV excess, as they would trace more-or-less the same region. Radiative transfer effects, such as optical depths or opacities, could be the explanation, but have not been modeled.
While the wings component would be emitted in a broadly optically thin medium (the part of the accretion column close to the planet), the post-shock region emission could be saturated, explaining the plateau-like shape at the peak of the core component in low-order lines.
This plateau diminishes with higher order lines (\Hbeta, \Hgamma, \Hdelta), and increases the correlation with the UV excess, as is the case from Fig.~\ref{fig:shapes_correlations}, although this pattern breaks down for \Halpha's core component, which shows stronger correlation with UV excess than the higher-order transitions.

While we did not try to fit any shock component to the wings component, it is also possible that part of the wings component may originate from a shock, which would not be surprising given its strong correlation with the UV excess.
However, whether this component would be significant relative to the total line flux is difficult to confirm, and adding free-parameters for a shock-column emission model would only accentuate the already existing degeneracies. Still, the wings component profiles are broadly reminiscent of that predicted by \cite{marleau_accreting_2022} for a shock-emission in a magnetospheric accretion scenario, when the column dust and gas opacities are included, although this was true for quite high accretion rates ($>10^{-8}$~\Msun/yr) compared to the values we find in this paper.

Figure~\ref{fig:slab_modeling} shows that the continuum of \delorme seems to be modulated by the accretion shock. In fact, we find that the continuum level around \Halpha increases by a factor $\sim 2.5$ in outburst epochs (2022-10-15/15/17). This strong contribution of the UV excess to higher wavelength is consistent with the apparent over-luminosity of the source. \cite{Almendros-Abad_2025_accretion_burst_free_floating_pmo} has recently identified a long-lived accretion burst in the free-floating planetary-mass object Cha1107-7626. They find that the flux levels around \Halpha increase by a factor 3--6 within the outburst, which is of the order of magnitude of our findings. However,the duration of their accretion burst is much longer that what we see for \delorme (a few days). It is possible that our UVES observations were captured within an accretion burst similar to Cha1107-7626, and that the burst observed on 2022-10-14/15/17 is only a smaller modulation of a longer general variation. While the \Halpha shape variability observed in \cite{Almendros-Abad_2025_accretion_burst_free_floating_pmo} is much different from that observed for \delorme, this may be due to different accretion geometry, viewing angle, or regime.


\subsubsection{Unresolved binarity}
\label{subsec:binarity}
The two components within the lines (wings and core) could also be caused by two tight planetary-mass objects. The overluminosity caused by binarity could have been left unnoticed in color-magnitude diagrams since the color variation along the late-M early-L dwarf sequence is not strong, especially at near-infrared wavelengths. One tight planetary-mass binary (2M0441) in Taurus has been identified as part of a quadruple system \cite{2010ApJ...714L..84T, 2015ApJ...811L..30B}. But the architecture of the system remains different from Delorme 1(AB)b. 
 
So far, AO-assisted imaging of \delorme\ \citep{delorme_direct-imaging_2013, eriksson_strong_2020} has not found such evidence (equal mass binaries with a separation greater than $\sim30$~mas or $\sim1.4$~au would have been detected). An equal-mass binary on a circular orbit seen edge-on with a semi-major axis smaller than 1.4~au would induce a semi-amplitude of radial velocity shift on the profiles or photospheric lines greater than $7.35$~\kms. We looked for cross-correlation signals in the mean UVES spectrum of the companion, but could not find any and our line profile decomposition does not allow -- by construction -- to identify shifts in wavelengths.
Alternatively, searching for such variability within the emission lines themselves would be quite difficult given the intricacy between the wings and core components.
Future observations of the companion with near-infrared echelle spectrographs, such as CRIRES \citep{2023A&A...671A..24D} or with the upgraded VLTI/GRAVITY \citep{2022Msngr.189...17A} will be more suitable to explore that scenario.

\subsubsection{Contribution from chromospheric activity}
\label{sec:chromospheric_contribution}

One possible explanation for the emission lines is chromospheric activity, where the lines (or one of the line components) would not be emitted through accretion processes, but rather in the chromosphere of the object.
Chromospheric activity is known to play an important role in the emission lines of late M dwarfs, and has been regarded as a noise in the context of accretion.
\cite{manara_x-shooter_2013,manara_extensive_2017} proposed that line brightness falls into two regimes: one where the line intensity is below a threshold where it is compatible with chromospheric activity, and one where it is above that threshold and most of the line flux comes from accretion. In the case of \delorme, \cite{eriksson_strong_2020} found that using common line-luminosity--\Mdot\ relations (although obtained from CTTS observations), \delorme\ mass accretion rate is far above the threshold for the lines to be originating from chromospheric activity. However, they used VLT/MUSE observations, in which the line profiles were unresolved.
The work presented here shows that Balmer line profiles have plateau-like shapes for \Halpha\ and \Hbeta, and symmetrical profiles for higher order lines. This plateau-like shape is surprisingly similar to that found for active M dwarfs \citep[M4--M5;][]{fuhrmeister_carmenes_2018,muheki_high-resolution_2020}, although they lack the extended wings observed in \delorme. The core-component obtained from line decomposition, however, does not display these wings, and serves as a good match to the profiles of active M-dwarfs, with a similar FWHM of $\sim50$ \kms.
If the core component is in fact a sign of chromospheric activity, it would still exceed the expected chromospheric noise floor, which is supposed to decrease with later spectral types, making its apparent brightness even more of an issue.
We also verified that the core components' converted accretion luminosity is above the threshold for chromospheric contamination from \cite{manara_extensive_2017}. Interestingly, the wings component does have fluxes compatible with chromospheric emission, although the line morphology is inconsistent with known chromospheric signatures \citep{fuhrmeister_carmenes_2018, muheki_high-resolution_2020}.

Activity on stellar and sub-stellar objects is often linked to their rotational velocity \citep{stauffer_rotational_1997_I,stauffer_rotational_1997_II,stauffer_keck_spectra_1998,newton_emission_2017}. However, the $v \sin i$, period or inclination of \delorme\ are currently unknown, and it is unclear whether these correlations extend to the planetary-mass domain.
It has also been shown that chromospheric \Halpha\ emission may be indicated by \Halpha-TiO features \citep{fang_stellar_2020} at $\sim7000$--$7100$~\AA, yet \cite{eriksson_strong_2020} did not find any of these features in their MUSE spectra.
If the UV excess contribution does extend to these wavelengths, it is possible that the TiO features have been obscured by continuum veiling associated with shock emission.

Overall, the origin of the core component is unclear. The fact that shock models provide such a good fit suggests that it is due to the shock, but its profile similarity with active M dwarfs suggests that it is due to activity. It is possible that the core component may be entirely be due to either accretion or activity, or a combination of the two. The number of class III objects with UV and optical spectra at such low masses, age, and spectral types (L0 and later) is quite low and a larger sample is required to refine the chromospheric noise limit and ensure it is not underestimated.

\subsection{Measuring accretion rates}

In this paper, we derived accretion rates following various methods: line-profiles fitting, line-flux conversion and UV excess fitting. The various estimates for the wings and core components do not provide a consistent accretion rate. For example, magnetospheric accretion models provide accretion rates that are strongly degenerated with other parameters (e.g., column temperature), and shock models assume (by design) that the lines are entirely emitted at the shock, which does not seem to apply to the case of \delorme. Instead, the UV excess offers a more direct measurement of the accretion luminosity.

Similarly to what has been found for stellar and sub-stellar objects, the accretion luminosity of the UV excess is well in line with the average of that derived from line-flux relationships \citep{alcala_x-shooter_2017}, at least for the wings component, which is linked to accretion. Overall, we find from UV excess modeling that over the course of our observation campaign, \delorme\ had a broadly stable accretion rate of $(1-3) \times 10^{-12}$~\Msun/yr, with an outburst epoch of $\sim 6.63 \times 10^{-12}$~\Msun/yr. This is to be compared with the results from \cite{betti_near-infrared_2022,betti_erratum_2022}, \cite{eriksson_strong_2020} and \cite{ringqvist_resolved_2023}.

Using stellar scalings \citep{alcala_x-shooter_2017}, \cite{eriksson_strong_2020} found an accretion rate of $3.4 \times 10^{-13}$~\Msun/yr on 2018-09-18.
\cite{betti_near-infrared_2022,betti_erratum_2022} found that between 2021-11-20 and 2022-01-24, \delorme\ had its accretion rate vary within the range $\sim 5 \times 10^{-13}-5 \times 10^{-12}$~\Msun/yr.
Both are broadly in line with our results, even without decomposing the line as we did here. Given that the line was quite faint in our epoch 2021-10-25 \citep{ringqvist_resolved_2023}, it is quite possible that the optical lines studied by \cite{eriksson_strong_2020} were even fainter, artificially reducing the importance of the core component, or that the core component was simply absent during these observations (although we do not find any evidence for its disappearance). As for the results of \cite{betti_near-infrared_2022,betti_erratum_2022}, they used near-infrared lines which are usually thought to be devoid of chromospheric emission. If the core component is in fact due to chromospheric contribution, it is possible that it may be absent from the \PaBeta\ line, and as such \cite{betti_near-infrared_2022,betti_erratum_2022} would have only computed accretion rates from the wings component. This would need to be checked with observations at resolutions similar to that of UVES, for example: VLT/CRIRES.
Alternatively, it is also possible that the agreement of the measurements over multiple years are a simple coincidence, where lines variability and intrinsic biases would have negated one another.

While not directly discussed in this paper, extinction may have some impact on the line fluxes, and results. Extinction is commonly studied for stars and their circum-stellar disk, but has also been studied in the vicinity of forming planets. For example, \cite{szulagyi_hydrogen_2020} found that observed spread of line variability increases by a few \%, up to 20\% depending on the planet mass, with stronger effects on lower mass planets. While \delorme is not located within the circumstellar disk of its host binary, the  dynamics within the CPD itself may result in variable extinction. This effect is however hard to to measure, as extinction should affect uniformly emission lines and their local continuum, and it was shown in Sect.~\ref{sec:UV_modeling} that said continuum variability is also linked to the variability of the UV excess (and accretion rate). Our modeling still accounted for extinction as a free parameter, and converged to $A_V=0$ for all epochs, which seems to indicate that extinction should not affect our results. However, should our findings be applied to the study of embedded planets, extinction should be considered, especially in light of the recent results from \cite{cugno2025_extinction}: they find that extinction may reach values of up to $A_V = 4.6$ in gap-opening planets.

\section{Conclusion}
\label{sec:conclusion}

This study presents the results of a high-resolution UV/Visible monitoring campaign of \delorme, conducted with VLT/UVES at $R\sim50,000$.
The main results and conclusions are:

\begin{itemize}
    \item[--] We detect Balmer emission lines, whose profiles partly resemble those seen in free-floating objects of similar mass, including 2M1115 \citep{viswanath2024_entropy_I}. 
    
    \item[--] \delorme\ lines have a limited profile variability that is mostly exhibited in the October 14 and 15, 2022 epochs. The line flux variability amplitude at hours timescales reaches up to $\sim10$~\%, while longer timescales reach up to $\sim100$~\%.
    
    \item[--] We detect a UV excess, blueward of 3750~\AA, near the location of the Balmer jump. We find the UV excess to be variable both in flux and shape. This is typical of ongoing accretion on the target, confirming the accreting nature of \delorme\ despite its estimated old age (30--45~Myr).
    
    \item[--] The UV excess modeling allows us to infer accretion rates of $\Mdot\approx (1$--$3) \times 10^{-12}$~\Msun/yr among most epochs with a peak at $\Mdot \sim 6.6 \times 10^{-12}$~\Msun/yr on the 2022-10-14 epoch. \delorme\ seems to be a weak accretor at these epochs.
    
    \item[--] The line ratios resemble the shape of Type~3 decrements from \cite{antoniucci_x-shooter_2017}, up to \Hepsilon/H7, which they could not interpret given the models available at the time of the study. However, the lines decrement matches that of Type~2 decrements at higher orders, that correspond to optically thin emitting regions. 
    
    \item[--] All lines can be decomposed into two static components whose relative intensities vary between epochs, which we named wings and core component throughout this paper. This two-component decomposition is able to explain all of the auto-correlation features of the lines and provides a fit to all lines of all epochs. The higher amplitude variability of the wings component is found to be responsible for variations in line profile shape, in particular in the October 10, 14, and 15 epochs.
    
    \item[--] The wings component shape for all lines is reminiscent of red-shifted absorptions and asymmetries induced by infalling accretion flows. The high correlation factor between the scaling of the wings components across all lines and the UV excess level strongly suggests that this wings component is produced by accretion. This is further supported by axisymmetric magnetospheric accretion models that are broadly able to reproduce the main features of the wings component, even though they cannot simultaneously reproduce the core and the wings of the lines.
    
    \item[--] The core component correlates with the UV excess flux, albeit to a lesser extent. We cannot exclude that this component originates from chromospheric activity. The core component explains the strongest features of the autocorrelation diagrams and dominates the emission lines presented in \cite{ringqvist_resolved_2023}. It can be reproduced qualitatively by 1D accretion shock models as well, but the lack of strong correlation with the UV excess remains to be explained. Overall, its origin is not completely clear.
    
    \item[--] Finally, unlike what was observed in \cite{demars_emission_2023}on two other younger companions, the \Halpha\ variability amplitude on hourly timescales is not able to explain the low detection rates of \Halpha\ surveys. Its variability on longer timescales may still provide a reasonable explanation for the lack of detections.
\end{itemize}

In this work, we presented the first analysis of variability in resolved line profiles of an accreting PMC. Our analysis complements that of \cite{ringqvist_resolved_2023} and \cite{viswanath2024_entropy_I}, where we presented the first high-resolution Balmer line profiles of accreting PMCs.
While this work mostly probed the hours and months timescales variability, it is still lacking the days-timescales sampling that would be required to assess whether rotational modulation is present in the data.
We also stress that the work presented here shows that the understanding of accretion processes on PMCs requires both multi-epoch and multi-wavelength analysis.
This is highlighted by the complementary nature of the line modeling and the correlations between the different accretion tracers.

\begin{acknowledgements}

We acknowledge support in France from the French National Research Agency (ANR) through project grant ANR-20-CE31-0012 and the Programmes Nationaux de Plan\'etologie et de Physique Stellaire (PNP and PNPS).
G.-D.~Marleau acknowledges the support of the DFG priority program SPP 1992 ``Exploring the Diversity of Extrasolar Planets'' (MA~9185/1), from the European Research Council (ERC) under the Horizon 2020 Framework Program via the ERC Advanced Grant ``ORIGINS'' (PI: Henning), Nr.~832428,
and via the research and innovation programme ``PROTOPLANETS'', grant agreement Nr.~101002188 (PI: Benisty).
Parts of this work have been carried out within the framework of the NCCR PlanetS supported by the Swiss National Science Foundation.
C.~F.~Manara acknowledges funding from the European Research Council (ERC WANDA: 101039452). Views and opinions expressed are however those of the author(s) only and do not necessarily reflect those of the European Union or the European Research Council Executive Agency. Neither the European Union nor the granting authority can be held responsible for them.
J.~Bouvier acknowledges funding from the European Research Council (ERC) under the European Union's Horizon 2020 research and innovation programme (grant agreement no. 742095; SPIDI: Star-Planets-Inner Disk Interactions, \url{https://www.spidi-eu.org})
S.P. is supported by the ANID FONDECYT Postdoctoral program No. 3240145.
This publication makes use of VOSA, developed under the Spanish Virtual Observatory (\url{https://svo.cab.inta-csic.es}) project funded by MCIN/AEI/10.13039/501100011033/ through grant PID2020-112949GB-I00. VOSA has been partially updated by using funding from the European Union's Horizon 2020 Research and Innovation Programme, under Grant Agreement Nr.~776403 (EXOPLANETS-A).
This research has made use of the SVO Filter Profile Service ``Carlos Rodrigo'', funded by MCIN/AEI/10.13039/501100011033/ through grant PID2020-112949GB-I00.

This research has made use the following Python libraries: numpy \citep{numpy}, scipy \citep{scipy}, astropy \citep{astropy_I,astropy_II,astropy_III}, matplotlib \citep{matplotlib}, pandas \citep{pandas_publication}.

\end{acknowledgements}

\bibliographystyle{yahapj}
\bibliography{ref}

\begin{appendix}

\section{Observation log}
\label{sec:appendix:observation_log}
The observation log can be found in Table~\ref{tab:obs}.

\begin{table}
\caption{Observation log}
\label{tab:obs}
\small
\begin{tabular}{ccccccc}
\hline\hline
Date & UT start & DIT & PWV & seeing & $\mathsf{{\uptau_{{0}}}}$ & airm \\
 & (h:m:s) & (s) & (mm) & ($''$) & (ms) &  \\
\hline

2021-10-25 & 05:42:33 & 1482 & 2.17 & 0.53 & 10.50 & 1.28 \\
2021-10-25 & 06:07:57 & 1482 & 2.05 & 0.49 & 11.30 & 1.34 \\
\hline
2022-10-13 & 02:52:03 & 1482 & 1.55 & 1.06 & 1.90 & 1.21 \\
2022-10-13 & 03:17:23 & 1482 & 1.52 & 1.25 & 2.20 & 1.19 \\
\hline
2022-10-14 & 06:52:50 & 1482 & 1.18 & 0.45 & 7.70 & 1.34 \\
2022-10-14 & 07:18:09 & 1482 & 1.13 & 0.52 & 8.10 & 1.42 \\
\hline
2022-10-15 & 04:40:38 & 1482 & 0.85 & 0.62 & 7.40 & 1.17 \\
2022-10-15 & 05:05:57 & 1482 & 0.87 & 0.56 & 10.20 & 1.18 \\
2022-10-15 & 06:30:49 & 1482 & 0.79 & 0.54 & 4.50 & 1.30 \\
2022-10-15 & 06:56:07 & 1482 & 0.77 & 0.53 & 6.90 & 1.36 \\
\hline
2022-10-17 & 03:02:29 & 1482 & 1.16 & 0.91 & 4.70 & 1.19 \\
2022-10-17 & 03:27:48 & 1482 & 1.10 & 0.81 & 4.50 & 1.17 \\
\hline
2022-11-01 & 01:21:16 & 1482 & 2.00 & 0.78 & 2.20 & 1.24 \\
2022-11-01 & 01:46:35 & 1482 & 2.29 & 0.69 & 3.80 & 1.20 \\
\hline
2022-11-04 & 00:49:41 & 1482 & 1.28 & 0.68 & 6.40 & 1.27 \\
2022-11-04 & 01:42:04 & 1482 & 1.47 & 0.54 & 5.10 & 1.20 \\
2022-11-04 & 02:07:24 & 1482 & 1.52 & 0.50 & 7.30 & 1.17 \\
\hline
2022-11-05 & 00:57:06 & 1482 & 2.57 & 0.90 & 6.10 & 1.25 \\
2022-11-05 & 01:22:25 & 1482 & 2.66 & 0.69 & 5.50 & 1.21 \\
2022-11-05 & 01:48:25 & 1482 & 2.71 & 0.68 & 4.40 & 1.19 \\
2022-11-05 & 02:13:44 & 1482 & 2.69 & 0.66 & 4.40 & 1.17 \\
2022-11-05 & 02:40:34 & 1482 & 2.68 & 0.68 & 6.10 & 1.16 \\
2022-11-05 & 03:05:51 & 1482 & 2.59 & 0.64 & 7.10 & 1.16 \\
2022-11-05 & 03:36:05 & 1482 & 2.41 & 0.73 & 5.80 & 1.18 \\
2022-11-05 & 04:01:24 & 1482 & 2.28 & 0.84 & 5.30 & 1.20 \\
\hline
2022-12-04 & 00:46:22 & 1482 & 4.70 & 0.64 & 8.40 & 1.16 \\
2022-12-04 & 01:11:41 & 1482 & 4.86 & 0.64 & 9.20 & 1.16 \\
2022-12-04 & 01:41:03 & 1482 & 4.89 & 0.63 & 6.60 & 1.18 \\
2022-12-04 & 02:06:22 & 1482 & 4.83 & 0.74 & 5.60 & 1.20 \\
\hline
2022-12-31 & 01:05:53 & 1482 & 1.19 & 0.61 & 4.60 & 1.26 \\
2022-12-31 & 01:31:12 & 1482 & 1.28 & 0.56 & 5.10 & 1.31 \\
\hline
2023-01-02 & 01:13:31 & 1482 & 3.77 & 0.45 & 7.00 & 1.29 \\
2023-01-02 & 01:38:50 & 1482 & 3.63 & 0.53 & 6.50 & 1.35 \\
\hline
\end{tabular}
\end{table}

\section{Primary lines and spectrum}
\begin{figure*}
    \centering
    \includegraphics[width=\textwidth]{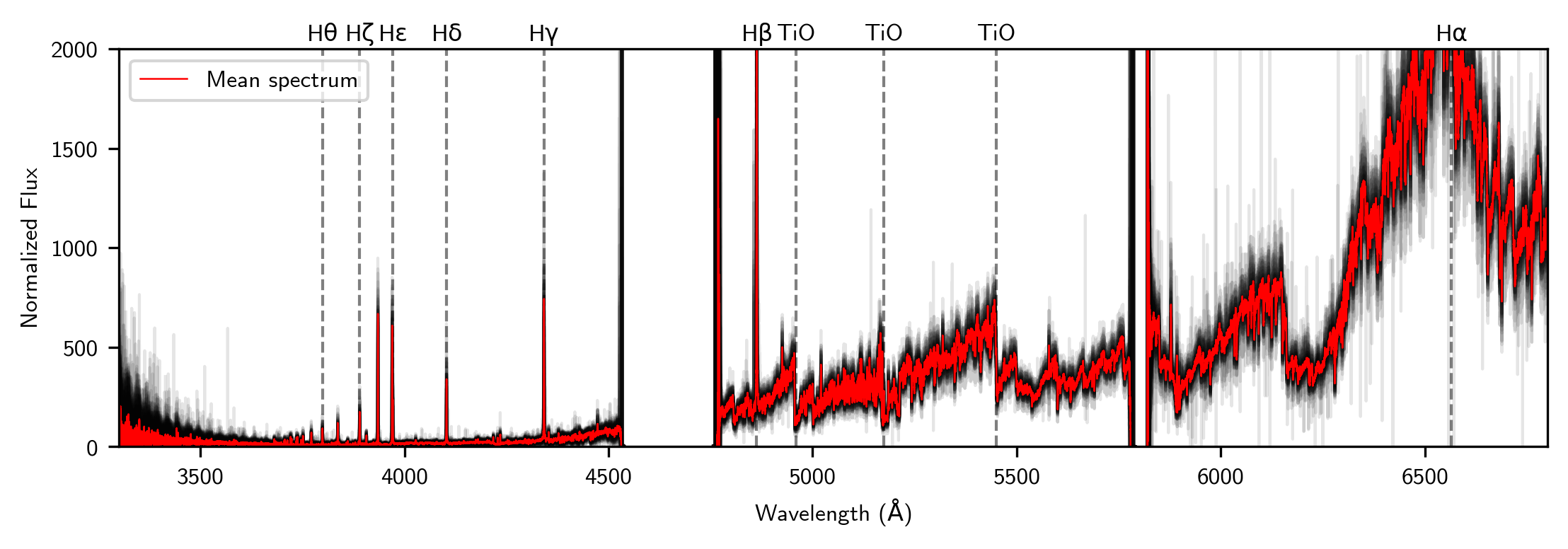}
    \caption{In black: primary spectra for all epochs, normalized to 1 around \Halpha. The mean spectrum is shown red. There are no apparent discrepancies between the spectral slopes of the various epochs.}
    \label{fig:appendix:prim_overlap}
\end{figure*}

\begin{figure*}
    \centering
    \includegraphics[width=\textwidth]{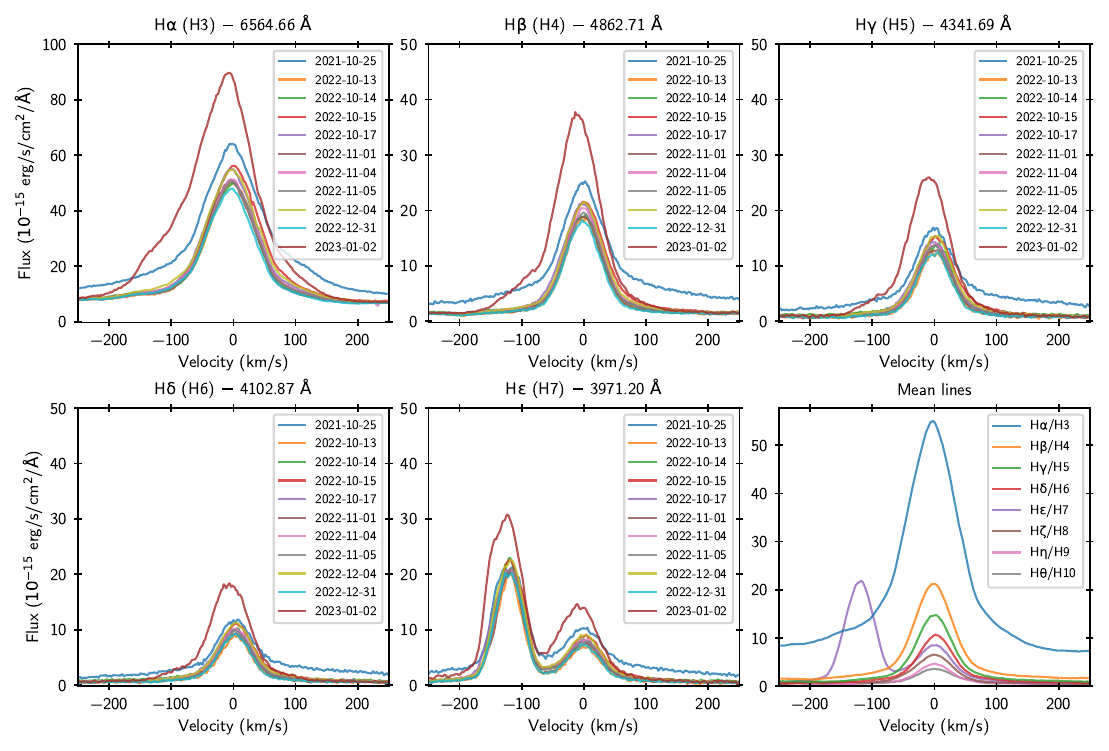}
    \caption{Balmer lines of the host binary of \delorme.}
    \label{fig:appendix:primary_lines}
\end{figure*}

The spectrum of the host binary Delorme~1~(AB) is shown in Fig.~\ref{fig:appendix:prim_overlap}, with all epochs overlapped in black.
Its Balmer lines are shown in Fig.~\ref{fig:appendix:primary_lines}. The line profiles are highly different from that of \delorme, which serves as an additional check for the de-blending of the lines of the companion from that of the primary.

\section{Companion individual lines}
\label{sec:appendix:all_lines}
\begin{figure*}
    \centering
    \includegraphics[width=\textwidth]{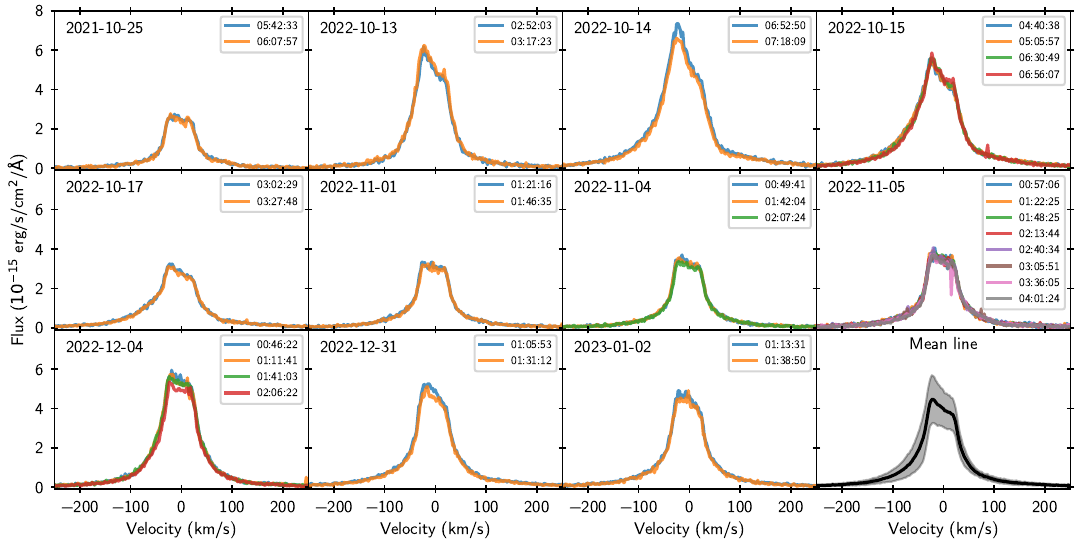}
    \caption{\Halpha\ line for all individual epochs. Times are UT times in hh:mm:ss format. In the last panel, the shaded region is the standard deviation among all epochs. $\uplambda_0=6564.66464$~\AA (vacuum).}
    \label{fig:appendix:all_halpha_lines}
\end{figure*}

\begin{figure*}
    \centering
    \includegraphics[width=\textwidth]{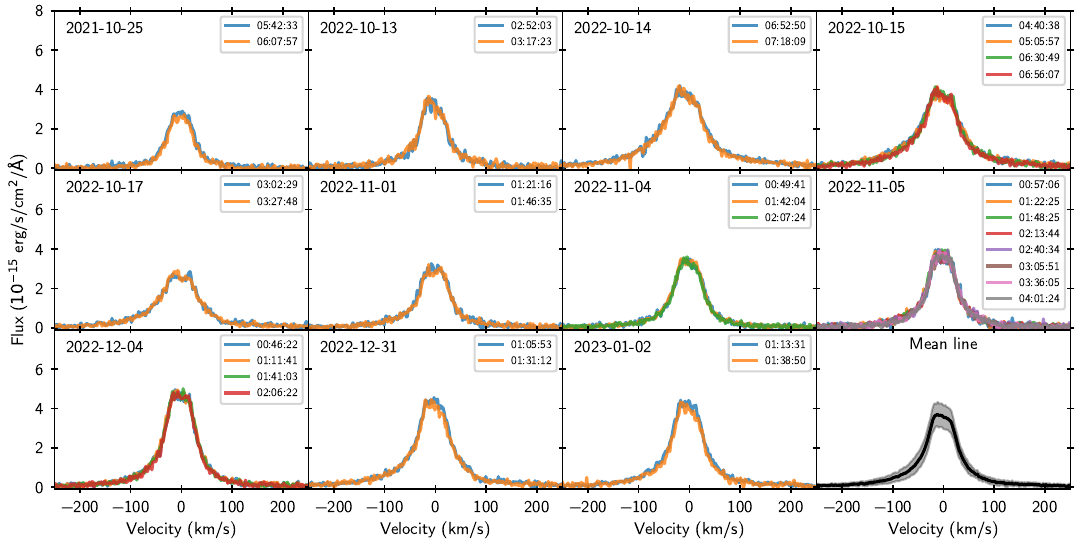}
    \caption{\Hbeta\ line for all individual epochs. Times are UT times in hh:mm:ss format. In the last panel, the shaded region is the standard deviation among all epochs. $\uplambda_0=4862.71$~\AA (vacuum).}
    \label{fig:appendix:all_hbeta_lines}
\end{figure*}

\begin{figure*}
    \centering
    \includegraphics[width=\textwidth]{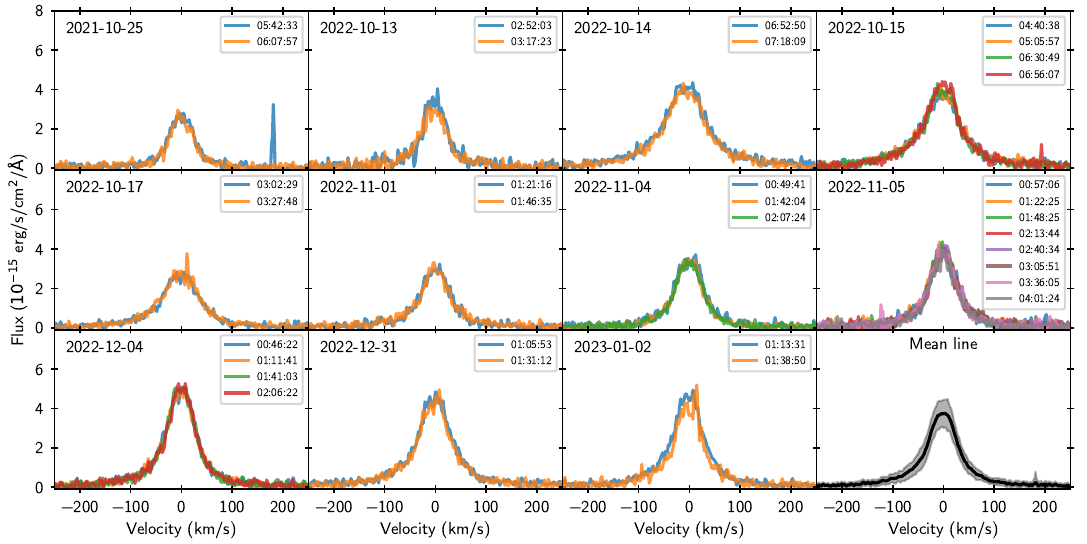}
    \caption{\Hgamma\ line for all individual epochs. Times are UT times in hh:mm:ss format. In the last panel, the shaded region is the standard deviation among all epochs. $\uplambda_0=4341.692$~\AA (vacuum).}
    \label{fig:appendix:all_hgamma_lines}
\end{figure*}

\begin{figure*}
    \centering
    \includegraphics[width=\textwidth]{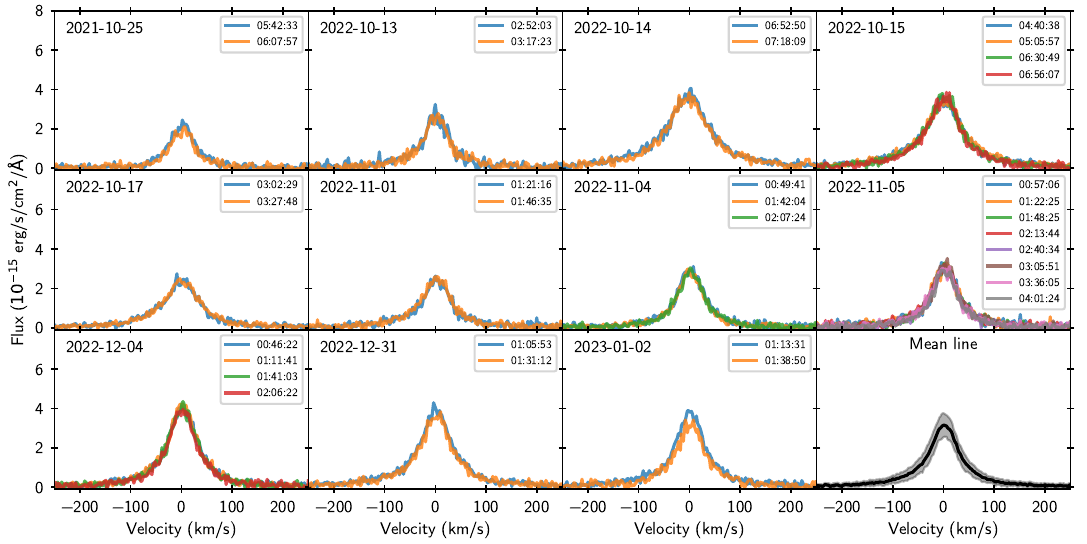}
    \caption{\Hdelta\ line for all individual epochs. Times are UT times in hh:mm:ss format. In the last panel, the shaded region is the standard deviation among all epochs. $\uplambda_0=4102.86503488$~\AA (vacuum).}
    \label{fig:appendix:all_hdelta_lines}
\end{figure*}

The lines for each individual exposure can be found in Figs.~\ref{fig:appendix:all_halpha_lines}, \ref{fig:appendix:all_hbeta_lines}, \ref{fig:appendix:all_hgamma_lines} and \ref{fig:appendix:all_hdelta_lines} for \Halpha, \Hbeta, \Hgamma\ and \Hdelta\ respectively. The lines fluxes for each line can be found in Table~\ref{tab:full_lines_fluxes}. The integrated flux of each wings/core component can be found in Table~\ref{tab:shapes_lines_fluxes}.


\begin{sidewaystable*}
\centering
\caption{Integrated lines fluxes of \delorme, where cgs stands for erg/s/cm$^2$ in the case of lines, and \ergsang\ for the UV excess mean flux, computed on the 3350--3600~\AA\ range.}
\label{tab:full_lines_fluxes}
    \begin{tabular}{c| ccccccccc}
    \hline
    Date & \Halpha\ (H3) & \Hbeta\ (H4) & \Hgamma\ (H5) & \Hdelta\ (H6) & \Hepsilon\ (H7) & \Hzeta\ (H8) & \Heta\ (H9) & \Htheta\ (H10) & UV mean\\
    & $10^{-15}$~cgs & $10^{-15}$~cgs & $10^{-15}$~cgs & $10^{-15}$~cgs & $10^{-15}$~cgs & $10^{-15}$~cgs & $10^{-15}$~cgs & $10^{-15}$~cgs & $10^{-17}$~cgs\\
    \hline
2021-10-25 - 05:42:33 & $5.57 \pm 0.03$ & $3.66 \pm 0.05$ & $2.89 \pm 0.07$ & $2.09 \pm 0.05$ & $2.07 \pm 0.09$ & $1.11 \pm 0.04$ & $0.86 \pm 0.06$ & $0.45 \pm 0.05$ & $3.76 \pm 0.42$\\
2021-10-25 - 06:07:57 & $5.45 \pm 0.03$ & $3.09 \pm 0.06$ & $2.83 \pm 0.09$ & $1.81 \pm 0.05$ & $1.78 \pm 0.14$ & $1.10 \pm 0.08$ & $0.95 \pm 0.05$ & $0.67 \pm 0.04$ & $2.81 \pm 0.43$\\
2022-10-13 - 02:52:03 & $12.09 \pm 0.05$ & $4.68 \pm 0.05$ & $3.91 \pm 0.09$ & $2.80 \pm 0.07$ & $2.51 \pm 0.14$ & $1.73 \pm 0.07$ & $1.18 \pm 0.08$ & $0.78 \pm 0.06$ & $1.81 \pm 0.61$\\
2022-10-13 - 03:17:23 & $12.63 \pm 0.04$ & $4.85 \pm 0.05$ & $3.66 \pm 0.08$ & $2.62 \pm 0.06$ & $2.82 \pm 0.13$ & $1.53 \pm 0.07$ & $0.94 \pm 0.07$ & $0.81 \pm 0.06$ & $1.72 \pm 0.51$\\
2022-10-14 - 06:52:50 & $17.49 \pm 0.03$ & $8.23 \pm 0.04$ & $7.89 \pm 0.07$ & $5.47 \pm 0.04$ & $5.54 \pm 0.08$ & $3.78 \pm 0.05$ & $2.48 \pm 0.06$ & $1.77 \pm 0.05$ & $8.25 \pm 0.40$\\
2022-10-14 - 07:18:09 & $16.12 \pm 0.03$ & $7.89 \pm 0.04$ & $7.23 \pm 0.05$ & $5.15 \pm 0.04$ & $4.92 \pm 0.08$ & $3.49 \pm 0.04$ & $2.45 \pm 0.05$ & $1.80 \pm 0.04$ & $6.90 \pm 0.39$\\
2022-10-15 - 04:40:38 & $14.65 \pm 0.04$ & $8.02 \pm 0.04$ & $6.82 \pm 0.06$ & $4.76 \pm 0.04$ & $4.73 \pm 0.10$ & $3.28 \pm 0.05$ & $1.98 \pm 0.05$ & $1.15 \pm 0.04$ & $6.93 \pm 0.36$\\
2022-10-15 - 05:05:57 & $14.55 \pm 0.02$ & $7.91 \pm 0.04$ & $6.94 \pm 0.05$ & $4.70 \pm 0.03$ & $4.25 \pm 0.08$ & $3.06 \pm 0.05$ & $1.93 \pm 0.05$ & $1.36 \pm 0.04$ & $6.72 \pm 0.30$\\
2022-10-15 - 06:30:49 & $13.98 \pm 0.03$ & $7.54 \pm 0.05$ & $6.55 \pm 0.08$ & $4.67 \pm 0.05$ & $4.06 \pm 0.10$ & $2.84 \pm 0.07$ & $1.82 \pm 0.06$ & $1.17 \pm 0.05$ & $6.67 \pm 0.37$\\
2022-10-15 - 06:56:07 & $13.73 \pm 0.03$ & $7.22 \pm 0.05$ & $6.72 \pm 0.08$ & $4.41 \pm 0.05$ & $4.54 \pm 0.12$ & $2.97 \pm 0.07$ & $2.07 \pm 0.05$ & $1.69 \pm 0.05$ & $5.86 \pm 0.42$\\
2022-10-17 - 03:02:29 & $8.15 \pm 0.03$ & $5.09 \pm 0.04$ & $4.24 \pm 0.05$ & $3.00 \pm 0.03$ & $2.81 \pm 0.07$ & $2.03 \pm 0.04$ & $1.42 \pm 0.04$ & $0.83 \pm 0.03$ & $3.73 \pm 0.33$\\
2022-10-17 - 03:27:48 & $7.96 \pm 0.02$ & $4.95 \pm 0.04$ & $4.36 \pm 0.04$ & $3.00 \pm 0.03$ & $2.77 \pm 0.07$ & $2.10 \pm 0.03$ & $1.37 \pm 0.04$ & $0.94 \pm 0.04$ & $4.57 \pm 0.30$\\
2022-11-01 - 01:21:16 & $7.45 \pm 0.03$ & $4.66 \pm 0.04$ & $3.87 \pm 0.07$ & $2.66 \pm 0.05$ & $2.45 \pm 0.13$ & $1.67 \pm 0.06$ & $0.94 \pm 0.06$ & $0.80 \pm 0.05$ & $3.47 \pm 0.49$\\
2022-11-01 - 01:46:35 & $7.27 \pm 0.05$ & $4.51 \pm 0.04$ & $3.99 \pm 0.07$ & $2.79 \pm 0.05$ & $2.63 \pm 0.08$ & $1.90 \pm 0.05$ & $1.15 \pm 0.05$ & $0.85 \pm 0.05$ & $3.49 \pm 0.42$\\
2022-11-04 - 00:49:41 & $7.58 \pm 0.03$ & $4.87 \pm 0.05$ & $4.43 \pm 0.06$ & $2.99 \pm 0.05$ & $2.53 \pm 0.12$ & $1.74 \pm 0.06$ & $1.25 \pm 0.05$ & $0.93 \pm 0.05$ & $3.26 \pm 0.42$\\
2022-11-04 - 01:42:04 & $7.62 \pm 0.02$ & $4.96 \pm 0.04$ & $4.07 \pm 0.06$ & $2.84 \pm 0.04$ & $2.33 \pm 0.07$ & $1.72 \pm 0.05$ & $0.96 \pm 0.04$ & $0.80 \pm 0.04$ & $3.75 \pm 0.31$\\
2022-11-04 - 02:07:24 & $7.29 \pm 0.02$ & $4.77 \pm 0.04$ & $4.00 \pm 0.04$ & $2.84 \pm 0.04$ & $2.42 \pm 0.08$ & $1.66 \pm 0.04$ & $1.19 \pm 0.04$ & $0.86 \pm 0.04$ & $2.71 \pm 0.33$\\
2022-11-05 - 00:57:06 & $8.26 \pm 0.10$ & $5.29 \pm 0.06$ & $4.64 \pm 0.09$ & $3.21 \pm 0.07$ & $3.09 \pm 0.14$ & $2.08 \pm 0.07$ & $1.21 \pm 0.07$ & $0.94 \pm 0.06$ & $2.66 \pm 0.50$\\
2022-11-05 - 01:22:25 & $8.40 \pm 0.07$ & $5.41 \pm 0.04$ & $4.99 \pm 0.07$ & $3.33 \pm 0.05$ & $3.03 \pm 0.11$ & $2.04 \pm 0.08$ & $1.26 \pm 0.06$ & $0.73 \pm 0.05$ & $4.58 \pm 0.42$\\
2022-11-05 - 01:48:25 & $8.54 \pm 0.04$ & $5.60 \pm 0.04$ & $4.90 \pm 0.06$ & $3.19 \pm 0.05$ & $3.06 \pm 0.10$ & $1.91 \pm 0.05$ & $1.25 \pm 0.06$ & $0.87 \pm 0.05$ & $3.40 \pm 0.38$\\
2022-11-05 - 02:13:44 & $8.31 \pm 0.05$ & $5.38 \pm 0.04$ & $4.54 \pm 0.07$ & $3.18 \pm 0.04$ & $2.92 \pm 0.12$ & $1.87 \pm 0.06$ & $1.21 \pm 0.06$ & $0.74 \pm 0.05$ & $2.43 \pm 0.43$\\
2022-11-05 - 02:40:34 & $8.48 \pm 0.06$ & $5.38 \pm 0.05$ & $4.67 \pm 0.06$ & $3.05 \pm 0.05$ & $2.82 \pm 0.10$ & $1.82 \pm 0.05$ & $1.34 \pm 0.06$ & $0.88 \pm 0.05$ & $2.25 \pm 0.45$\\
2022-11-05 - 03:05:51 & $8.03 \pm 0.03$ & $5.19 \pm 0.03$ & $4.69 \pm 0.07$ & $3.21 \pm 0.04$ & $3.06 \pm 0.10$ & $1.96 \pm 0.05$ & $1.39 \pm 0.06$ & $0.92 \pm 0.05$ & $3.23 \pm 0.40$\\
2022-11-05 - 03:36:05 & $7.86 \pm 0.03$ & $5.26 \pm 0.04$ & $4.59 \pm 0.08$ & $3.03 \pm 0.04$ & $3.04 \pm 0.13$ & $1.97 \pm 0.06$ & $1.09 \pm 0.05$ & $0.99 \pm 0.05$ & $3.47 \pm 0.46$\\
2022-11-05 - 04:01:24 & $7.88 \pm 0.02$ & $5.21 \pm 0.04$ & $4.23 \pm 0.06$ & $2.82 \pm 0.04$ & $2.32 \pm 0.08$ & $1.42 \pm 0.04$ & $0.83 \pm 0.05$ & $0.59 \pm 0.04$ & $2.26 \pm 0.36$\\
2022-12-04 - 00:46:22 & $13.34 \pm 0.02$ & $7.39 \pm 0.04$ & $6.54 \pm 0.07$ & $4.46 \pm 0.05$ & $3.60 \pm 0.10$ & $2.72 \pm 0.06$ & $1.64 \pm 0.05$ & $1.02 \pm 0.05$ & $4.65 \pm 0.34$\\
2022-12-04 - 01:11:41 & $13.14 \pm 0.02$ & $7.39 \pm 0.04$ & $6.32 \pm 0.05$ & $4.52 \pm 0.04$ & $4.04 \pm 0.10$ & $2.62 \pm 0.04$ & $1.61 \pm 0.04$ & $1.21 \pm 0.04$ & $4.64 \pm 0.30$\\
2022-12-04 - 01:41:03 & $12.96 \pm 0.03$ & $7.21 \pm 0.04$ & $6.41 \pm 0.07$ & $4.35 \pm 0.05$ & $3.81 \pm 0.11$ & $2.51 \pm 0.06$ & $1.66 \pm 0.06$ & $1.17 \pm 0.05$ & $5.32 \pm 0.40$\\
2022-12-04 - 02:06:22 & $12.00 \pm 0.03$ & $6.98 \pm 0.04$ & $6.42 \pm 0.07$ & $4.13 \pm 0.05$ & $3.74 \pm 0.11$ & $2.48 \pm 0.05$ & $1.41 \pm 0.06$ & $1.15 \pm 0.05$ & $4.39 \pm 0.42$\\
2022-12-31 - 01:05:53 & $12.08 \pm 0.02$ & $7.59 \pm 0.03$ & $6.89 \pm 0.04$ & $4.73 \pm 0.03$ & $4.19 \pm 0.08$ & $2.82 \pm 0.05$ & $1.70 \pm 0.05$ & $1.26 \pm 0.04$ & $6.16 \pm 0.35$\\
2022-12-31 - 01:31:12 & $11.28 \pm 0.02$ & $7.26 \pm 0.04$ & $6.44 \pm 0.06$ & $4.34 \pm 0.04$ & $3.85 \pm 0.09$ & $2.80 \pm 0.05$ & $1.65 \pm 0.05$ & $1.19 \pm 0.04$ & $5.21 \pm 0.41$\\
2023-01-02 - 01:13:31 & $10.86 \pm 0.03$ & $7.00 \pm 0.05$ & $6.38 \pm 0.09$ & $4.27 \pm 0.08$ & $3.71 \pm 0.13$ & $2.57 \pm 0.09$ & $1.57 \pm 0.07$ & $1.13 \pm 0.05$ & $5.23 \pm 0.32$\\
2023-01-02 - 01:38:50 & $10.31 \pm 0.05$ & $6.43 \pm 0.09$ & $5.42 \pm 0.13$ & $3.50 \pm 0.09$ & $2.87 \pm 0.21$ & $2.16 \pm 0.09$ & $1.37 \pm 0.09$ & $1.04 \pm 0.08$ & $5.41 \pm 0.49$\\
    \end{tabular}
\end{sidewaystable*}

\begin{sidewaystable*}
\centering
\caption{Integrated fluxes for each wings/core component of the \Halpha, \Hbeta, \Hgamma\ and \Hdelta\ lines of \delorme, where cgs stands for erg/s/cm$^2$.}
\label{tab:shapes_lines_fluxes}
    \begin{tabular}{c| cccccccc}
    \hline
    Date & \Halpha\ (H3) wings & \Halpha\ (H3) core & \Hbeta\ (H4) wings & \Hbeta\ (H4) core & \Hgamma\ (H5) wings & \Hgamma\ (H5) core & \Hdelta\ (H6) wings & \Hdelta\ (H6) core\\
    & $10^{-15}$~cgs & $10^{-15}$~cgs & $10^{-15}$~cgs & $10^{-15}$~cgs & $10^{-15}$~cgs & $10^{-15}$~cgs & $10^{-15}$~cgs & $10^{-15}$~cgs\\
    \hline
2021-10-25 - 05:42:33 & $1.22 \pm 0.03$ & $4.35 \pm 0.03$ & $0.15 \pm 0.06$ & $3.48 \pm 0.06$ & $0.00 \pm 0.09$ & $3.02 \pm 0.09$ & $0.00 \pm 0.06$ & $2.35 \pm 0.06$\\
2021-10-25 - 06:07:57 & $1.17 \pm 0.03$ & $4.22 \pm 0.03$ & $0.00 \pm 0.06$ & $3.19 \pm 0.06$ & $0.07 \pm 0.09$ & $2.73 \pm 0.09$ & $0.00 \pm 0.06$ & $2.04 \pm 0.06$\\
2022-10-13 - 02:52:03 & $3.62 \pm 0.03$ & $8.44 \pm 0.03$ & $0.72 \pm 0.06$ & $4.00 \pm 0.06$ & $0.48 \pm 0.09$ & $3.29 \pm 0.09$ & $0.46 \pm 0.06$ & $2.65 \pm 0.06$\\
2022-10-13 - 03:17:23 & $3.87 \pm 0.03$ & $8.79 \pm 0.03$ & $0.89 \pm 0.06$ & $3.98 \pm 0.06$ & $0.51 \pm 0.09$ & $3.04 \pm 0.09$ & $0.29 \pm 0.06$ & $2.81 \pm 0.06$\\
2022-10-14 - 06:52:50 & $8.39 \pm 0.03$ & $8.79 \pm 0.03$ & $3.77 \pm 0.06$ & $4.57 \pm 0.06$ & $3.19 \pm 0.09$ & $4.82 \pm 0.09$ & $2.76 \pm 0.06$ & $4.15 \pm 0.06$\\
2022-10-14 - 07:18:09 & $7.69 \pm 0.03$ & $8.12 \pm 0.03$ & $3.47 \pm 0.06$ & $4.54 \pm 0.06$ & $2.74 \pm 0.09$ & $4.64 \pm 0.09$ & $2.43 \pm 0.06$ & $4.05 \pm 0.06$\\
2022-10-15 - 04:40:38 & $7.14 \pm 0.03$ & $7.00 \pm 0.03$ & $3.47 \pm 0.06$ & $4.58 \pm 0.06$ & $2.61 \pm 0.09$ & $4.24 \pm 0.09$ & $2.18 \pm 0.06$ & $3.71 \pm 0.06$\\
2022-10-15 - 05:05:57 & $6.99 \pm 0.03$ & $7.05 \pm 0.03$ & $3.33 \pm 0.06$ & $4.57 \pm 0.06$ & $2.53 \pm 0.09$ & $4.43 \pm 0.09$ & $2.17 \pm 0.06$ & $3.74 \pm 0.06$\\
2022-10-15 - 06:30:49 & $6.25 \pm 0.03$ & $7.39 \pm 0.03$ & $2.62 \pm 0.06$ & $4.93 \pm 0.06$ & $2.01 \pm 0.09$ & $4.67 \pm 0.09$ & $1.65 \pm 0.06$ & $4.05 \pm 0.06$\\
2022-10-15 - 06:56:07 & $6.02 \pm 0.03$ & $7.52 \pm 0.03$ & $2.52 \pm 0.06$ & $4.80 \pm 0.06$ & $1.75 \pm 0.09$ & $5.05 \pm 0.09$ & $1.30 \pm 0.06$ & $4.12 \pm 0.06$\\
2022-10-17 - 03:02:29 & $3.38 \pm 0.03$ & $4.58 \pm 0.03$ & $1.57 \pm 0.06$ & $3.59 \pm 0.06$ & $0.98 \pm 0.09$ & $3.37 \pm 0.09$ & $0.88 \pm 0.06$ & $2.68 \pm 0.06$\\
2022-10-17 - 03:27:48 & $3.27 \pm 0.03$ & $4.47 \pm 0.03$ & $1.39 \pm 0.06$ & $3.72 \pm 0.06$ & $0.95 \pm 0.09$ & $3.52 \pm 0.09$ & $0.84 \pm 0.06$ & $2.83 \pm 0.06$\\
2022-11-01 - 01:21:16 & $2.20 \pm 0.03$ & $5.14 \pm 0.03$ & $0.79 \pm 0.06$ & $3.91 \pm 0.06$ & $0.43 \pm 0.09$ & $3.49 \pm 0.09$ & $0.61 \pm 0.06$ & $2.54 \pm 0.06$\\
2022-11-01 - 01:46:35 & $2.07 \pm 0.03$ & $5.05 \pm 0.03$ & $0.76 \pm 0.06$ & $3.75 \pm 0.06$ & $0.68 \pm 0.09$ & $3.30 \pm 0.09$ & $0.45 \pm 0.06$ & $2.85 \pm 0.06$\\
2022-11-04 - 00:49:41 & $1.85 \pm 0.03$ & $5.63 \pm 0.03$ & $0.53 \pm 0.06$ & $4.42 \pm 0.06$ & $0.22 \pm 0.09$ & $4.29 \pm 0.09$ & $0.19 \pm 0.06$ & $3.31 \pm 0.06$\\
2022-11-04 - 01:42:04 & $1.92 \pm 0.03$ & $5.54 \pm 0.03$ & $0.52 \pm 0.06$ & $4.46 \pm 0.06$ & $0.17 \pm 0.09$ & $3.95 \pm 0.09$ & $0.15 \pm 0.06$ & $3.07 \pm 0.06$\\
2022-11-04 - 02:07:24 & $1.78 \pm 0.03$ & $5.39 \pm 0.03$ & $0.52 \pm 0.06$ & $4.25 \pm 0.06$ & $0.02 \pm 0.09$ & $4.04 \pm 0.09$ & $0.01 \pm 0.06$ & $3.26 \pm 0.06$\\
2022-11-05 - 00:57:06 & $2.36 \pm 0.03$ & $5.78 \pm 0.03$ & $0.77 \pm 0.06$ & $4.50 \pm 0.06$ & $0.01 \pm 0.09$ & $4.77 \pm 0.09$ & $0.38 \pm 0.06$ & $3.39 \pm 0.06$\\
2022-11-05 - 01:22:25 & $2.41 \pm 0.03$ & $5.83 \pm 0.03$ & $0.68 \pm 0.06$ & $4.80 \pm 0.06$ & $0.52 \pm 0.09$ & $4.53 \pm 0.09$ & $0.33 \pm 0.06$ & $3.62 \pm 0.06$\\
2022-11-05 - 01:48:25 & $2.33 \pm 0.03$ & $6.09 \pm 0.03$ & $0.79 \pm 0.06$ & $4.89 \pm 0.06$ & $0.22 \pm 0.09$ & $4.66 \pm 0.09$ & $0.24 \pm 0.06$ & $3.54 \pm 0.06$\\
2022-11-05 - 02:13:44 & $2.21 \pm 0.03$ & $5.93 \pm 0.03$ & $0.76 \pm 0.06$ & $4.61 \pm 0.06$ & $0.25 \pm 0.09$ & $4.36 \pm 0.09$ & $0.23 \pm 0.06$ & $3.52 \pm 0.06$\\
2022-11-05 - 02:40:34 & $2.27 \pm 0.03$ & $6.07 \pm 0.03$ & $0.62 \pm 0.06$ & $4.78 \pm 0.06$ & $0.11 \pm 0.09$ & $4.67 \pm 0.09$ & $0.09 \pm 0.06$ & $3.49 \pm 0.06$\\
2022-11-05 - 03:05:51 & $2.06 \pm 0.03$ & $5.83 \pm 0.03$ & $0.60 \pm 0.06$ & $4.63 \pm 0.06$ & $0.15 \pm 0.09$ & $4.70 \pm 0.09$ & $0.25 \pm 0.06$ & $3.58 \pm 0.06$\\
2022-11-05 - 03:36:05 & $2.05 \pm 0.03$ & $5.70 \pm 0.03$ & $0.49 \pm 0.06$ & $4.85 \pm 0.06$ & $0.01 \pm 0.09$ & $4.69 \pm 0.09$ & $0.27 \pm 0.06$ & $3.29 \pm 0.06$\\
2022-11-05 - 04:01:24 & $1.83 \pm 0.03$ & $5.95 \pm 0.03$ & $0.52 \pm 0.06$ & $4.72 \pm 0.06$ & $0.19 \pm 0.09$ & $4.04 \pm 0.09$ & $0.06 \pm 0.06$ & $3.19 \pm 0.06$\\
2022-12-04 - 00:46:22 & $3.83 \pm 0.03$ & $9.35 \pm 0.03$ & $1.26 \pm 0.06$ & $6.22 \pm 0.06$ & $0.84 \pm 0.09$ & $5.75 \pm 0.09$ & $0.77 \pm 0.06$ & $4.52 \pm 0.06$\\
2022-12-04 - 01:11:41 & $3.92 \pm 0.03$ & $9.05 \pm 0.03$ & $1.15 \pm 0.06$ & $6.29 \pm 0.06$ & $0.65 \pm 0.09$ & $5.80 \pm 0.09$ & $0.61 \pm 0.06$ & $4.70 \pm 0.06$\\
2022-12-04 - 01:41:03 & $3.80 \pm 0.03$ & $9.05 \pm 0.03$ & $0.97 \pm 0.06$ & $6.31 \pm 0.06$ & $0.51 \pm 0.09$ & $5.91 \pm 0.09$ & $0.49 \pm 0.06$ & $4.62 \pm 0.06$\\
2022-12-04 - 02:06:22 & $3.40 \pm 0.03$ & $8.56 \pm 0.03$ & $0.82 \pm 0.06$ & $6.21 \pm 0.06$ & $0.66 \pm 0.09$ & $5.82 \pm 0.09$ & $0.41 \pm 0.06$ & $4.44 \pm 0.06$\\
2022-12-31 - 01:05:53 & $3.78 \pm 0.03$ & $7.94 \pm 0.03$ & $2.07 \pm 0.06$ & $5.59 \pm 0.06$ & $1.49 \pm 0.09$ & $5.42 \pm 0.09$ & $1.25 \pm 0.06$ & $4.40 \pm 0.06$\\
2022-12-31 - 01:31:12 & $3.43 \pm 0.03$ & $7.47 \pm 0.03$ & $1.98 \pm 0.06$ & $5.31 \pm 0.06$ & $1.41 \pm 0.09$ & $5.12 \pm 0.09$ & $1.17 \pm 0.06$ & $4.07 \pm 0.06$\\
2023-01-02 - 01:13:31 & $3.32 \pm 0.03$ & $7.40 \pm 0.03$ & $1.55 \pm 0.06$ & $5.47 \pm 0.06$ & $1.05 \pm 0.09$ & $5.27 \pm 0.09$ & $1.04 \pm 0.06$ & $3.95 \pm 0.06$\\
2023-01-02 - 01:38:50 & $3.02 \pm 0.03$ & $7.11 \pm 0.03$ & $1.32 \pm 0.06$ & $5.01 \pm 0.06$ & $0.52 \pm 0.09$ & $5.17 \pm 0.09$ & $0.52 \pm 0.06$ & $3.72 \pm 0.06$\\
    \end{tabular}
\end{sidewaystable*}

\section{Lines decomposition}
\label{sec:appendix:lines_decomposition}
The lines decomposition for each individual line is shown in Figs.~\ref{fig:two_comp_Halpha_allepochs}, \ref{fig:two_comp_Hbeta_allepochs}, \ref{fig:two_comp_Hgamma_allepochs} and \ref{fig:two_comp_Hdelta_allepochs} for \Halpha, \Hbeta, \Hgamma\ and \Hdelta\ respectively.

\begin{figure*}
    \includegraphics[width=\linewidth]{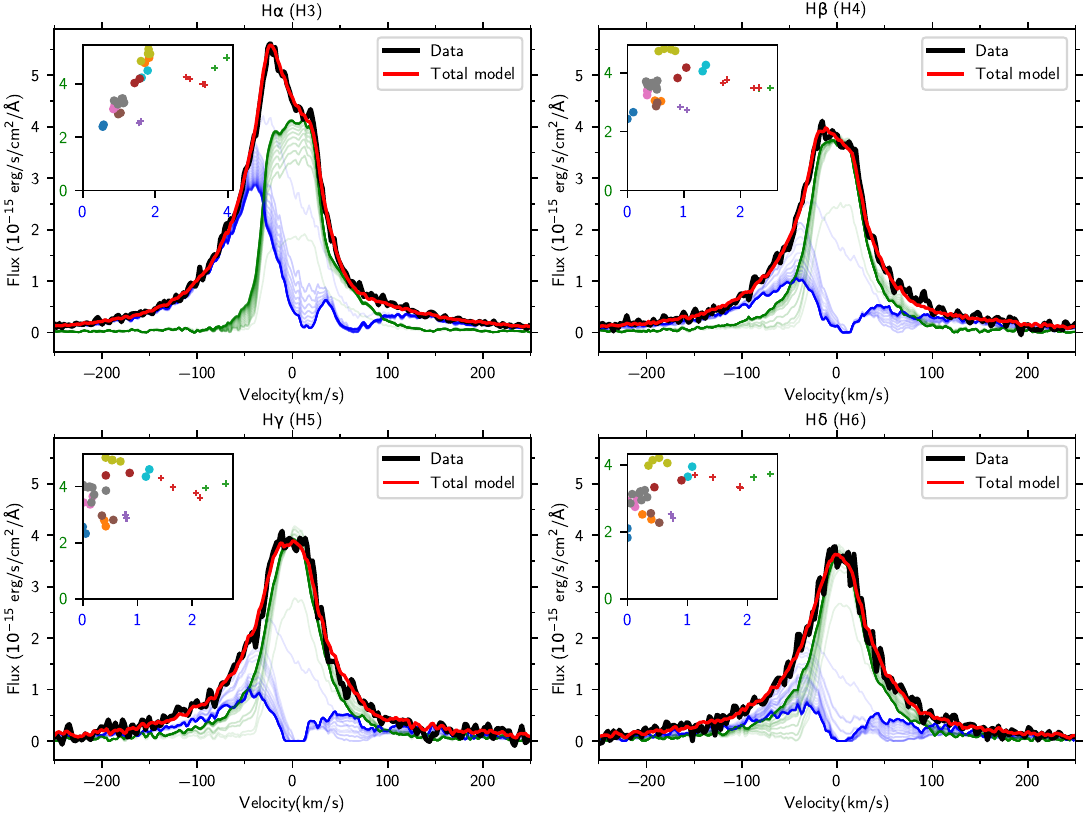}
    \caption{Wings/core components decomposition of the \Halpha, \Hbeta, \Hgamma\ and \Hdelta\ lines for epoch 2022-10-15 -- 06:30:49 (black). The blue-green shaded lines correspond to successive iterations of the decomposition. The solid blue and green lines correspond to the final iteration. The red line is the total model at the last iteration.
    The inset corresponds to the scaling factor of each of the wings/core components, depending on the epoch. The points are color-coded for each epoch, as is done in Fig.~\ref{fig:all_lines}. The crosses correspond to the outburst epochs.}
    \label{fig:shapes_evolution}
\end{figure*}

\begin{figure*}
    \centering
    \includegraphics[width=0.99\textwidth]{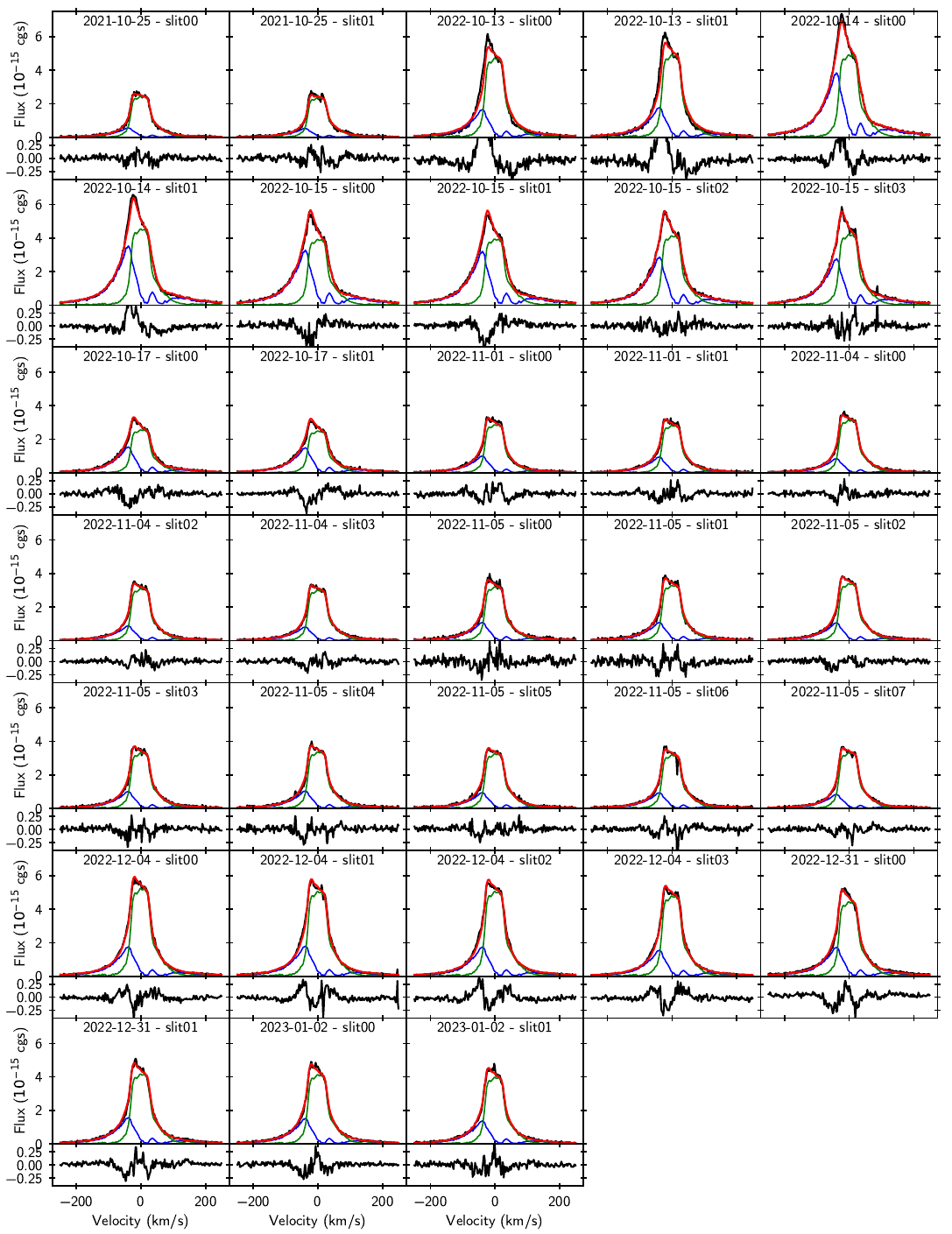}
    \caption{Two-components fits of \Halpha, for all epochs, at the last iteration. Each sub-panel is divided in the fit and the corresponding residuals.}
    \label{fig:two_comp_Halpha_allepochs}
\end{figure*}

\begin{figure*}
    \centering
    \includegraphics[width=\textwidth]{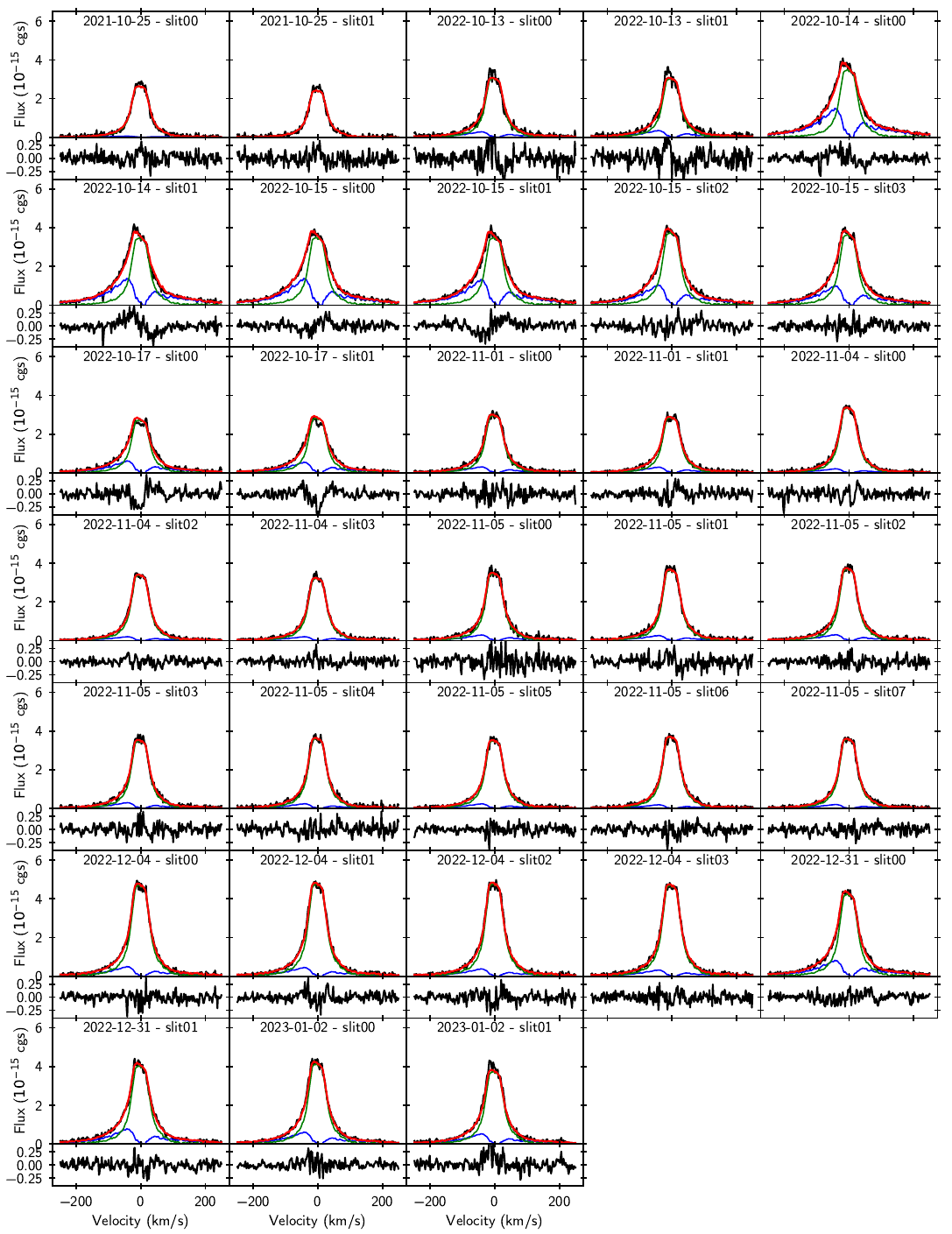}
    \caption{Two-components fits of \Hbeta, for all epochs, at the last iteration. Each sub-panel is divided in the fit and the corresponding residuals.}
    \label{fig:two_comp_Hbeta_allepochs}
\end{figure*}

\begin{figure*}
    \centering
    \includegraphics[width=\textwidth]{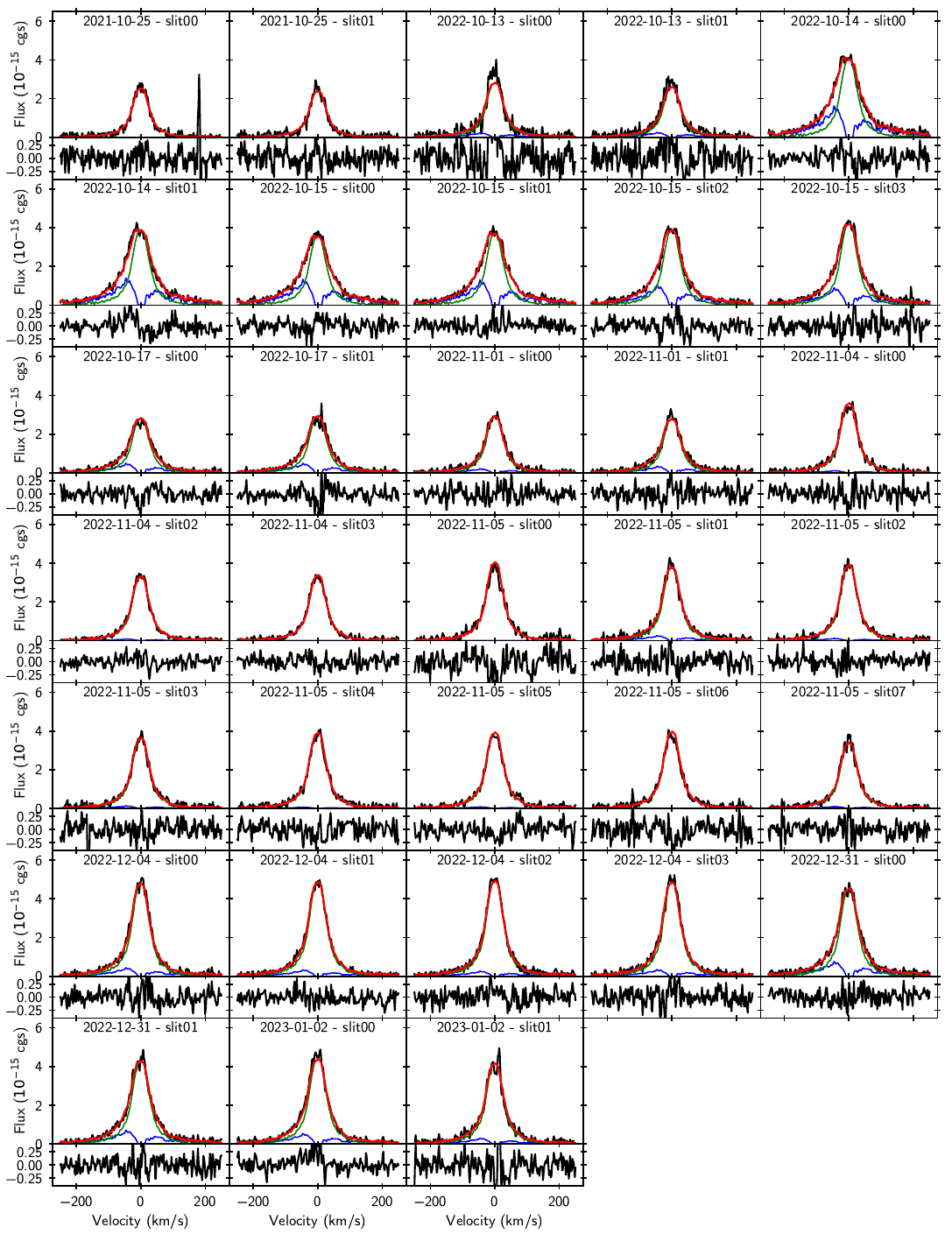}
    \caption{Two-components fits of \Hgamma, for all epochs, at the last iteration. Each sub-panel is divided in the fit and the corresponding residuals.}
    \label{fig:two_comp_Hdelta_allepochs}
\end{figure*}

\begin{figure*}
    \centering
    \includegraphics[width=\textwidth]{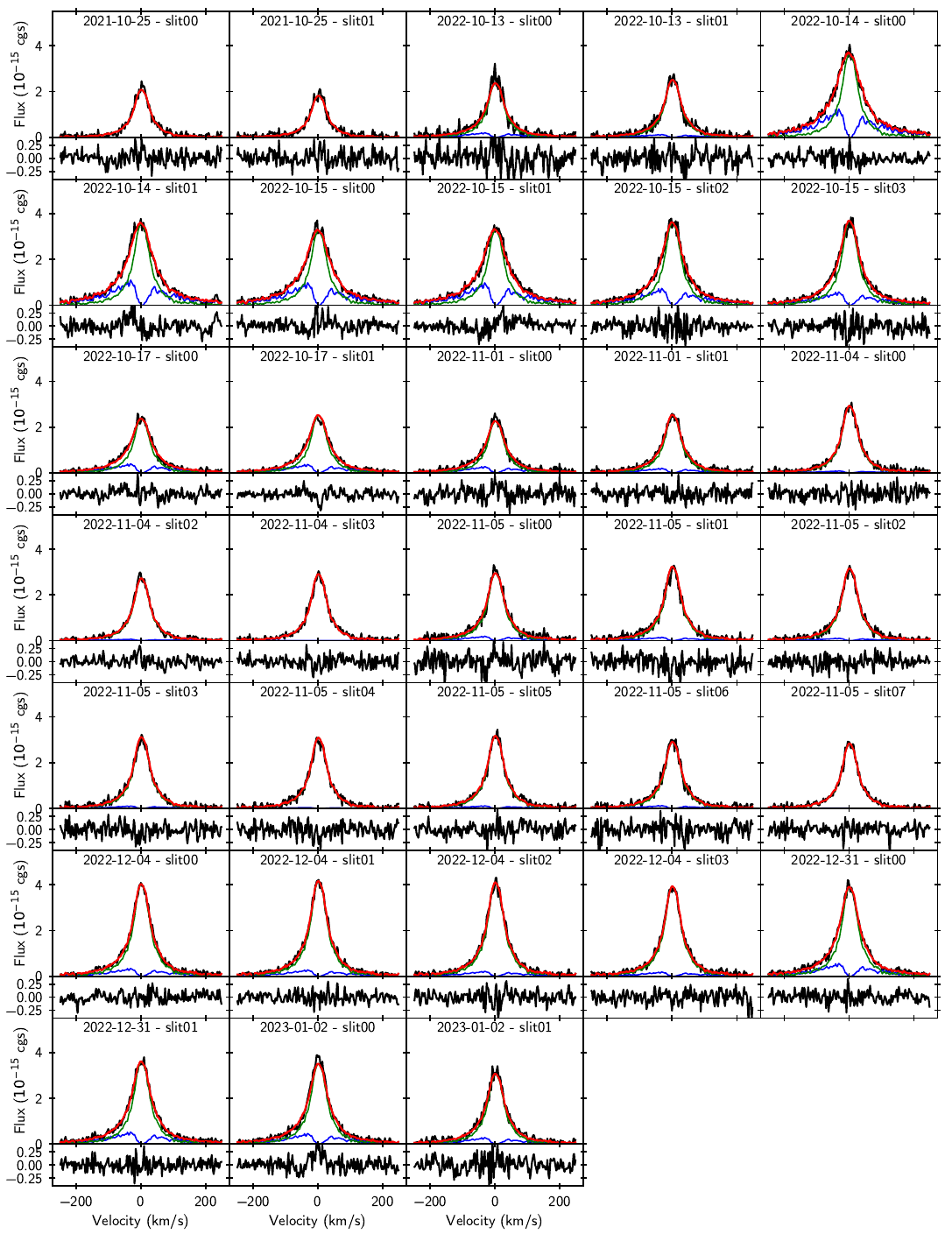}
    \caption{Two-components fits of \Hdelta, for all epochs, at the last iteration. Each sub-panel is divided in the fit and the corresponding residuals.}
    \label{fig:two_comp_Hgamma_allepochs}
\end{figure*}

\section{Non-negative matrix factorization}
\label{sec:appendix:nmf}

Our decomposition of the H {\sc i} emission line profiles with the L-BFGS-B approach is in fact similar to a dimension reduction problem encountered in signal processing. We therefore considered using principle component analysis to check whether we could retrieve similar components as with the L-BFGS-B approach which may still be sensitive to local minima in the parameter space of the fit. But one must impose the principal components to be additive. We therefore considered non-negative matrix factorization (NMF) using the \texttt{sklearn} implementation of the method \citep{1999Natur.401..788L} with Non-negative Double Singular Value Decomposition (NNDSVD) initialization \citep{2008PatRe..41.1350B}.

\begin{figure*}
    \includegraphics[width=\linewidth]{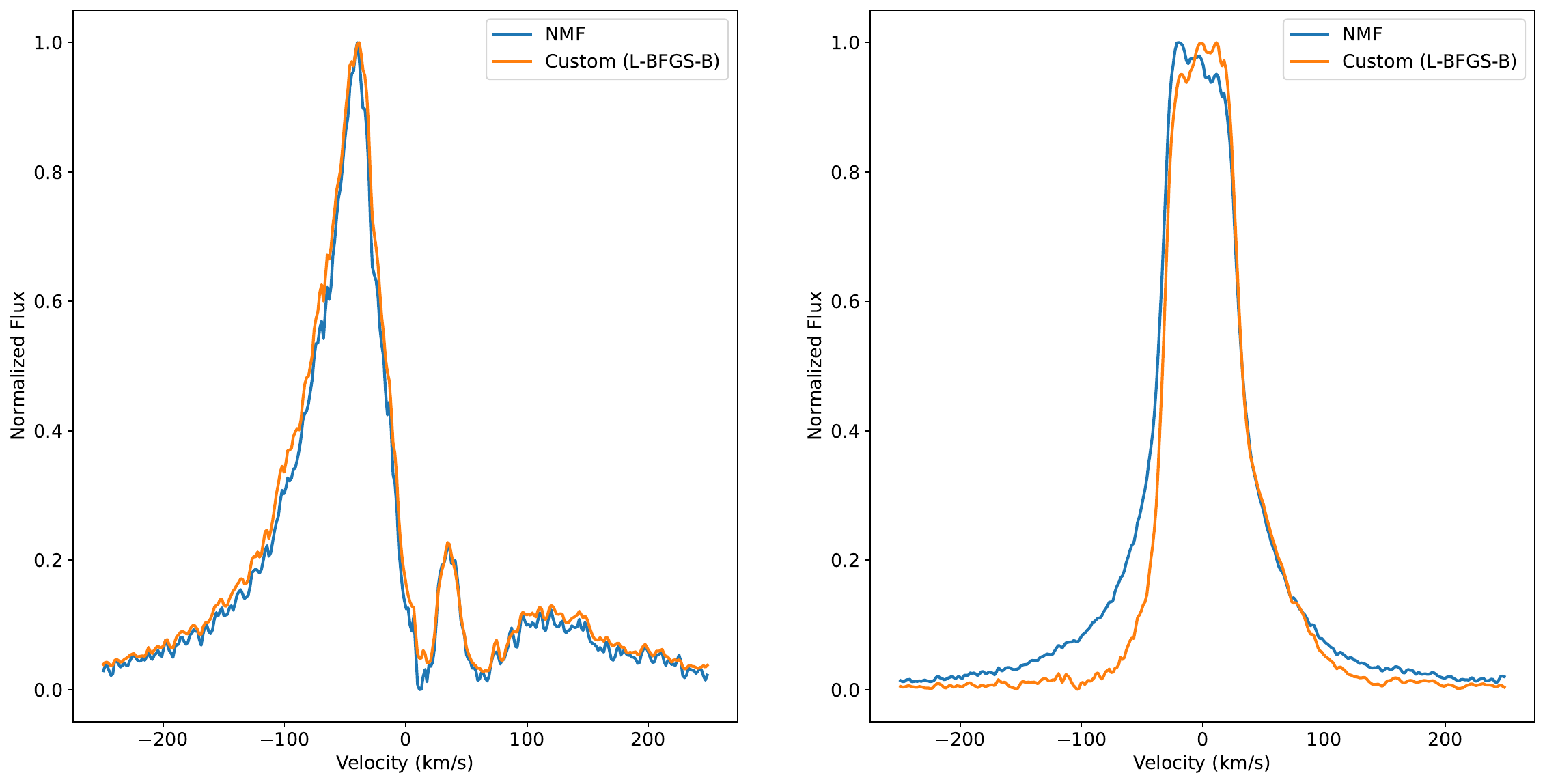}
    \label{fig:appendix:compare_nmf_to_mine}
    \caption{Comparison of the decomposition presented in this work (based on the L-BFGS-B algorithm) and the corresponding NMF decomposition.}
\end{figure*}

The NMF method assumes that all data have positive values. For that purpose, we added a constant value to bring any negative value in our profile to zero before running the NMF decomposition. 
A comparison between the decomposition presented in this work, and the NMF results (for \Halpha) is shown in Fig.~\ref{fig:appendix:compare_nmf_to_mine}.

The decomposition with two components broadly confirm the one achieved with the L-BFGS-B. The wings component is well reproduced for \Halpha to \Hbeta. We only notice for \Halpha a slight change in the blue wing (more extended in the case of the NMF) and a change of slope at the peak of the core component. We also find the second component (wings) to be systematically fainter at \Halpha\ than with the L-BFGS-B decomposition at all epochs, and even null in the case of the faintest epochs.

The NMF allows to explore more easily the possibility of a larger number of components. We find that \Halpha\ can require an additional component but its shape is nearly identical to the shape of the profile at the brightest epochs. However, the two main components still conserves qualitatively the same shape. Conversely, we find the results (even with two components) to be much more sensitive to the initialization method of the algorithm.
It should be noted that the NMF decomposition requires the addition of a constant ``additive'' component which likely biases the fine structure of the derived components.



\section{Modeling results}

\subsection{Lines modeling}
\label{sec:appendix:modeling_results}
\begin{figure*}
    \includegraphics[width=0.5\linewidth]{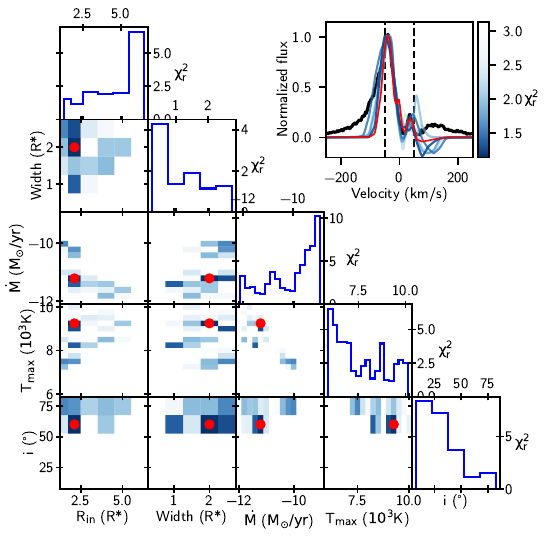}%
    \includegraphics[width=0.5\linewidth]{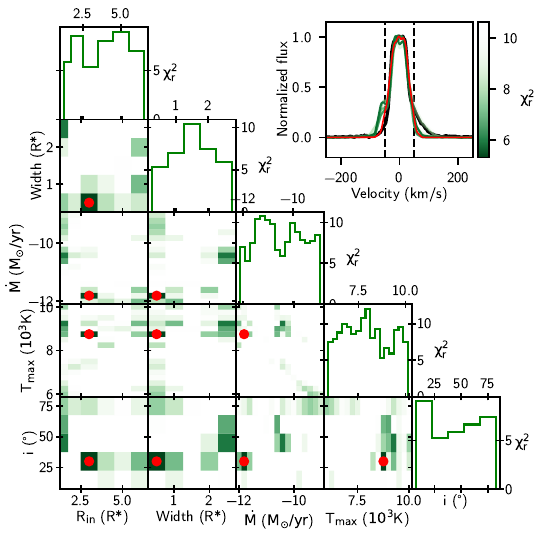}\\
    \includegraphics[width=0.5\linewidth]{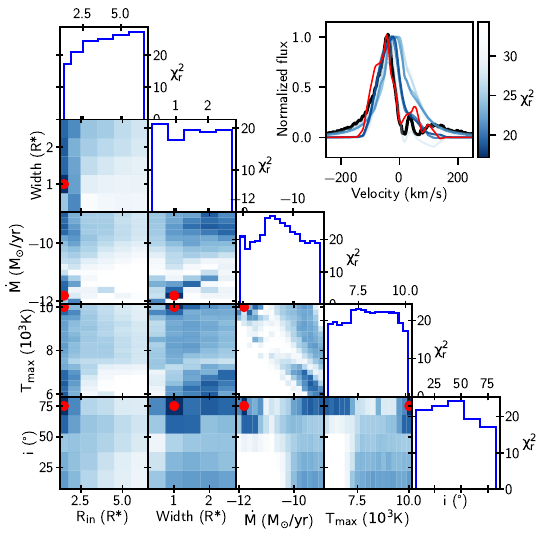}%
    \includegraphics[width=0.5\linewidth]{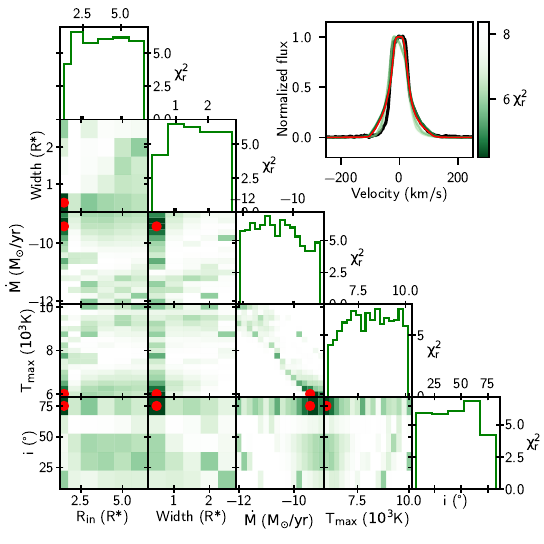}
    \caption{$\chi^{2}_r$ maps for the various combination of free parameters in the magnetospheric accretion models for the wings (blue, left) and core (green, right) components of the \Halpha line. The top row shows the case when fitting only on center of the component, whereas the bottom row shows the case when fitting the full line. In the insets, the various colored lines represent some of the best models with the corresponding $\chi^{2}_r$ shown in the colorbar. The red line corresponds to the best model indicated by the red dot in the corner plot.}
    \label{fig:modeling_Halpha_corner_plots}
\end{figure*}

\begin{figure*}
    \includegraphics[width=0.5\linewidth]{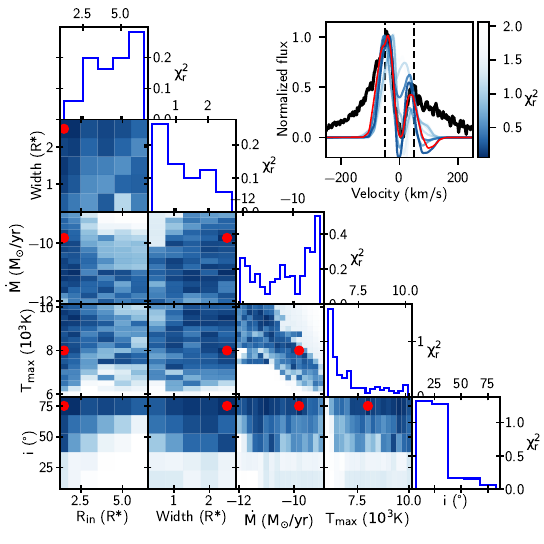}%
    \includegraphics[width=0.5\linewidth]{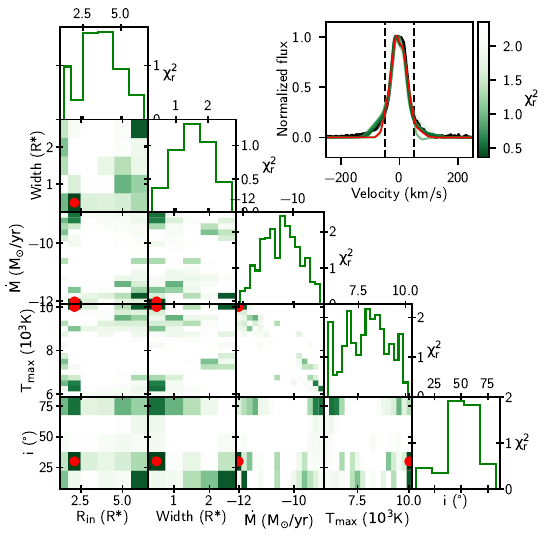}\\
    \includegraphics[width=0.5\linewidth]{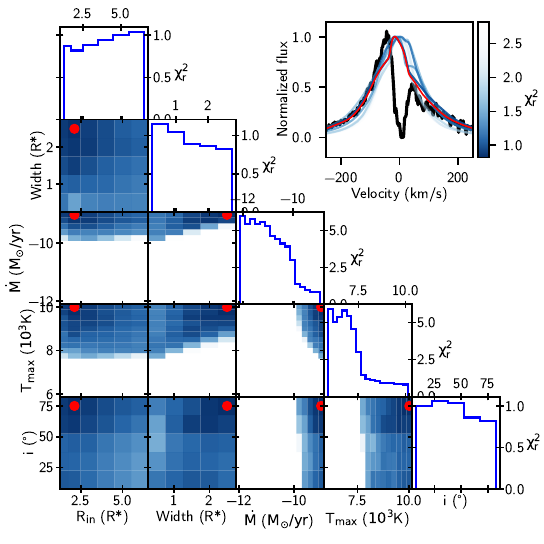}%
    \includegraphics[width=0.5\linewidth]{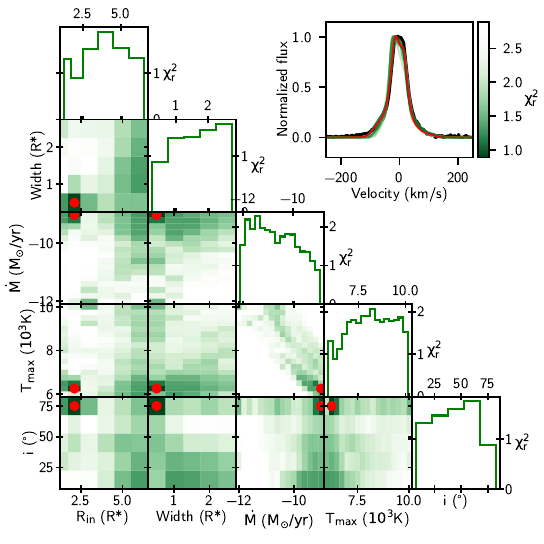}
    \caption{Same as Figure \ref{fig:modeling_Halpha_corner_plots} for \Hbeta.}
    \label{fig:modeling_Hbeta_corner_plots}
\end{figure*}

\begin{figure*}
    \includegraphics[width=\linewidth]{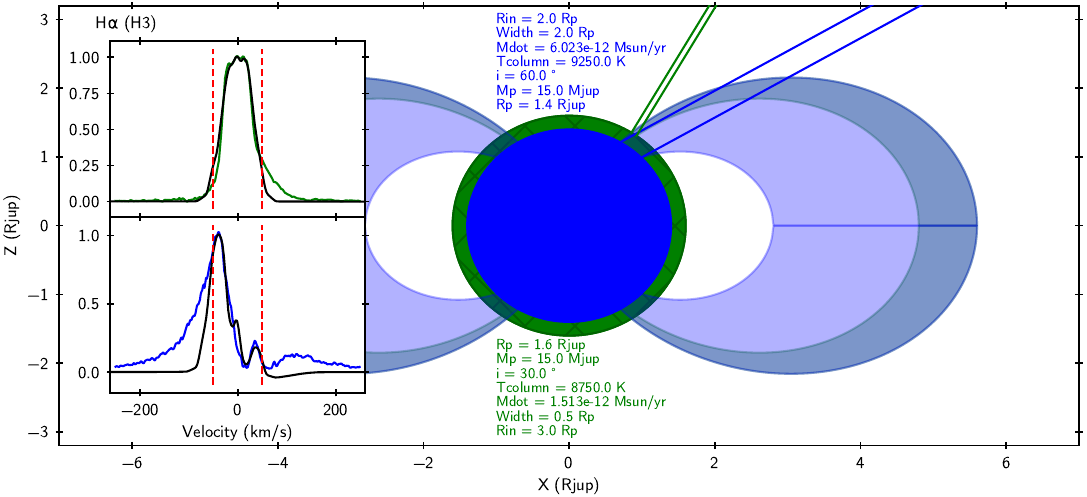}\\
    \includegraphics[width=\linewidth]{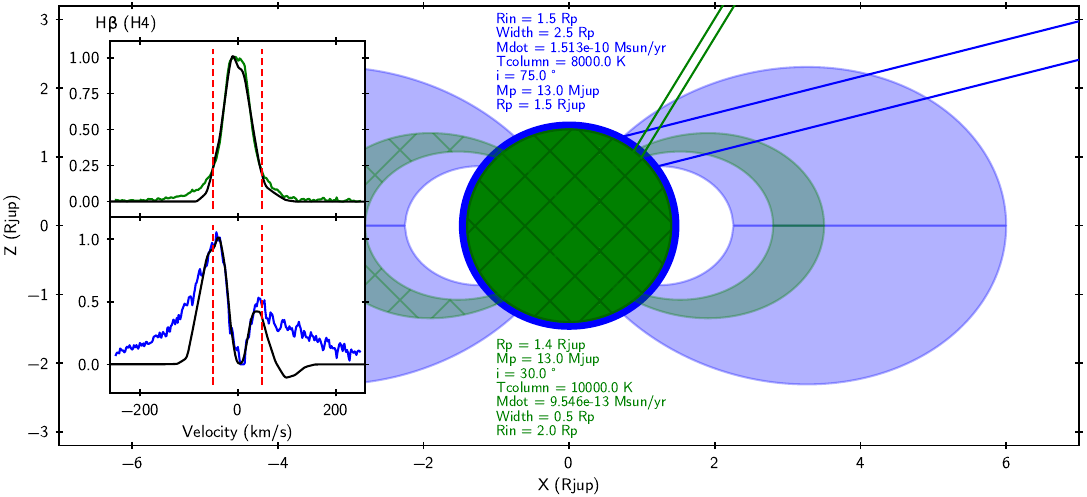}
    \caption{Sketch of the best fit model for the core of both the wings and core components of \Halpha\ (top) and \Hbeta\ (bottom).
    The blue/green line in the top right of the image corresponds to the line of sight intercepting the shock. The different inclinations for the blue/green lines are due to two local minimas in the grid (see Figs.~\ref{fig:modeling_Halpha_corner_plots} \& \ref{fig:modeling_Hbeta_corner_plots}). The planet in the center, and its "apparent shell" of a different color represent the sizer of the planet in the wings (blue) and core (green) components modeling.
    The insets on the left shows the data for the wings/core (blue/green) component, and the best fit (black line), along with the fitting limits (red dashed lines).}
    \label{fig:modeling_sketch_core}
\end{figure*}

\begin{figure*}
    \includegraphics[width=\linewidth]{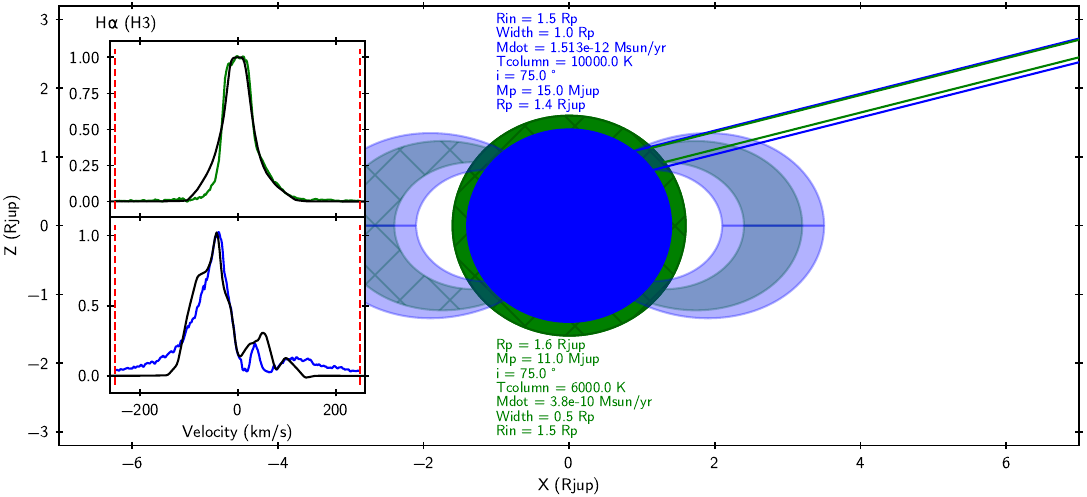}\\
    \includegraphics[width=\linewidth]{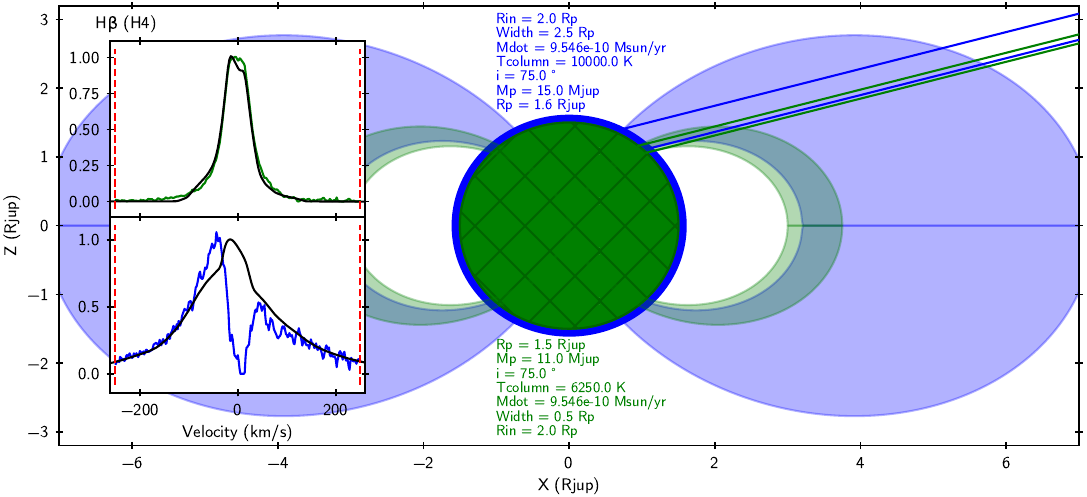}
    \caption{Same as Fig.~\ref{fig:modeling_sketch_core}, for the full fitting range. Sketch of the best fit model for the full range of both the wings and core components of \Halpha\ (top) and \Hbeta\ (bottom). The blue/green line in the top right of the image corresponds to the line of sight intercepting the shock.
    The planet in the center, and its "apparent shell" of a different color represent the sizer of the planet in the wings (blue) and core (green) components modeling.
    The insets on the left shows the data for the wings/core (blue/green) component, and the best fit (black line), along with the fitting limits (red dashed lines).}
    \label{fig:modeling_sketch_full}
\end{figure*}

This section presents the $\chi^2_r$ maps of the lines modeling (Sect.~\ref{sec:modeling}). Figures~\ref{fig:modeling_Halpha_corner_plots} and \ref{fig:modeling_Hbeta_corner_plots} show the $\chi^2_r$ maps over the Magnetospheric Accretion models grid \citep{thanathibodee_magnetospheric_2019}, for \Halpha\ and \Hbeta\ respectively, both in the core-fitting and full-line fitting. Figures~\ref{fig:modeling_sketch_core} and \ref{fig:modeling_sketch_full} show a sketch of the best fits for \Halpha\ and \Hbeta, in both the core-fitting and full-line fitting cases.

\subsection{UV modeling}

\begin{figure*}[t]
    \includegraphics[width=\linewidth]{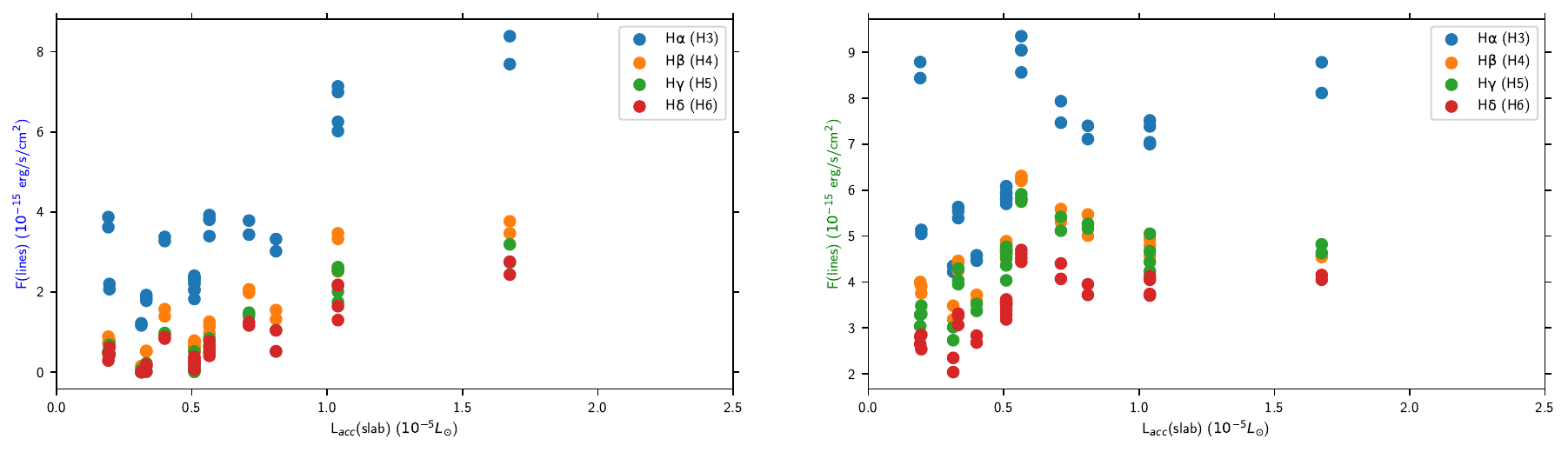}
    \caption{Comparison between the total line flux of each component (wings/core) and the accretion luminosity derived from the UV excess slab-modeling (see Sect.~\ref{sec:UV_modeling}).}
    \label{fig:appendix:uv_lacc_to_line_flux_comparison}
\end{figure*}

Figure~\ref{fig:appendix:uv_lacc_to_line_flux_comparison} shows the comparison between the total line flux of each component (wings/core) and the accretion luminosity derived from the UV excess slab-modeling, allowing to compare to relations from \cite{alcala_x-shooter_2017}.

\section{Looking for remaining uncertainties on the flux calibration of \delorme}
\label{sec:fluxcal}
\begin{figure}
    \centering
    \includegraphics[width=\columnwidth]{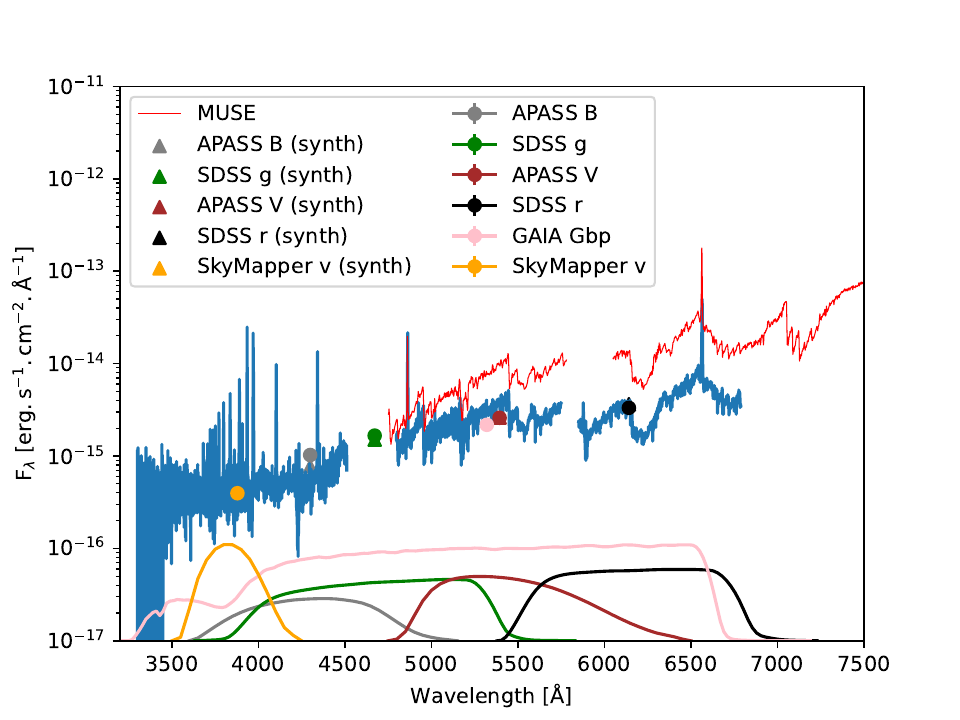}
    \caption{UVES spectrum of Delorme 1AB (blue) compared to the available photometry of the binary at optical and near-UV wavelengths.}
    \label{fig:compsynthphot}
\end{figure}

We show in Figure \ref{fig:compsynthphot} a comparison of the spectrum of Delorme 1AB to the published photometry at optical and near-UV wavelengths converted to flux densities with the VOSA\footnote{\url{http://svo2.cab.inta-csic.es/theory/vosa/}} tool \citep{2008A&A...492..277B}. We used the corresponding filter transmissions available on the Spanish Virtual Observatory\footnote{\url{http://svo2.cab.inta-csic.es/theory/fps/}} to compute synthetic fluxes from the UVES spectrum. The transmission curves of each filter is shown at bottom in the figure. The blind zones corresponding to the detector gaps and split between the red and BLUE arms were replaced by linear interpolation. 

The difference between the synthetic and observed flux densities were used to compute renormalization factors for the three separate spectra collected by the three detectors of UVES ($1.70\pm0.12$ for the BLUE arm based on the APASS B photometry, $0.86\pm0.02$ for the BLUE arm based on the SkyMapper v-band photometry, $1.27\pm0.05$ for the ship operating at short wavelengths in the RED arm based on APASS V, and $1.17\pm0.02$ for the large wavelengths of the RED arm based on SDSS r). We overlay the MUSE spectrum of the binary star extracted from the flux-calibrated datacubes over a circular aperture of 30 pixels (0.75$''$) \citep[calibration achieved with the ESO MUSE pipeline; see][]{eriksson_strong_2020}. The UVES and MUSE spectra have similar spectral features, but the MUSE spectrum is redder and overluminous compared to UVES and to the reference photometry. 

We can not exclude however that Delorme 1AB has a more variable photometry in the near-UV than in the V-band on long timescales because it is an active star and this regions concentrates bright emission lines. This is in fact suggested by the different factors found for the SkyMapper and APASS photometry. 

\begin{figure}
    \centering
    \includegraphics[width=\columnwidth]{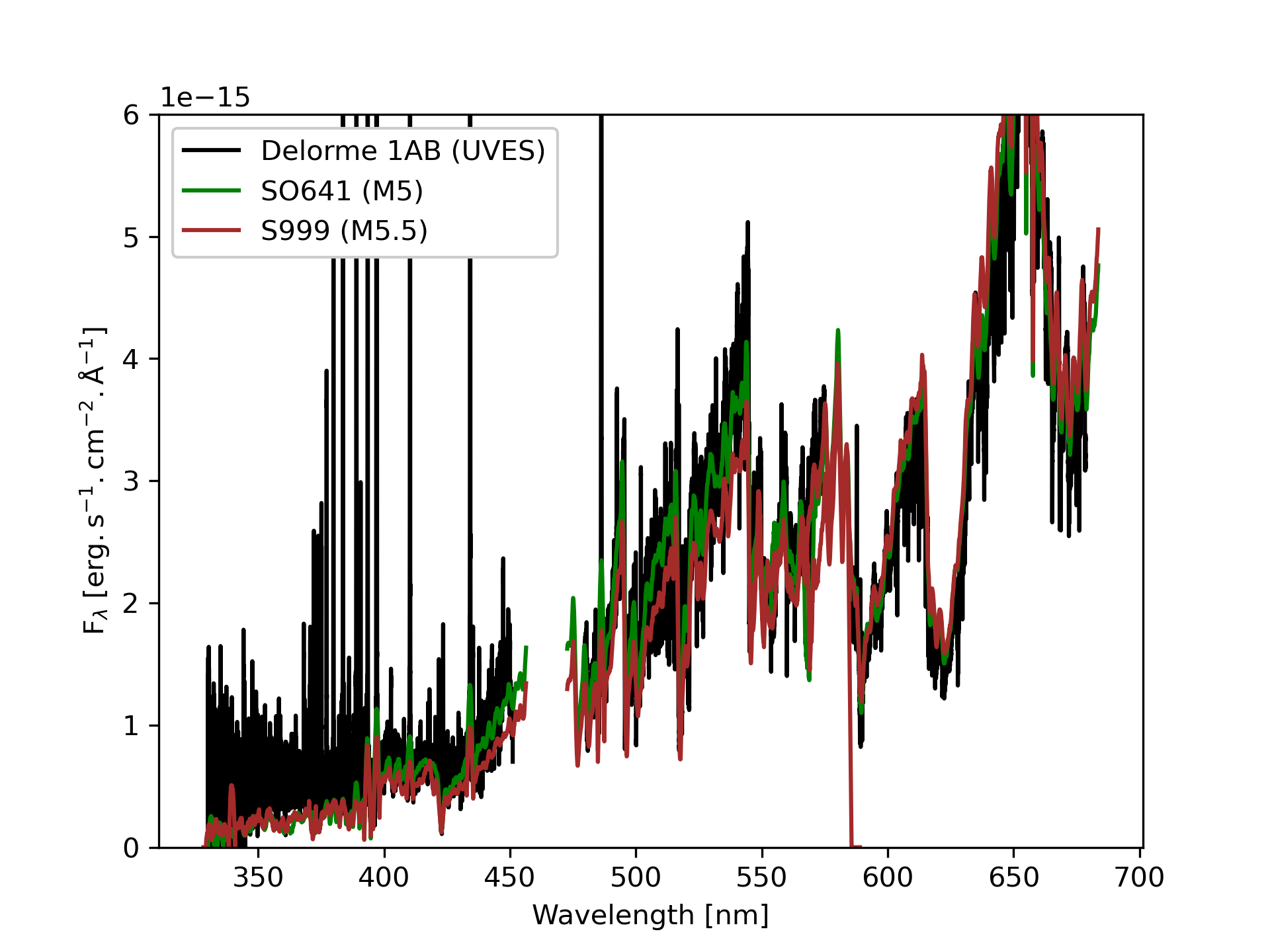}
    \caption{Comparison of the UVES spectrum of Delorme 1 AB to the VLT/X-SHOOTER spectra of class III dwarfs renormalized to Delorme 1 AB mean flux from 500 to 650nm.}
    \label{Fig:M5template}
\end{figure}

We show in Figure \ref{Fig:M5template} the comparison of the UVES spectrum to the X-SHOOTER spectrum of the young M5 class III star SO641 from \cite{manara_x-shooter_2013}. The comparison reveal close spectral features, and a good correspondence of the spectral slopes longward of 430nm, confirming the M5 spectroscopic spectral type of Delorme 1AB quoted in \cite{2014AJ....147..146K} and that the scalings quoted above may be due to variability. The X-SHOOTER spectrum of the class III M5.5 star S999 from \cite{manara_x-shooter_2013} also provide a good fit to the data, although the overtone of TiO at 540nm is not as deep in the template and our BLUE arm data appears overluminous by 48\% shortward of 450nm. 

We conclude that these comparison rather confirm the reliability of the flux calibration of our data and have decided not correct them from any extra scaling parameters to avoid introducing biases.  

\end{appendix}
\end{document}

%% file: macros.tex
\newcommand{\Rjup}{\ensuremath{\mathrm{R}_{\mathrm{Jup}}}\xspace}
\newcommand{\MJup}{\ensuremath{\mathrm{M}_{\mathrm{Jup}}}\xspace}
\newcommand{\Mjup}{\MJup} 

\newcommand{\Msun}{\ensuremath{\mathrm{M}_{\odot}}}

\newcommand{\Lsun}{\ensuremath{\mathrm{L}_{\odot}}}

\newcommand{\kms}{\ensuremath{\mathrm{km/s}}\xspace}
\newcommand{\um}{$\upmu$m\xspace}
\newcommand{\Mdot}{\ensuremath{\dot{M}}\xspace}

\newcommand{\ergsang}{erg/s/cm$^2$/\AA}

\newcommand{\MdotUJ}{\ensuremath{\Mjup/\mbox{yr}}\xspace}
\newcommand{\MdotUS}{\ensuremath{\Msun/\mbox{yr}}\xspace}

\newcommand{\Lacc}{\ensuremath{L_{\textrm{acc}}}\xspace}

\newcommand{\Tcol}{\ensuremath{T_{\textrm{col}}}\xspace}
\newcommand{\ffill}{\ensuremath{f_{\textrm{fill}}}\xspace}
\newcommand{\Mp}{\ensuremath{M_{\textrm{p}}}\xspace}
\newcommand{\Rp}{\ensuremath{R_{\textrm{p}}}\xspace}
\newcommand{\Rin}{\ensuremath{R_{\textrm{in}}}\xspace}
\newcommand{\Teff}{\ensuremath{T_{\textrm{eff}}}\xspace}

\newcommand{\PaBeta}{\ensuremath{\mathrm{Pa}{\beta}}\xspace}
\newcommand{\Halpha}{\ensuremath{\mathrm{H}{\alpha}}\xspace}
\newcommand{\Hbeta}{\ensuremath{\mathrm{H}{\beta}}\xspace}
\newcommand{\Hgamma}{\ensuremath{\mathrm{H}{\gamma}}\xspace}
\newcommand{\Hdelta}{\ensuremath{\mathrm{H}{\delta}}\xspace}
\newcommand{\Hepsilon}{\ensuremath{\mathrm{H}{\epsilon}}\xspace}
\newcommand{\Hzeta}{\ensuremath{\mathrm{H}{\zeta}}\xspace}
\newcommand{\Heta}{\ensuremath{\mathrm{H}{\eta}}\xspace}
\newcommand{\Htheta}{\ensuremath{\mathrm{H}{\theta}}\xspace}

\newcommand{\BrGamma}{\ensuremath{\mathrm{Br}{\gamma}}\xspace}
\newcommand{\PaGamma}{\ensuremath{\mathrm{Pa}{\gamma}}\xspace}

\newcommand{\HI}{\ensuremath{\mathrm{H\,\textsc{i}}}\xspace}

\newcommand{\Ha}{\Halpha} 

\newcommand{\gqlupb}{GQ~Lup~b\xspace}
\newcommand{\gsc}{GSC~06214-00210~b\xspace}

\newcommand{\delorme}{Delorme~1~(AB)b\xspace}
\newcommand{\dlrb}{\delorme} 

\renewcommand{\deg}{\ensuremath{^{\circ}}}

\newcommand{\GBP}{\ensuremath{G_{\mathrm{BP}}}\xspace}

\makeatletter
\let\jnl@style=\rm
\def\ref@jnl#1{{\jnl@style#1}}

\def\aj{\ref@jnl{AJ}}                   
\def\actaa{\ref@jnl{Acta Astron.}}      
\def\araa{\ref@jnl{ARA\&A}}             
\def\apj{\ref@jnl{ApJ}}                 
\def\apjl{\ref@jnl{ApJ}}                
\def\apjs{\ref@jnl{ApJS}}               
\def\ao{\ref@jnl{Appl.~Opt.}}           
\def\apss{\ref@jnl{Ap\&SS}}             
\def\aap{\ref@jnl{A\&A}}                
\def\aapr{\ref@jnl{A\&A~Rev.}}          
\def\aaps{\ref@jnl{A\&AS}}              
\def\azh{\ref@jnl{AZh}}                 
\def\baas{\ref@jnl{BAAS}}               
\def\bac{\ref@jnl{Bull. astr. Inst. Czechosl.}}
\def\caa{\ref@jnl{Chinese Astron. Astrophys.}}
\def\cjaa{\ref@jnl{Chinese J. Astron. Astrophys.}}
\def\icarus{\ref@jnl{Icarus}}           
\def\jcap{\ref@jnl{J. Cosmology Astropart. Phys.}}
\def\jrasc{\ref@jnl{JRASC}}             
\def\memras{\ref@jnl{MmRAS}}            
\def\mnras{\ref@jnl{MNRAS}}             
\def\na{\ref@jnl{New A}}                
\def\nar{\ref@jnl{New A Rev.}}          
\def\pra{\ref@jnl{Phys.~Rev.~A}}        
\def\prb{\ref@jnl{Phys.~Rev.~B}}        
\def\prc{\ref@jnl{Phys.~Rev.~C}}        
\def\prd{\ref@jnl{Phys.~Rev.~D}}        
\def\pre{\ref@jnl{Phys.~Rev.~E}}        
\def\prl{\ref@jnl{Phys.~Rev.~Lett.}}    
\def\pasa{\ref@jnl{PASA}}               
\def\pasp{\ref@jnl{PASP}}               
\def\pasj{\ref@jnl{PASJ}}               
\def\rmxaa{\ref@jnl{Rev. Mexicana Astron. Astrofis.}}%
\def\qjras{\ref@jnl{QJRAS}}             
\def\skytel{\ref@jnl{S\&T}}             
\def\solphys{\ref@jnl{Sol.~Phys.}}      
\def\sovast{\ref@jnl{Soviet~Ast.}}      
\def\ssr{\ref@jnl{Space~Sci.~Rev.}}     
\def\zap{\ref@jnl{ZAp}}                 
\def\nat{\ref@jnl{Nature}}              
\def\iaucirc{\ref@jnl{IAU~Circ.}}       
\def\aplett{\ref@jnl{Astrophys.~Lett.}} 
\def\apspr{\ref@jnl{Astrophys.~Space~Phys.~Res.}}
\def\bain{\ref@jnl{Bull.~Astron.~Inst.~Netherlands}} 
\def\fcp{\ref@jnl{Fund.~Cosmic~Phys.}}  
\def\gca{\ref@jnl{Geochim.~Cosmochim.~Acta}}   
\def\grl{\ref@jnl{Geophys.~Res.~Lett.}} 
\def\jcp{\ref@jnl{J.~Chem.~Phys.}}      
\def\jgr{\ref@jnl{J.~Geophys.~Res.}}    
\def\jqsrt{\ref@jnl{J.~Quant.~Spec.~Radiat.~Transf.}}
\def\memsai{\ref@jnl{Mem.~Soc.~Astron.~Italiana}}
\def\nphysa{\ref@jnl{Nucl.~Phys.~A}}   
\def\physrep{\ref@jnl{Phys.~Rep.}}   
\def\physscr{\ref@jnl{Phys.~Scr}}   
\def\planss{\ref@jnl{Planet.~Space~Sci.}}   
\def\procspie{\ref@jnl{Proc.~SPIE}}   

\def\ptp{\ref@jnl{Prog.~Th.~Phys.}}   

\makeatother